\documentclass[a4paper,aps,prd,onecolumn,preprintnumbers,showpacs,nofootinbib]{revtex4}
\usepackage{amsmath,amssymb,graphics,epsfig,subfigure}
\usepackage{color}
\usepackage{float}
\usepackage{hyperref}
\hypersetup{colorlinks=true,linkcolor=blue,citecolor=green}

\newcommand{\xmark}{\ding{55}}

\usepackage{enumerate}
\usepackage{subfigure}
\usepackage{float}
\usepackage{amssymb}
\usepackage{booktabs}
\usepackage{tabularx}

\usepackage{orcidlink}
\usepackage{pifont} %

\begin{document}

\title{A new type of multi-branch periodic orbits in dyonic black holes}

\author{Chao-Hui Wang$^{a,b,c}$}
\author{Yu-Peng Zhang$^{a,b,c}$}
\author{Tao Zhu$^{d,e}$}
\email{zhut05@zjut.edu.cn, corresponding author}
\author{Shao-Wen Wei$^{a,b,c}$}
\email{weishw@lzu.edu.cn, corresponding author}
\affiliation{
$^{a}$Lanzhou Center for Theoretical Physics, Key Laboratory of Theoretical Physics of Gansu Province, School of Physical Science and Technology, Lanzhou University, Lanzhou 730000, People's Republic of China.\\
$^{b}$Institute of Theoretical Physics $\&$ Research Center of Gravitation, Lanzhou University, Lanzhou 730000, People's Republic of China.\\
$^{c}$ Gansu Provincial Research Center for Basic Disciplines of Quantum Physics, Lanzhou University, Lanzhou 730000, People's Republic of China.\\
$^{d}$Institute for Theoretical Physics and Cosmology, Zhejiang University of Technology, Hangzhou, 310023, People's Republic of China.\\
$^{e}$United Center for Gravitational Wave Physics (UCGWP), Zhejiang University of Technology, Hangzhou, 310023, People's Republic of China.
}

\begin{abstract}
We investigate bound timelike periodic orbits in dyonic black hole spacetimes arising from quasi-topological electromagnetism.
By varying the coupling parameter $\alpha_1$, we show that the exterior monotonicity of the metric function, rather than the number of horizons, controls the topology of the radial effective potential, which can exhibit either a single well or multiple wells separated by potential barriers.
When $f(r)$ is non-monotonic outside the event horizon, the effective potential develops multiple wells, leading to multiple MBO branches and several coexisting periodic-orbit branches with the same rational number $q$.
These branches are topologically equivalent but geometrically distinct, because they correspond to different energies or angular momenta, leading to different radial extents and eccentricities.
In particular, bound periodic orbits with $E>1$ can occur, and up to three branches may coexist. We also find an inverted radial response: the innermost branch becomes more circular as the energy or angular momentum increases, whereas the outer branches become more eccentric. By contrast, when $f(r)$ is monotonic outside the event horizon, the effective potential has a single well and only one periodic orbit branch exists, even for black holes with multiple horizons. Our results identify metric non-monotonicity as the geometric origin of multi-branch periodic motion and suggest a timelike counterpart to the multiple photon ring signatures of nonstandard black hole geometries.
\end{abstract}



\maketitle

\section{Introduction}\label{sec1}

The dynamics of test particle motion in black hole spacetimes provide a fundamental
probe of the underlying geometry and play a central role in observable strong-gravity
phenomena, including gravitational waves \cite{LIGOScientific:2016aoc, LIGOScientific:2016sjg} and
black hole shadows \cite{EventHorizonTelescope:2019dse, EventHorizonTelescope:2022wkp}. In Kerr
spacetime, the introduction of Boyer--Lindquist coordinates \cite{Boyer:1966qh} and the separability
of the Hamilton--Jacobi equation \cite{Carter:1968ks} establish complete integrability of geodesic
motion, forming the foundation of relativistic orbital dynamics. Characteristic structures such as
the innermost stable circular orbit (ISCO) and the marginally bound orbit (MBO) \cite{Bardeen:1972fi},
together with the classification of bound trajectories \cite{Wilkins:1972rs, Schmidt:2002qk, Fujita:2009bp}, play an
essential role in black hole astrophysics \cite{Momynov:2024bhn}, accretion physics \cite{Boshkayev:2021chc}, and
extreme mass ratio inspirals (EMRIs) \cite{Healy:2009zm}. Comprehensive treatments of black hole geodesics can be found in the
classic references \cite{Misner:1973prb, Chandrasekhar:1984siy}.

Beyond Kerr geometry, geodesic motion has been extensively investigated in charged and non-Kerr spacetimes, including Reissner--Nordstr\"om and Kerr--Newman black holes \cite{Pugliese:2010ps, Pugliese:2011py, Pugliese:2011xn, Pugliese:2013zma}, rotating boson stars \cite{Grandclement:2014msa, Grould:2017rzz, Teodoro:2020kok}, wormholes \cite{Muller:2008zza}, hairy black holes \cite{Teodoro:2021ezj, Cunha:2015yba}, and higher-dimensional black holes \cite{Gibbons:1999uv, Herdeiro:2000ap, Diemer:2013fza}. These studies revealed a variety of novel orbital structures absent in Kerr spacetime, including semi-orbits, pointy-petal trajectories, and static light points \cite{Delgado:2021jxd, Lehebel:2022yyz}.
These studies show that departures from Kerr geometry can substantially enrich geodesic dynamics. The present work explores a complementary possibility: whether the non-monotonic exterior metric function of dyonic black holes in quasi-topological electromagnetism can produce multiple radial effective-potential wells and hence qualitatively multi-branch families of bound timelike periodic orbits.

Among all bound trajectories, periodic orbits occupy a particularly important
position because they provide a discrete skeleton of phase space and a systematic
classification of relativistic motion through rational frequency ratios
\cite{Levin:2008mq}. Such orbits are characterized by rational relations between the
radial and azimuthal frequencies and can be uniquely labeled by a rational number $q$.
In standard black hole spacetimes such as Schwarzschild and Kerr, periodic orbits
form a single continuous branch for fixed conserved quantities \cite{Levin:2008mq,
Levin:2008ci, Levin:2008yp, Levin:2009sk, Grossman:2011ps, Lim:2024mkb}, reflecting
the underlying single-well structure of the effective potential. Periodic orbit
theory naturally captures relativistic zoom--whirl behavior and has been widely
applied to EMRI dynamics and gravitational wave modeling \cite{Glampedakis:2002ya,
Grossman:2011im}. In particular, burst-like waveform features are closely related to
the rational parameter $q$ \cite{Healy:2009zm}, providing a direct connection
between orbital topology and gravitational wave observables \cite{Wang:2025hla}.
Periodic orbit analyses have also been extended to charged black
holes \cite{Misra:2010pu}, black bounces \cite{Zhou:2020zys}, Kehagias--Sfetsos
black holes in deformed Ho\v{r}ava--Lifshitz gravity \cite{Wei:2019zdf},
quantum-corrected black holes \cite{Jiang:2024cpe, Yang:2024lmj, Tu:2023xab}, naked
singularities \cite{Babar:2017gsg}, and other modified gravity
scenarios \cite{Chen:2025aqh, Meng:2025kzx, Meng:2024cnq, Lu:2025cxx, Alloqulov:2025bxh,
Deng:2025wzz, Haroon:2025rzx, Gogoi:2026obd}.

The existence of only a single branch of periodic orbits in these geometries is closely related to a fundamental geometric property of the spacetime, namely the monotonicity of the metric function outside the event horizon. In standard black hole spacetimes, the metric function remains monotonic outside the event horizon, leading to a single photon sphere and a single-well effective potential for both null and timelike geodesics.
Consequently, null geodesics give rise to a single photon-ring structure
\cite{Cunha:2020azh}, whereas timelike circular orbits are organized into
continuous stable and unstable branches \cite{Wei:2022mzv}.
This correspondence suggests that the orbital structure of both null and timelike particles is fundamentally controlled by the global behavior of the metric function outside the evevt horizon.

A qualitatively different situation arises when the metric function becomes non-monotonic outside the event horizon. In such cases, the effective potential can develop multiple extrema and multiple potential wells, giving rise to several photon spheres and disconnected families of bound motion. From the observational perspective, multiple photon spheres can produce multiple bright rings in black hole shadows and significantly enhance the total observed flux \cite{Gan:2021pwu, Gan:2021xdl, Guo:2022muy, Zhou:2023yaj, Ramadhan:2026ekb}. These optical signatures provide direct probes of nonstandard spacetime geometry and may impose constraints on black hole parameters through imaging observations.
While the null-geodesic manifestations of multiple photon spheres and their
associated optical signatures \cite{Gan:2021pwu, Gan:2021xdl, Guo:2022muy, Zhou:2023yaj, Ramadhan:2026ekb} and Echoes \cite{Hui:2023ibl}  have been extensively investigated
,
recent work has shown that multi-well geometries can support multiple
coexisting branches of periodic orbits and that topologically equivalent
branches may generate distinguishable gravitational-wave signatures in EMRI
systems \cite{Wang:2026ncv}. Nevertheless, a systematic analysis of how the
exterior monotonicity of the metric function and the horizon structure control
the existence of such multi-branch bound timelike periodic orbits in dyonic
black hole spacetimes is still needed.

A particularly suitable framework for addressing these questions is provided by dyonic black holes in quasi-topological electromagnetism \cite{Liu:2019rib}. More generally, dyonic black hole solutions have been constructed in theories involving dilaton fields, multiple Abelian gauge sectors, and nonlinear electrodynamics \cite{Abishev:2015pqa, Abishev:2017hmd, Bronnikov:2026kgl}. These systems exhibit rich geometrical and dynamical structures beyond conventional charged black holes. In the quasi-topological electromagnetic theory considered in Ref.~\cite{Liu:2019rib}, the interplay between electric and magnetic charges and higher-order electromagnetic couplings can render the metric function non-monotonic outside the event horizon. Remarkably, depending on the model parameters, these black holes may admit multiple horizons and up to three photon spheres \cite{Liu:2019rib, Wei:2020rbh}. The photon sphere structure of related dilatonic dyonic black holes has also been investigated independently in Ref.~\cite{Ivashchuk:2025rnc}. 
Electromagnetic duality can relate electric, magnetic, and dyonic configurations with identical spacetime dynamics in the source-free Einstein--Maxwell theory \cite{Ferreira:2026ldq}. However, this degeneracy does not directly apply to the quasi-topological dyonic black hole considered here, since its metric function depends explicitly on the electric charge, magnetic charge, and quasi-topological coupling. Therefore, the multi-branch periodic orbits studied below are controlled by the radial behavior of this specific metric function.

In addition to their optical properties, dyonic black holes exhibit a variety of unconventional orbital phenomena, including
static timelike orbits \cite{Wei:2023bgp}, the Aschenbach effect \cite{Wei:2023fkn}, paired stable and unstable circular
orbits \cite{Ye:2023gmk}, geometrically thick equilibrium tori \cite{Zhou:2023yaj}, and nontrivial stability structures of
circular geodesics \cite{Boshkayev:2024zxq}.  These systems may also violate the universal chaos bound \cite{Lei:2020clg}, generate
gravitational echoes without artificial boundary conditions \cite{Huang:2021qwe}, and exhibit a triple point phase structure from
a thermodynamic perspective \cite{Li:2022vcd, Chen:2024sow}. 
Thermodynamic properties of dyonic black holes in nonlinear electrodynamics, including Born--Infeld theory, have also been investigated in detail \cite{Panahiyan:2018gzv}. These studies provide complementary information about the thermodynamic phase structure of dyonic black holes, while the present work focuses on the geodesic dynamics of massive test particles in a fixed quasi-topological dyonic black hole background.
These results indicate a strong interplay between spacetime geometry and particle dynamics and suggest that periodic orbits in such geometries may display qualitatively new features absent in standard black hole spacetimes. Related studies of periodic orbits and their gravitational-wave radiation in modified black hole backgrounds further show that such orbital structures can encode strong-field information in observable waveforms \cite{Yang:2024lmj, Tu:2023xab, Chen:2025aqh, Meng:2025kzx, Lu:2025cxx, Alloqulov:2025bxh,
Deng:2025wzz, Haroon:2025rzx, Gogoi:2026obd, Wang:2026ncv}.

Motivated by these observations, in this work we investigate bound timelike periodic orbits in dyonic black hole spacetimes with
non-monotonic metric functions outside the event horizon. We demonstrate that the associated multi-well effective potential gives
rise to multiple coexisting periodic orbits for the same rational number $q$, leading to distinct dynamical branches of periodic
motion.
Together with recent geometric analyses of photon-sphere distributions in black hole spacetimes
\cite{Qiao:2024qsf, Qiao:2025ojr}, our results add to the growing evidence that exterior metric non-monotonicity is closely related to multiple photon spheres in the null sector and, in quasi-topological dyonic black holes, to multi-branch bound timelike periodic orbits.
Our results reveal a qualitatively new regime of orbital dynamics beyond the standard single-branch picture. In particular,
periodic orbits that are topologically equivalent can nevertheless exhibit distinct geometrical and dynamical properties, providing
new theoretical insight into gravitational wave signatures from EMRI systems and offering a potential observational probe of
non-Kerr black hole geometries \cite{Wang:2026ncv}.

This paper is organized as follows. In Sec.~\ref{sec2}, we present the dyonic black hole solutions in quasi-topological electromagnetic theory, classify the horizon structures, and derive the geodesic equations and effective potentials for massive particles. Sec.~\ref{sec3} briefly reviews the classification of periodic orbits. In Sec.~\ref{sec4}, we present our main results, distinguishing between multi-branch and single-branch periodic orbits according to whether the metric function is non-monotonic outside the event horizon. The multi-branch case (subsec.~\ref{Multibranch}), unique to dyonic black holes, is analyzed in detail, with particular emphasis on the relationships among different branches. We show that for a fixed rational number $q$ associated with a given triplet $(z,w,v)$, periodic orbits from different branches are topologically equivalent when either the particle energy or angular momentum is held fixed. Moreover, the eccentricities of the innermost and outermost periodic orbits exhibit opposite trends as these parameters vary. For comparison, Sec.~\ref{singlebranch} examines the single-branch case, demonstrating that monotonic metric functions yield a unique branch of periodic orbits, as in Schwarzschild, Reissner--Nordstr\"om, and Kerr spacetimes. Sec.~\ref{sec5} concludes with a summary and discussion.

\section{Black Hole Solution and Geodesics}\label{sec2}

\subsection{Black hole with quasi-topological electromagnetism}\label{BHs}
We consider a static and spherically symmetric black hole spacetime described by
\begin{eqnarray}\label{linele}
	ds^2=-f(r)dt^2+\frac{1}{g(r)}dr^2+r^2(d\theta^2+\sin^2\theta d\phi^2),
\end{eqnarray}
where the metric functions $f(r)$ and $g(r)$ depend only on the radial coordinate $r$.

In this work, we focus on the dyonic black hole solution proposed in Ref.~\cite{Liu:2019rib}, which arises in quasi-topological electromagnetism. The corresponding action is
\begin{equation}
\label{action}
S=\frac{1}{16\pi}\int\sqrt{-g}\,d^{4}x
\left[
R-\alpha_{1}F^{2}
-\alpha_{2}\left((F^{2})^{2}-2F^{(4)}\right)
\right],
\end{equation}
where $F^{2}=F^{\mu\nu}F_{\mu\nu}$ is the Maxwell term. The higher order invariants are
defined as
$F^{(2)}=F^{\mu}_{\;\nu}F^{\nu}_{\;\mu}=-F^{2}$ and
$F^{(4)}=F^{\mu}_{\;\nu}F^{\nu}_{\;\rho}F^{\rho}_{\;\sigma}F^{\sigma}_{\;\mu}$.
Throughout this work, we adopt geometric units with $G=c=1$.

The solution is characterized by five independent parameters,
\[
(M,Q,P,\alpha_1,\alpha_2),
\]
whose physical meanings are summarized as follows:

\begin{itemize}
\item[(1)]
$M$ is the black hole mass and sets the overall length scale of the spacetime.

\item[(2)]
$Q$ and $P$ denote the electric and magnetic charges, respectively. The simultaneous presence of both charges makes the solution dyonic. In the limit $P=0$, the magnetic sector disappears and the solution reduces to a purely electrically charged configuration.

\item[(3)]
$\alpha_1$ controls the strength of the standard Maxwell term. In the limit $\alpha_1=1$, the electromagnetic sector reduces to the usual Maxwell normalization.

\item[(4)]
$\alpha_2$ measures the strength of the quasi-topological nonlinear electromagnetic correction. The dimensionless combination $\alpha_2/M^2$ determines the deviation from the standard Einstein--Maxwell theory. In the dyonic sector, where electric and magnetic charges coexist, this coupling can modify the spacetime geometry and qualitatively change the causal structure.
\end{itemize}

It should be noted that the effect of the quasi-topological term depends on the electromagnetic configuration. For special globally polarized configurations, the higher order terms do not contribute to the Maxwell equations or to the energy--momentum tensor \cite{Liu:2019rib}. The same correction is also absent in purely electric or purely magnetic limits. However, the black hole considered here is genuinely dyonic, with both electric and magnetic charges present. In this case, the quasi-topological coupling contributes nontrivially to the solution and modifies the spacetime geometry through nonlinear electromagnetic backreaction. The resulting metric function is
\begin{eqnarray}\label{metric}
f(r)=g(r)
=
1-\frac{2M}{r}
+\frac{\alpha_1 P^2}{r^2}
+\frac{Q^2}{\alpha_1 r^2}
\,{}_2F_1
\left[
\frac14,1;
\frac54;
-\frac{4P^2\alpha_2}{r^4\alpha_1}
\right],
\end{eqnarray}
where ${}_2F_1$ is the hypergeometric function.

Several familiar black hole solutions are recovered in appropriate limits. For $\alpha_2=0$ and $\alpha_1=1$, Eq.~(\ref{metric}) reduces to the Reissner--Nordstr\"om solution. The Schwarzschild spacetime is further recovered by setting $Q=P=0$. Therefore, the parameters $(Q,P,\alpha_1,\alpha_2)$ collectively quantify deviations from standard Einstein--Maxwell black holes.

As shown in Ref.~\cite{Liu:2019rib}, the null, weak, and dominant energy conditions are satisfied for $\alpha_1>0$ and $\alpha_2/M^2>0$, whereas the strong energy condition is violated. In the present work, we focus on the parameter region satisfying these physically reasonable conditions.

\begin{figure}[http]
	\center{
	\subfigure[$\alpha_2$ vs $\alpha_1$]{\label{Fig:para}
	\includegraphics[width=6.0cm]{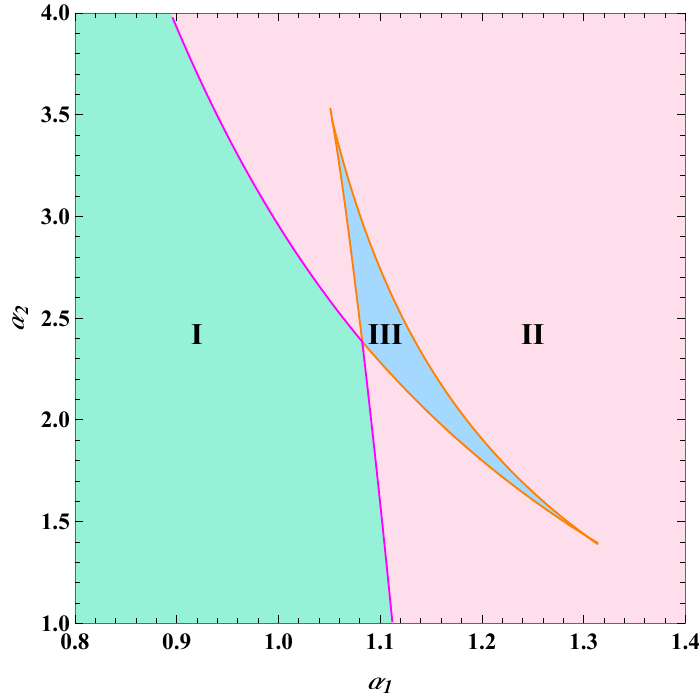}}
	\subfigure[$f(r)$ vs $r$]{\label{Fig:frcases}
	\includegraphics[width=6.18cm]{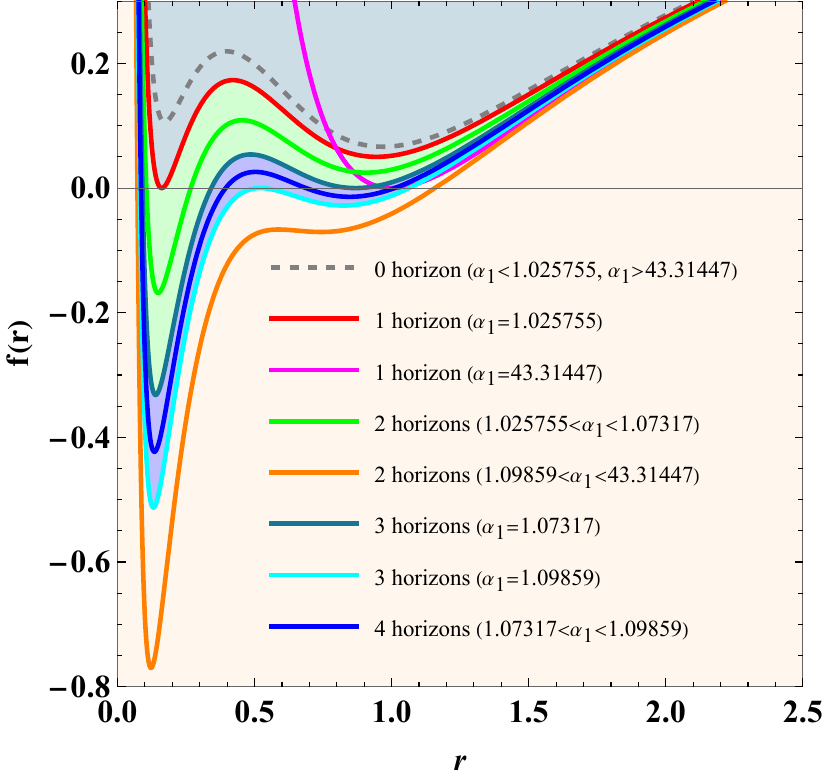}}
	}
\caption{
Fixed $M=1$, $P=0.15$, and $Q=1.05$.
(a) Parameter space structure in the $(\alpha_1,\alpha_2)$ plane classified by the number of horizons.
Region I corresponds to naked singularities, the purple curve to degenerate one-horizon configurations, region II to two-horizon black holes, the orange curve to critical three-horizon configurations, and region III to four-horizon black holes.
(b) Representative profiles of the metric function $f(r)$ for $\alpha_2=2.76$ and different values of $\alpha_1$.
}
\label{solutionscases}
\end{figure}

The spacetime structure depends sensitively on the parameters
$(\alpha_1,\alpha_2)$, especially on the Maxwell-sector coupling
$\alpha_1$ and the quasi-topological electromagnetic coupling
$\alpha_2$.
In the present work, our primary goal is to investigate how the
exterior behavior of the metric function influences the structure of
bound periodic orbits. To isolate this effect and avoid unnecessary
degeneracies in the parameter space, we focus on a representative
one-parameter slice by fixing
\[
M=1,\qquad P=0.15,\qquad Q=1.05,\qquad \alpha_2=2.76,
\]
and varying only the parameter $\alpha_1$ throughout the subsequent
analysis.
The resulting parameter space structure and representative profiles of
$f(r)$ are shown in Fig.~\ref{solutionscases}.

It is important to distinguish the horizon number from the monotonicity of the metric function outside the event horizon. These two properties are related but not equivalent. In the parameter region where the spacetime has three or four horizons, the metric function is monotonic outside the outer event horizon. By contrast, in the one- and two-horizon sectors, both monotonic and non-monotonic exterior behaviors can occur, depending on the specific values of $(\alpha_1,\alpha_2)$.

Along the representative slice $\alpha_2=2.76$ shown in Fig.~\ref{Fig:frcases},
the metric function exhibits the following horizon structure:
\begin{itemize}
\item[(1)] \textit{No horizon}: for $\alpha_1<1.025755$ or $\alpha_1>43.31447$, the equation $f(r)=0$ has no positive root, and the spacetime contains a naked singularity.

\item[(2)] \textit{One horizon}: at $\alpha_1=1.025755$ and $\alpha_1=43.31447$, two roots merge and the spacetime has a degenerate horizon.

\item[(3)] \textit{Two horizons}: for $1.025755<\alpha_1<1.07317$ and for $1.09859<\alpha_1<43.31447$, the spacetime has two horizons.

\item[(4)] \textit{Three horizons}: at $\alpha_1=1.07317$ and $\alpha_1=1.09859$, the metric function has three positive roots.

\item[(5)] \textit{Four horizons}: for $1.07317<\alpha_1<1.09859$, the spacetime has four horizons.
\end{itemize}

The exterior behavior of the metric function along this slice is more subtle. For
\[
1.025755\leq \alpha_1<1.07317,
\]
the metric function is non-monotonic outside the event horizon. This includes the degenerate one-horizon point at $\alpha_1=1.025755$ and the adjacent two-horizon black holes. In this interval, $f(r)$ develops exterior extrema, which will later generate a multi-well effective potential.

For
\[
\alpha_1\geq 1.07317,
\]
the metric function is monotonic outside the outer event horizon. This includes the three- and four-horizon cases in the narrow interval
\[
1.07317\leq \alpha_1\leq 1.09859,
\]
as well as the second two-horizon branch and the degenerate one-horizon endpoint for
\[
1.09859<\alpha_1\leq 43.31447.
\]
Therefore, the existence of three or four horizons does not imply exterior non-monotonicity. Rather, for the representative slice considered here, the multi-well orbital structure arises only in the one- and two-horizon sector satisfying $1.025755\leq\alpha_1<1.07317$.

This distinction is central to the present work. The number of horizons characterizes the causal structure of the spacetime, whereas the exterior monotonicity of $f(r)$ controls the topology of the effective potential for timelike motion. As will be shown below, non-monotonic exterior behavior leads to multiple effective potential wells and hence to multi-branch periodic orbits, while monotonic exterior behavior leads to a single-well potential and only one periodic orbit branch.

\subsection{Geodesics and effective potential}\label{Geo}

The motion of a massive test particle in the dyonic black hole spacetime is governed by the Lagrangian
\begin{eqnarray}\label{Lagrangian}
\mathcal{L}=\frac{1}{2}g_{\mu\nu}\frac{dx^{\mu}}{d\tau}\frac{dx^{\nu}}{d\tau}=-\frac{1}{2},
\end{eqnarray}
where $\tau$ denotes the proper time. Owing to spherical symmetry, the motion can be restricted to the equatorial plane $\theta=\pi/2$ without loss of generality. The generalized momenta are defined as
\begin{eqnarray}\label{momentum}
p_{\mu}=\frac{\partial \mathcal{L}}{\partial \dot{x}^{\mu}}.
\end{eqnarray}
The spacetime admits two Killing vectors, $\xi^{\mu}=(\partial_t)^{\mu}$ and $\psi^{\mu}=(\partial_{\phi})^{\mu}$, leading to two conserved quantities along geodesics: the energy $E$ and angular momentum $L$ per unit mass,
\begin{eqnarray}\label{conqua}
p_t=-f(r) \dot{t}=-E, \quad p_{\phi}=r^2 \dot{\phi}=L,
\end{eqnarray}
The normalization of the four-velocity,
\begin{eqnarray}
g_{\mu\nu} \dot{x}^\mu \dot{x}^\nu = -1.
\end{eqnarray}
yields the first order equations of motion:
\begin{eqnarray}\label{geo}
\dot{t}=\frac{E}{f(r)}, \quad \dot{\phi}=\frac{L}{r^2}, \quad \dot{r}^2=E^2-V_{\mathrm{eff}},
\end{eqnarray}
where the effective potential is given by
\begin{equation}\label{vefpen}
V_{\mathrm{eff}}=f(r)\left(1+\frac{L^2}{r^2}\right).
\end{equation}

Bound motion is determined by the structure of the effective potential $V_{\mathrm{eff}}$, particularly by the existence and stability of circular orbits. Among them, two characteristic orbits play a central role: the marginally bound orbit (MBO) and the innermost stable circular orbit (ISCO).

The marginally bound orbit corresponds to the critical unstable circular orbit separating bound and unbound trajectories. It is determined by
\begin{equation}\label{mbo}
V_{\mathrm{eff}}=1, \qquad \partial_r V_{\mathrm{eff}}=0.
\end{equation}
Physically, the MBO marks the onset of gravitational capture and therefore provides important information about particle capture processes and capture cross sections in strong gravitational fields. Similar applications have recently been explored in studies of gravitational capture in compact-object spacetimes \cite{Momynov:2024bhn}.

The innermost stable circular orbit (ISCO) defines the inner boundary of stable circular motion and satisfies
\begin{equation}\label{riso}
V_{\mathrm{eff}}=E^2, \qquad \partial_r V_{\mathrm{eff}}=0, \qquad \partial_{r,r}V_{\mathrm{eff}}=0.
\end{equation}
The ISCO plays a fundamental role in black hole astrophysics, since it determines the inner edge of thin accretion disks and strongly affects accretion efficiency, emitted luminosity, and disk morphology. Related applications have been extensively discussed in the context of compact object accretion models \cite{Boshkayev:2021chc}.

From the perspective of orbital dynamics, the MBO and ISCO define important reference boundaries for the parameter region supporting bound timelike motion. For a fixed angular momentum $L$, circular orbits are determined by the extrema of the same effective-potential curve $V_{\mathrm{eff}}(r,L)$, namely
$\partial_r V_{\mathrm{eff}}=0$.
A local maximum of $V_{\mathrm{eff}}$ corresponds to an unstable circular orbit, whereas a local minimum corresponds to a stable circular orbit. Bound radial motion is possible when the particle energy lies inside a potential well, with the radial turning points determined by
$E^2=V_{\mathrm{eff}}(r)$.
In a standard single-well geometry, the bound radial interval is confined by the potential barrier associated with an unstable circular orbit and by an outer radial turning point, as discussed for periodic orbits in regular hairy black hole spacetimes \cite{Meng:2024cnq}.

The MBO and ISCO should therefore not be interpreted as the two extrema of every fixed-$L$ effective-potential curve. Rather, they are characteristic circular orbits that delimit important regions in the $(E,L)$ parameter space. The MBO corresponds to the marginally bound circular orbit with $E=1$, while the ISCO is the marginally stable circular orbit separating stable and unstable circular-orbit branches through $\partial_{r,r}V_{\mathrm{eff}}=0$. Together, they provide useful reference limits for selecting the angular-momentum and energy domains in which bound periodic orbits are searched for below. This point is particularly important in the present dyonic black hole spacetime. When the metric function becomes non-monotonic outside the event horizon, $V_{\mathrm{eff}}$ may develop several extrema and multiple wells on the same fixed-$L$ curve. As a result, several disconnected radial intervals can satisfy the bound-orbit condition, giving rise to multiple branches of bound timelike periodic orbits.

This distinction is important for the study of periodic orbits. Since periodic timelike geodesics are a special class of bound motion, their existence is determined by the full effective-potential profile and by the radial turning points satisfying \(E^2=V_{\mathrm{eff}}(r)\). In practice, the MBO and ISCO provide useful reference scales for selecting the relevant energy and angular-momentum domains, while the actual radial intervals of periodic motion are fixed by the potential wells of \(V_{\mathrm{eff}}(r,L)\). As will be shown below, when the metric function is monotonic outside the event horizon, the effective potential contains a single bound radial interval for admissible parameters, leading to one periodic-orbit branch. By contrast, when the metric function becomes non-monotonic outside the event horizon, multiple extrema of \(V_{\mathrm{eff}}\) can emerge on the same fixed-\(L\) curve, generating disconnected bound radial intervals and hence multiple periodic-orbit branches.

As will be shown below, when the metric function is monotonic outside the event horizon, the effective potential contains a single bound radial interval for admissible parameters, leading to one periodic-orbit branch. By contrast, when the metric function becomes non-monotonic outside the event horizon, multiple extrema of \(V_{\mathrm{eff}}\) can emerge on the same fixed-\(L\) curve, generating disconnected bound radial intervals and hence multiple periodic-orbit branches.

\section{Periodic orbits}\label{sec3}

Periodic orbits form a special class of bound geodesics in which a test particle returns exactly to its initial position after a finite proper time \cite{Levin:2008mq}. For a static, spherically symmetric black hole, the motion of a massive particle can be confined to the equatorial plane $\theta=\pi/2$ without loss of generality, so that the dynamics involve only the radial ($r$) and azimuthal ($\phi$) directions. Any non-circular equatorial bound orbit is characterized by two fundamental frequencies: the radial frequency $\omega_r$ and the azimuthal frequency $\omega_{\phi}$.

The radial frequency $\omega_r$, which quantifies the oscillatory motion between the periastron and apastron, is defined as
\begin{equation}\label{radfre}
\omega_r=\frac{2\pi}{T_r},
\end{equation}
where $T_r$ denotes the radial period. The azimuthal frequency $\omega_{\phi}$ measures the average rate of angular advance during one radial cycle and is given by
\begin{equation}\label{angfre}
\omega_{\phi}=\frac{1}{T_r}\int_{0}^{T_{r}}\frac{d\phi}{dt}\,dt
=\frac{\Delta \phi_{r}}{T_{r}},
\end{equation}
where $\Delta \phi_{r}$ is the accumulated azimuthal angle over a single radial oscillation.

For generic bound orbits, the ratio $\omega_{\phi}/\omega_r$ is irrational. Periodic orbits arise only when this ratio is rational and can be written as \cite{Levin:2008mq}
\begin{equation}\label{grerat}
\frac{\omega_{\phi}}{\omega_r}
=\frac{\Delta \phi_r}{2\pi}
=1+q,
\end{equation}
where $q$ is a positive rational number that uniquely labels the periodic orbit. The azimuthal advance during one radial period is
\begin{eqnarray}\label{deltaphi}
\Delta\phi_r
=\oint d\phi
=2\int_{r_{p}}^{r_{a}}\frac{\dot{\phi}}{\dot{r}}\,dr
=2\int_{r_{p}}^{r_{a}}
\frac{L}{r^2\sqrt{E^2-f(r)\left(1+\frac{L^2}{r^2}\right)}}\,dr,
\end{eqnarray}
where $r_a$ and $r_p$ denote the apastron and periastron, respectively.

The rational number $q$ admits a unique decomposition of the form
\begin{equation}\label{ratnum}
q=w+\frac{v}{z},
\end{equation}
where $w\geq0$ is an integer and $v/z$ is an irreducible fraction with coprime integers $v$ and $z$ satisfying
\begin{equation}\label{vnum}
1\leq v\leq z-1.
\end{equation}
Each periodic-orbit is therefore uniquely characterized by the triplet $(z,w,v)$, or equivalently by the rational number $q$. The total azimuthal advance over a complete periodic-orbit obeys the geometric relation \cite{Levin:2008mq}
\begin{equation}\label{Accaziang}
\Delta \phi_r
=2\pi\left(1+w+\frac{v}{z}\right)
=\frac{\Delta \phi}{z},
\end{equation}
where $\Delta \phi$ denotes the total azimuthal angle accumulated during one full orbit. Consequently, the particle executes exactly $z$ radial oscillations while completing an integer number of revolutions in the azimuthal direction.
It is important to note that the rational value \(q\) is not only a frequency-ratio label, but also depends on the radial interval explored by the particle. For fixed black hole parameters, \(q\) is determined by the turning points \(r_p\) and \(r_a\) through the integral expression for \(\Delta\phi_r\). In a single-well effective potential, a given rational value usually selects one radial interval and hence one periodic-orbit branch. In a multi-well potential, however, different radial intervals may yield the same rational value \(q\). Thus, the same resonance label \(q=q(z,w,v)\) can be realized by distinct periodic orbits located at different radial distances from the black hole.

Periodic orbits play a fundamental role in the organization of bound geodesic motion in strong gravitational fields. They provide a compact description of relativistic orbital dynamics and offer valuable insight into phenomena such as zoom--whirl behavior, gravitational wave emission in EMRIs, and geometrical signatures in black hole shadows.
In spacetimes with a single-well effective potential, each rational number $q$ corresponds to a unique periodic-orbit
\cite{Levin:2008mq, Glampedakis:2002ya, Grossman:2011im, Healy:2009zm,
Wang:2025hla, Lim:2024mkb, Levin:2008ci, Levin:2008yp, Levin:2009sk,
Grossman:2011ps, Misra:2010pu, Haroon:2025rzx, Zhou:2020zys,
Wei:2019zdf, Jiang:2024cpe, Yang:2024lmj, Tu:2023xab, Babar:2017gsg,
Chen:2025aqh, Meng:2025kzx, Lu:2025cxx, Alloqulov:2025bxh, Deng:2025wzz}. However, as we will show, this one-to-one correspondence can break down in non-monotonic geometries.

\section{Result}\label{sec4}
In this section, we study periodic orbits of massive test particles in a dyonic black hole spacetime by varying the coupling parameter $\alpha_1$, which determines the number of black hole horizons. As illustrated in Fig.~\ref{Fig:frcases}, different intervals of $\alpha_1$ correspond to spacetimes with one, two, three, or four horizons, leading to qualitative changes in the underlying causal structure and geodesic dynamics.

For each horizon configuration, we systematically analyze the properties of bound geodesic motion, including the horizon locations, the radii and angular momenta of the MBO and the ISCO, as well as the allowed region in the $(E,L)$ parameter space for the bound orbits. We further characterize periodic orbits through the associated rational number $q$ and present representative orbital trajectories.

This classification elucidates how horizon transitions induced by variations in $\alpha_1$ govern the existence and structure of periodic orbits. The results provide a coherent picture of relativistic dynamics in quasi-topological dyonic black hole spacetimes and offer potential theoretical guidance for identifying observable signatures in strong-gravity environments.

\subsection{Multi-branch periodic orbits }\label{Multibranch}

\subsubsection{One and Two-Horizon case A: $1.025755\leq \alpha_1<1.07317$ }\label{12horizona}

We begin with the parameter range in which the dyonic black hole admits either a single horizon or two distinct horizons. As shown in Fig.~\ref{Fig:frcases}, the spacetime possesses a single event horizon at $\alpha_1=1.025755$, while for $1.025755<\alpha_1<1.07317$ a second horizon appears. In this interval, the metric function $f(r)$ becomes non-monotonic outside the outer horizon, leading to a qualitatively richer geodesic structure.

\begin{figure}[htbp]
\includegraphics[width=7cm]{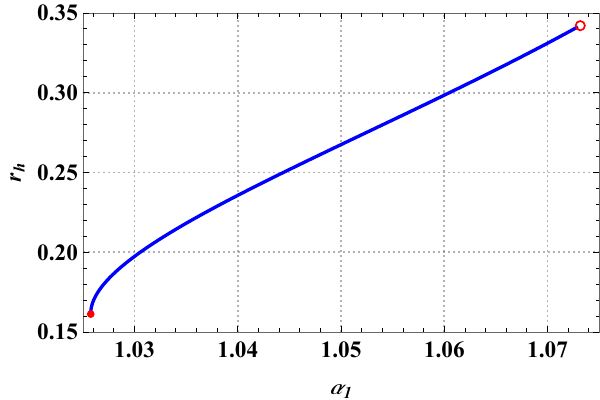}
\caption{
The outer event horizon radius $r_h$ as a function of $\alpha_1$
for fixed
$\alpha_2=2.76$,
$P=0.15$,
and
$Q=1.05$.
The red point denotes the degenerate one-horizon configuration at
$\alpha_1=1.025755$.
The blue curve corresponds to the two-horizon branch.
The open circle marks the critical three-horizon configuration at
$\alpha_1=1.07317$,
which separates the two-horizon and four-horizon phases.}
\label{12horizons}
\end{figure}

Figure.~\ref{12horizons} shows the evolution of the outer event horizon radius
with respect to $\alpha_1$. The horizon structure undergoes several qualitative
transitions as $\alpha_1$ varies, including degenerate one-horizon
configurations, two-horizon black holes, and critical three-horizon states
separating the two- and four-horizon phases. As will be shown later, these
transitions are closely related to changes in the exterior behavior of the
metric function and consequently to the topology of the effective potential
governing periodic orbits. The appearance of the critical three-horizon
configurations signals bifurcation points in the causal structure of the
spacetime.

\begin{figure}[http]
	\center{
	\subfigure[$r_{\mathrm{MBO}}$ vs $\alpha_1$]{\label{Fig:12hrmbo}
	\includegraphics[width=5.7cm]{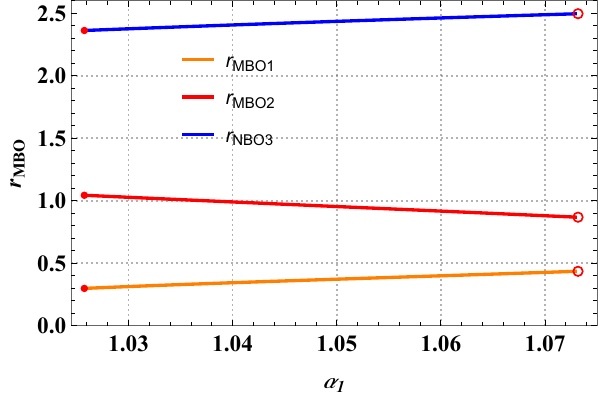}}
	\subfigure[$L_{\mathrm{MBO}}$ vs $\alpha_1$]{\label{Fig:12hLmbo}
	\includegraphics[width=5.7cm]{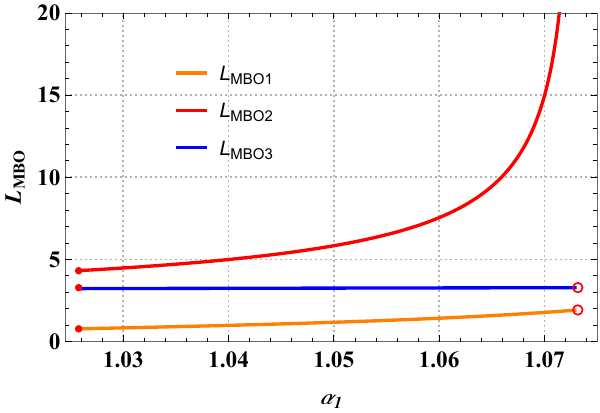}}
	}
\caption{The radius and angular momentum of MBOs as functions of $\alpha_1$ within the range $1.025755 \leq \alpha_1 < 1.07317$.}
\label{Fig:12hrLmbo}
\end{figure}

With the horizon structure established, we turn to bound geodesic motion. The MBOs, determined by Eq.~\eqref{mbo}, are shown in Fig.~\ref{Fig:12hrLmbo}. In the interval $1.025755\leq\alpha_1<1.07317$, the non-monotonicity of $f(r)$ gives rise to three distinct MBO branches. The first branch corresponds to the smallest $r_{\mathrm{MBO}}$ and $L_{\mathrm{MBO}}$, while the third branch has the largest radius. Notably, the second branch exhibits a larger radius than the first but a larger angular momentum than the third. As $\alpha_1$ increases, $r_{\mathrm{MBO1}}$ and $r_{\mathrm{MBO3}}$ shift outward, whereas $r_{\mathrm{MBO2}}$ decreases. In contrast, all three angular momenta increase monotonically.

\begin{figure*}[htbp]
	\center{
	\subfigure[$r_{\mathrm{ISCO}}$ vs $\alpha_1$ ]{\label{Fig:12hrisco}
	\includegraphics[width=5.7cm]{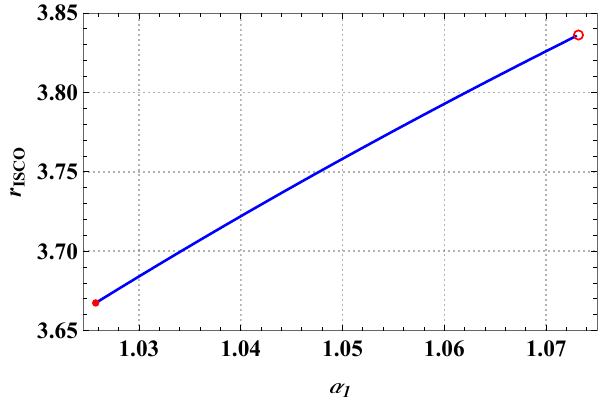}}
	\subfigure[$L_{\mathrm{ISCO}}$ vs $\alpha_1$ ]{\label{Fig:12hLisco}
	\includegraphics[width=5.7cm]{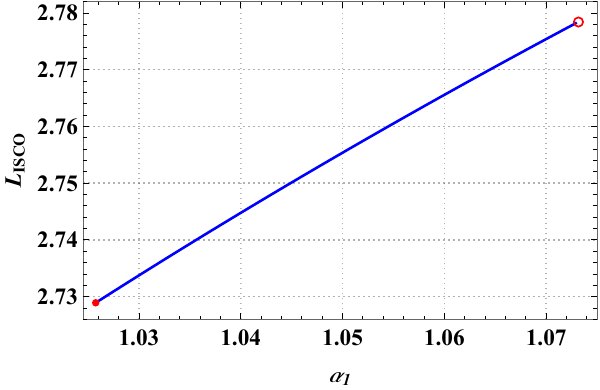}}
	\subfigure[$E_{\mathrm{ISCO}}$ vs $\alpha_1$]{\label{Fig:12hEisco}
	\includegraphics[width=5.7cm]{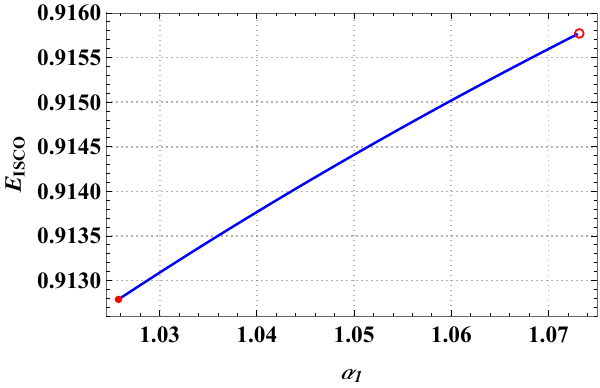}}
	}
\caption{The radius, angular momentum, and energy of ISCOs as a function of $\alpha_1$ within the range $1.025755 \leq \alpha_1 < 1.07317$.}
\label{Fig:12hrLEriso}
\end{figure*}

The properties of the ISCO, obtained from Eq.~\eqref{riso}, are shown in Fig.~\ref{Fig:12hrLEriso}. For the single-horizon case, we find
$r_{\mathrm{ISCO}}=3.667453494$,
$L_{\mathrm{ISCO}}=2.728918373$, and
$E_{\mathrm{ISCO}}=0.912789070$.
As $\alpha_1$ increases and the second horizon forms, the ISCO radius, angular momentum, and energy all increase monotonically.

These results show that the most direct consequence of the non-monotonic exterior metric function is the appearance of multiple MBO branches, which signal multiple capture or escape thresholds in the effective potential. The ISCO, on the other hand, marks the inner edge of the stable circular-orbit branch and varies smoothly with \(\alpha_1\).
Therefore, the MBO and ISCO play complementary roles: the former characterizes the marginally bound thresholds associated with the multi-branch effective-potential structure, while the latter sets the inner edge of the stable circular-orbit branch. Together, they provide the characteristic limits used below to determine the allowed parameter domain for bound periodic orbits. This combination determines the angular-momentum ranges and $(E,L)$ domains in which periodic orbits can exist.

\begin{figure*}[htbp]
	\subfigure[$\alpha_1=1.025755$]{\label{1hveff1}
	\includegraphics[width=5.7cm]{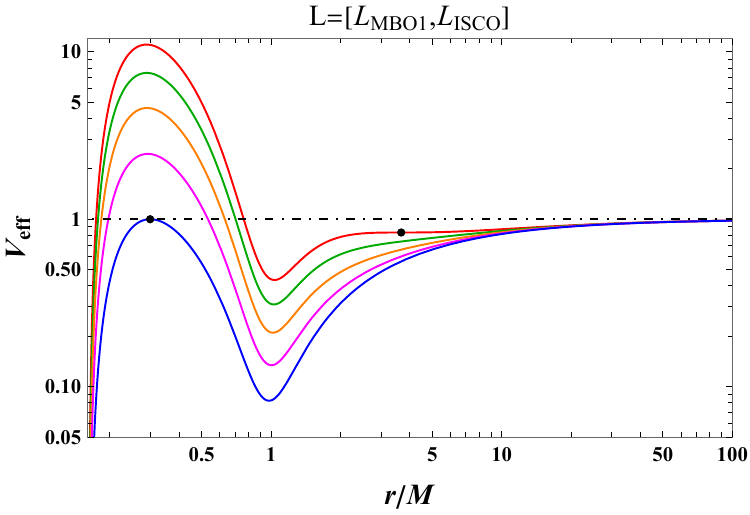}}
    \subfigure[$\alpha_1=1.025755$]{\label{1hveff2}
	\includegraphics[width=5.7cm]{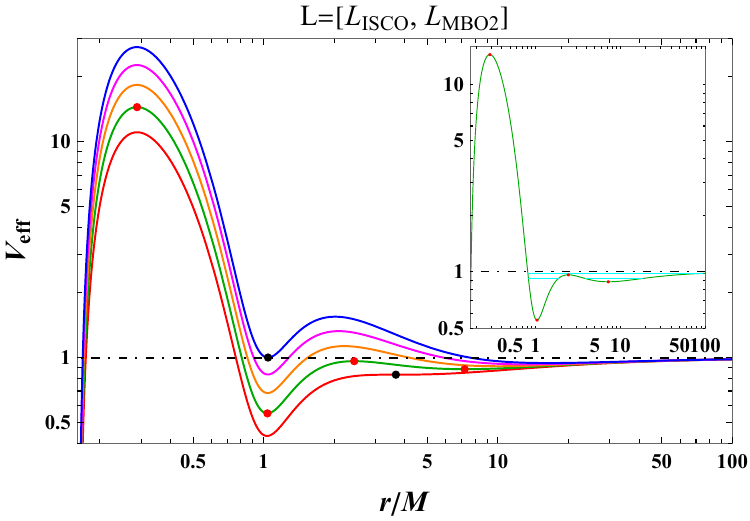}}
    \subfigure[$\alpha_1=1.025755$]{\label{1hveff3}
	\includegraphics[width=5.7cm]{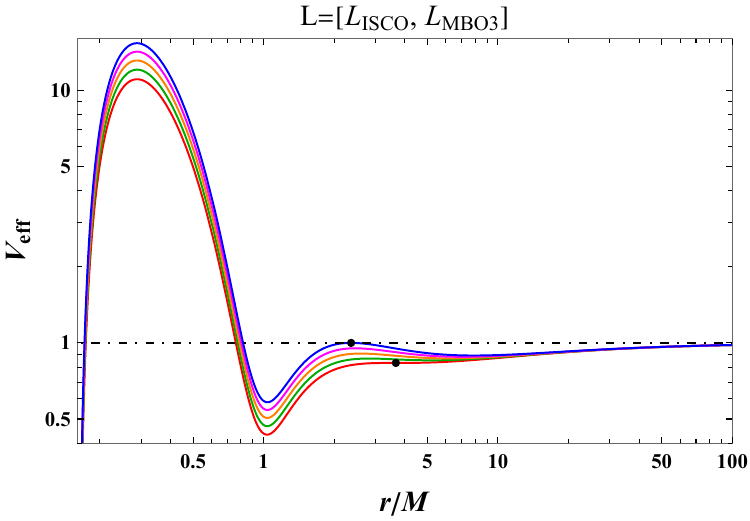}}
	\subfigure[$\alpha_1=1.0494625$]{\label{2hveff1}
	\includegraphics[width=5.7cm]{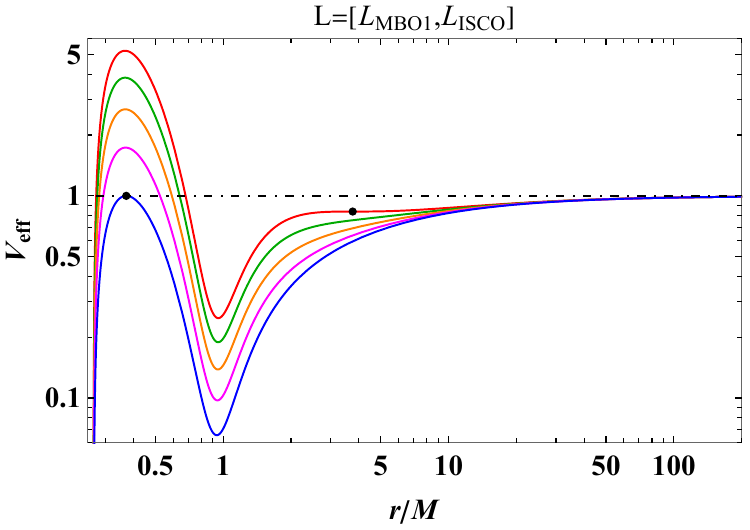}}
	\subfigure[$\alpha_1=1.0494625$]{\label{2hveff2}
	\includegraphics[width=5.7cm]{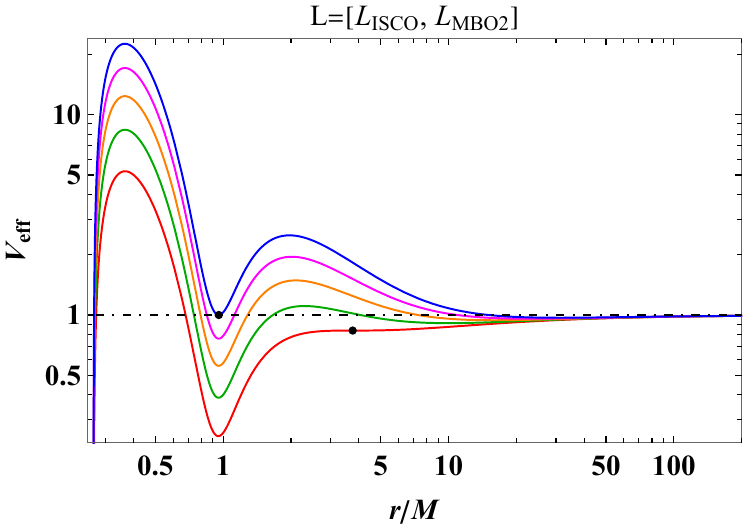}}
	\subfigure[$\alpha_1=1.0494625$]{\label{2hveff3}
	\includegraphics[width=5.7cm]{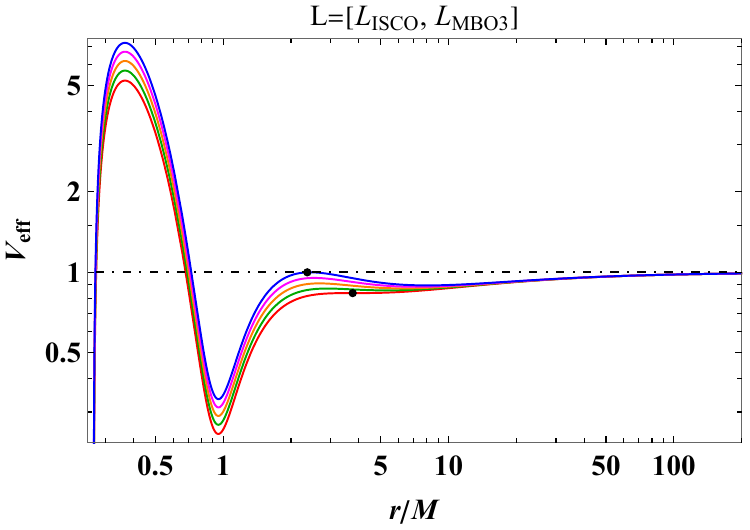}}
\caption{
Log--log plots of the effective potential $V_{\mathrm{eff}}(r)$ for massive particles moving along timelike geodesics in the dyonic black hole spacetime.
The top row corresponds to the single-horizon case $\alpha_1=1.025755$, while the bottom row corresponds to the two-horizon case $\alpha_1=1.0494625$.
The left, middle, and right columns represent the first, second, and third angular-momentum branches, respectively.
The red curves denote $L=L_{\mathrm{ISCO}}$, the blue curves denote $L=L_{\mathrm{MBO}}$, and intermediate curves correspond to $L_{\mathrm{ISCO}}<L<L_{\mathrm{MBO}}$.
Black dots mark the locations of the ISCO and MBO radii.
Extrema of $V_{\mathrm{eff}}$ correspond to timelike circular orbits, with
local maxima representing unstable circular orbits and local minima
representing stable circular orbits.
}
\label{Fig:12hveff}
\end{figure*}

The effective potential $V_{\mathrm{eff}}$, defined in Eq.~\eqref{vefpen}, depends on both the metric function $f(r)$ and the particle angular momentum $L$. Once $L_{\mathrm{MBO}}$ and $L_{\mathrm{ISCO}}$ are specified, the full structure of $V_{\mathrm{eff}}$ is fixed. Fig.~\ref{Fig:12hveff} shows the effective potentials corresponding to the three MBO branches for representative single- and two-horizon configurations.

Each column of Fig.~\ref{Fig:12hveff} corresponds to one branch of $L_{\mathrm{MBO}}$, while the top and bottom rows represent the single- and two-horizon cases, respectively. In the first branch, the ordering $L_{\mathrm{MBO1}}<L_{\mathrm{ISCO}}$ leads to an inverted dependence of $V_{\mathrm{eff}}$ on $L$, a behavior absent in spacetimes with monotonic $f(r)$. We therefore focus in the following on the second and third branches, which display standard monotonic behavior.

The resulting energy landscape exhibits a double-well structure separated by two potential barriers. For $L \in [L_{\mathrm{ISCO}}, L_{\mathrm{MBO2}}]$, the outer barrier can exceed unity, allowing bound orbits with $E>1$ to exist in the inner well. In contrast, for $L \in [L_{\mathrm{ISCO}}, L_{\mathrm{MBO3}}]$, the outer barrier reaches unity only at $L=L_{\mathrm{MBO3}}$, leading to a narrower energy window for bound motion in the outer well. For sufficiently small $L$, the potential may support up to three disconnected bound regions, implying the coexistence of multiple bound orbits.

The origin of this structure can be understood directly from the non-monotonic behavior of the metric function $f(r)$ outside the event horizon. Unlike Schwarzschild or Kerr spacetimes, where the exterior metric function is monotonic and the effective potential contains only a single well, the dyonic black hole can develop multiple extrema of $V_{\mathrm{eff}}$, producing a double-well structure.
As shown in the inset of Fig.~\ref{1hveff2}, the local maxima of the effective potential correspond to unstable timelike circular orbits, while the local minima correspond to stable timelike circular orbits. Consequently, stable and unstable circular orbits appear in pairs, in agreement with the topological analysis of Ref.~\cite{Ye:2023gmk}.
The inner minimum does not satisfy $\partial_{r,r}V_{\mathrm{eff}}=0$ and is thus identified as a \textit{marginally stable circular orbit}.

This mechanism is independent of the number of horizons itself. Instead, the essential ingredient is whether the metric function remains monotonic outside the outer event horizon. When $f(r)$ becomes non-monotonic, multiple effective-potential wells and barriers can emerge, allowing several disconnected bound regions and hence multiple branches of periodic motion.

A similar analysis applies to null geodesics by replacing the normalization condition in Eq.~\eqref{Lagrangian} from $-1$ to $0$. In this case, the non-monotonic exterior metric function can likewise generate multiple photon spheres and therefore multiple photon rings \cite{Gan:2021pwu, Gan:2021xdl, Guo:2022muy, Zhou:2023yaj,  Ramadhan:2026ekb}, in contrast to Schwarzschild or Kerr black holes, which possess only a single standard photon sphere outside the event horizon \cite{Wei:2020rbh}. These multiple photon rings are geometric properties of the spacetime itself and do not depend on the location of the observer. However, resolving them observationally would require sufficiently high angular resolution, potentially achievable through future VLBI or space-VLBI observations with long projected baselines.

Although the instability timescale of unstable circular or null geodesics can be characterized by Lyapunov exponents \cite{Cardoso:2008bp, Diaz-Guerra:2025jip}, such an analysis is distinct from the purpose of the present work. Here we focus on the existence and classification of bound timelike periodic-orbit branches generated by the multi-well effective potential, rather than on perturbation decay, quasinormal ringing, or the instability timescale of unstable photon orbits. A detailed Lyapunov-exponent analysis of the unstable circular orbits in this spacetime is therefore left for future study.

The existence of multiple effective potential wells therefore provides the geometric foundation for the emergence of multiple branches of bound timelike periodic orbits, which will be analyzed in detail in the following sections.

\begin{figure}[htbp]
	\subfigure[$L_{\mathrm{ISCO}}\leq L \leq L_{\mathrm{MBO2}}$]{\label{12hELl2}
	\includegraphics[width=5.7cm]{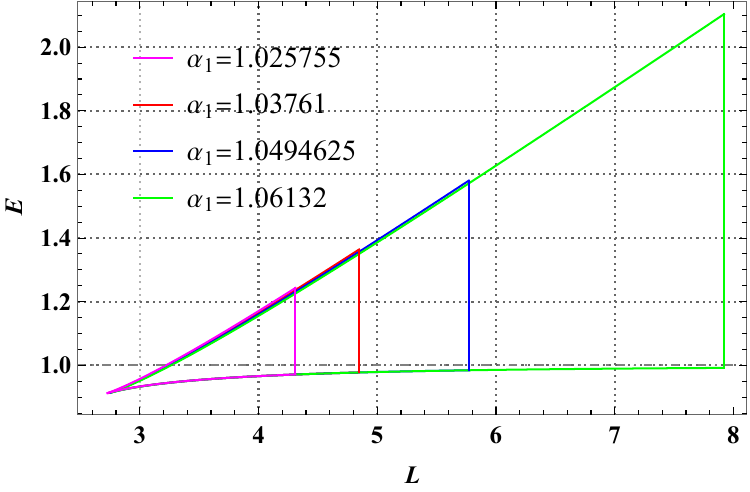}}
	\subfigure[$L_{\mathrm{ISCO}}\leq L \leq L_{\mathrm{MBO3}}$]{\label{12hELl3}
	\includegraphics[width=5.7cm]{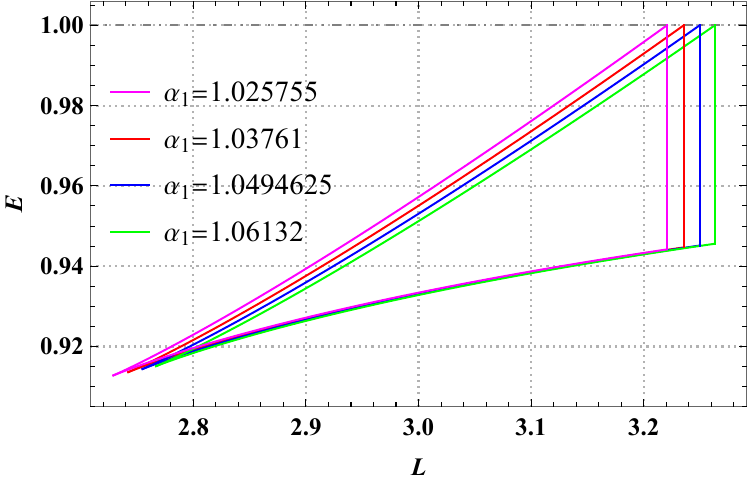}}
\caption{Allowed regions in the $(L, E)$ plane for the bound orbits around dyonic black holes for different values of $\alpha_1$.
(a) $L \in [ L_{\mathrm{ISCO}},L_{\mathrm{MBO2}} ]$. (b) $L \in [ L_{\mathrm{ISCO}},L_{\mathrm{MBO3}} ]$.}
\label{Fig:12hELp}
\end{figure}

The allowed regions for bound motion in the $(E,L)$ plane are shown in Fig.~\ref{Fig:12hELp}.
Based on conditions $\dot{r}=0$ and $\partial_r \dot{r}=0$, we numerically extract the domains permissible $(E, L)$ for the two branches defined by $L\in [L_{\mathrm{ISCO}}, L_{\mathrm{MBO2}}]$ and $L\in [L_{\mathrm{ISCO}}, L_{\mathrm{MBO3}}]$, as illustrated in Fig.~\ref{Fig:12hELp}.
Bound orbits exist only within triangular domains bounded by the ISCO and the corresponding MBO branch. These regions are consistent with the effective-potential structure and confirm the existence of bound states with $E>1$ in the second branch.

We now turn to the behavior of periodic orbits in the dyonic black hole spacetime with one or two horizons. In particular, we focus on the dependence of the rational number $q$, which characterizes periodic orbits, on the particle’s energy $E$ and angular momentum $L$. According to Eq.~\eqref{grerat}, $q$ can be written as $q=\Delta \phi_r/2\pi-1$, where $\Delta \phi_r$ is given by Eq.~\eqref{deltaphi} and depends explicitly on the conserved quantities of the particle and the background metric function $f(r)$. Since all other parameters are fixed, variations in $f(r)$ are controlled solely by the parameter $\alpha_1$.

\begin{figure}[htbp]
	\subfigure[$\alpha_1=1.025755$]{\label{1hEqb2}
	\includegraphics[width=4.1cm]{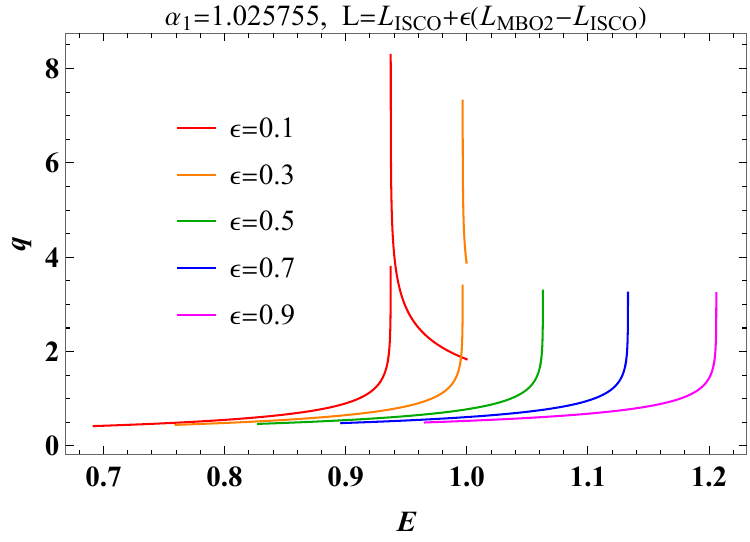}}
	\subfigure[$\alpha_1=1.025755$]{\label{1hEqb3}
	\includegraphics[width=4.1cm]{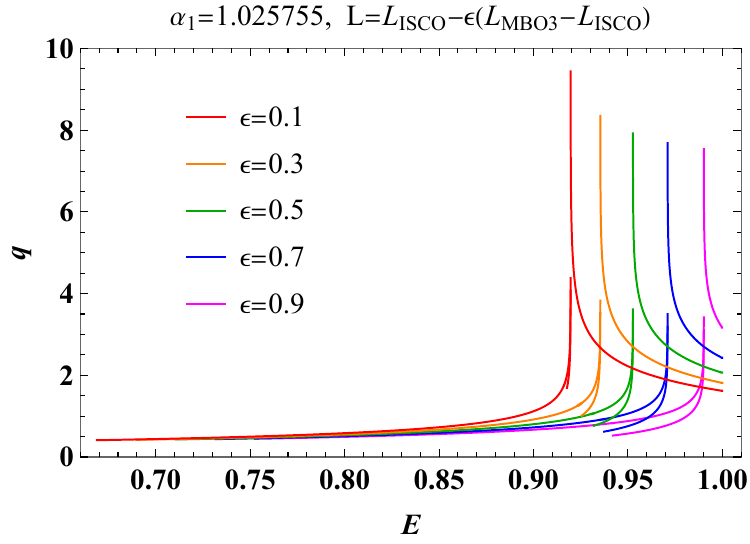}}
	\subfigure[$\alpha_1=1.0494625$]{\label{2hEqb2}
	\includegraphics[width=4.1cm]{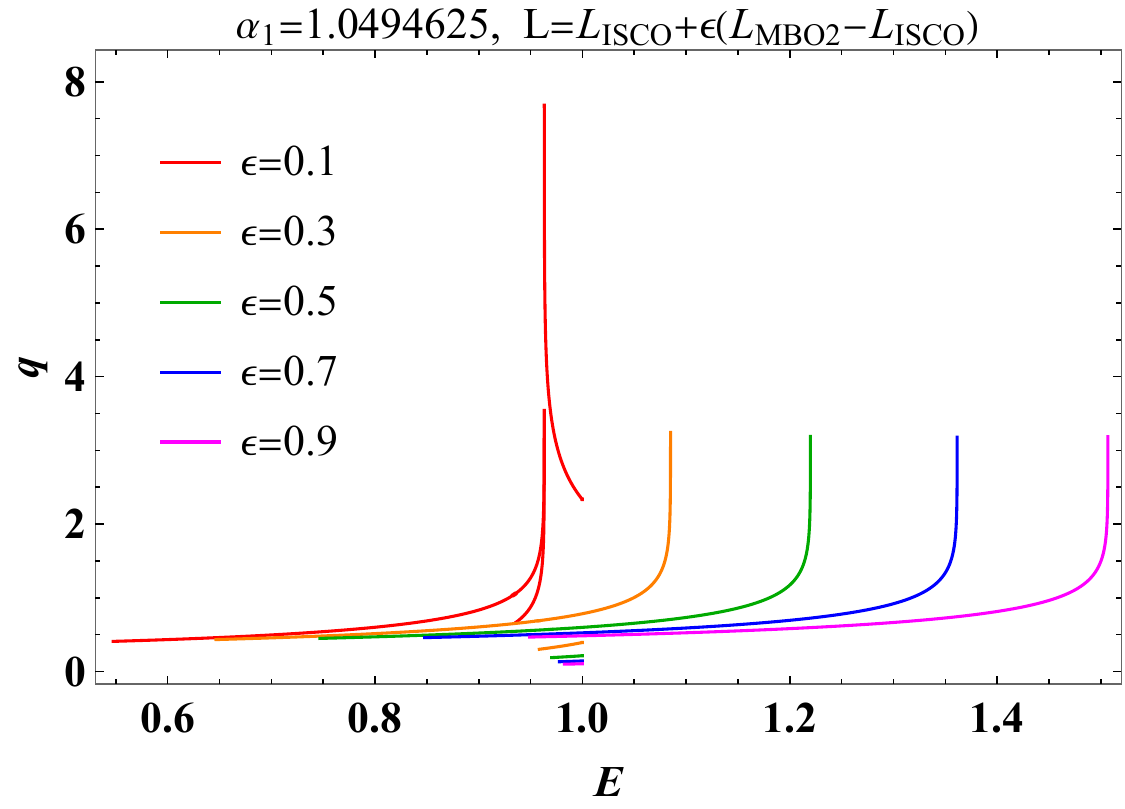}}
	\subfigure[$\alpha_1=1.0494625$]{\label{2hEqb3}
	\includegraphics[width=4.1cm]{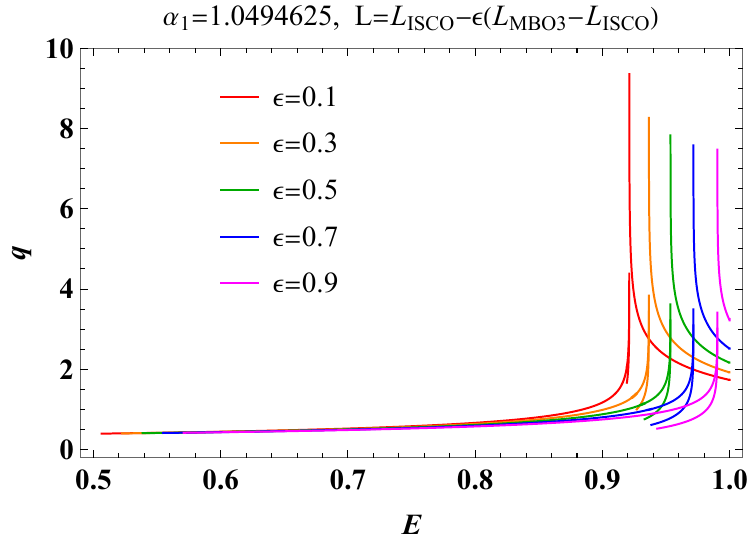}}
\caption{Energy $E$ vs. rational number $q$ for dyonic black holes. (a) and (b) Single-horizon case with $\alpha_1 = 1.025755$ and
angular momenta of MBO $L_{\mathrm{MBO2}}$ and $L_{\mathrm{MBO3}}$, respectively. (c) and (d) Double-horizon case with $\alpha_1 = 1.0494625$
and the same angular momentum.}
\label{Fig:12hEq}
\end{figure}

We now examine periodic orbits, characterized by the rational number $q$ defined in Eq.~\eqref{grerat}. The angular momentum is parametrized as
\begin{equation}\label{amp}
L=L_{\mathrm{ISCO}}+\epsilon (L_{\mathrm{MBO}}-L_{\mathrm{ISCO}}),
\end{equation}
with $\epsilon\in(0,1)$ corresponding to bound motion. Fig.~\ref{Fig:12hEq} shows the dependence of $q$ on the particle energy for representative single- and two-horizon configurations.
For small $\epsilon$, multiple branches of $q(E)$ appear, particularly for $E<1$. This behavior reflects the double-well structure of the effective potential and implies the existence of multiple periodic orbits with the same angular momentum. In the branch $L\in[L_{\mathrm{ISCO}},L_{\mathrm{MBO2}}]$, bound periodic orbits with $E>1$ are also allowed. In addition to the standard monotonic growth of $q$ with $E$, inverted branches appear in which $q$ decreases as $E$ increases, clustering toward a limiting energy as $q\to\infty$.

\begin{figure}[htbp]
	\subfigure[$\alpha_1=1.012577$]{\label{1hLq}
	\includegraphics[width=5.7cm]{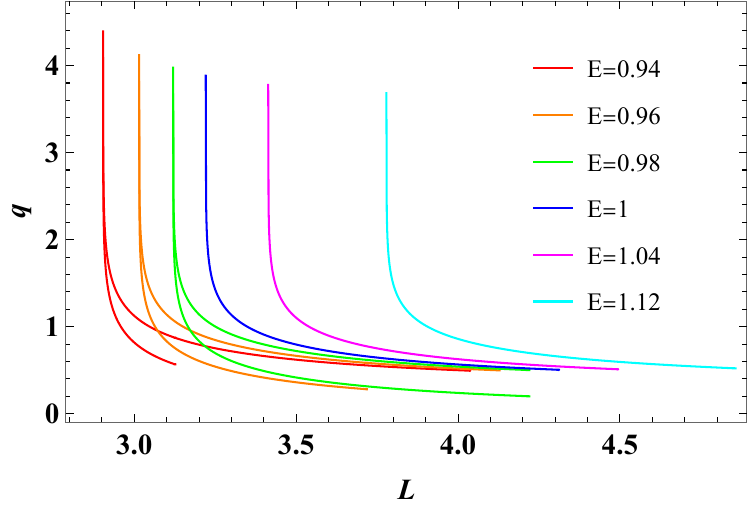}}
	\subfigure[$\alpha_1=1.0494625$]{\label{2hLq}
	\includegraphics[width=5.7cm]{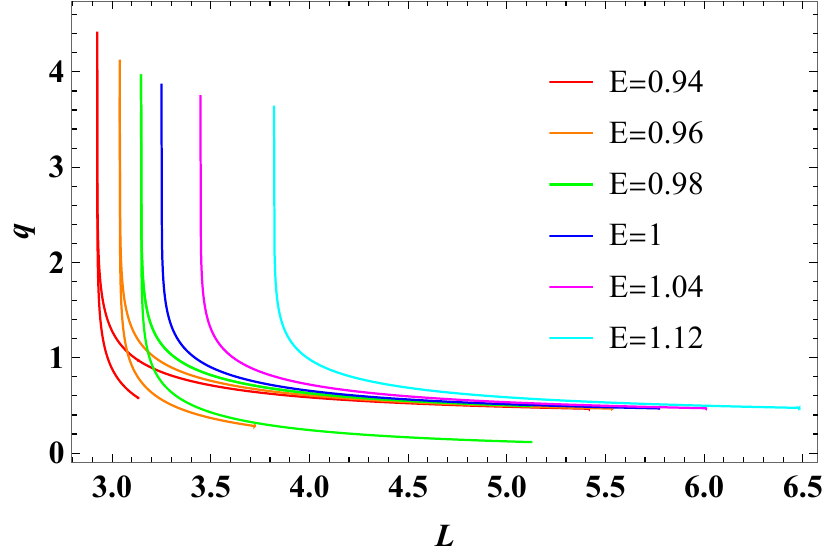}}
\caption{Angular momentum $L$ vs. rational number $q$ for dyonic black holes. (a) Single-horizon case. (b) Double-horizon case.}
\label{Fig:12hLq}
\end{figure}

The complementary dependence of $q$ on $L$ at fixed energy is shown in Fig.~\ref{Fig:12hLq}. For $E<1$, two branches of $q(L)$ are present, while for $E\geq1$ only a single branch remains. Together, these results demonstrate that $q$ encodes detailed information about both the spacetime geometry and the structure of bound motion.

Finally, specific periodic orbits are obtained by selecting rational values of $q=w+v/z$ according to Eq.~\eqref{ratnum}.
Each value of $q$ corresponds to a distinct zoom–whirl–vertex structure \cite{Levin:2008mq}. In the following subsections,
we present explicit examples of bound timelike periodic orbits for angular momentum ranges
$L\in[L_{\mathrm{ISCO}},L_{\mathrm{MBO2}}]$ and $L\in[L_{\mathrm{ISCO}},L_{\mathrm{MBO3}}]$, as well as for fixed energy, thereby illustrating the rich phenomenology induced by the non-monotonic metric function.\\

\paragraph{Periodic orbits with  single horizon}

\begin{enumerate}
  \item[a.1] Periodic orbits for  $L\in[L_{\mathrm{ISCO}},L_{\mathrm{MBO2}}]$ in the single-horizon case
\end{enumerate}

We first consider periodic orbits of massive particles in the single-horizon dyonic black hole spacetime with $\alpha_1=1.025755$. We focus on the angular-momentum interval $L\in[L_{\mathrm{ISCO}},L_{\mathrm{MBO2}}]$, for which bound orbits with $E>1$ are allowed, as indicated by Figs.~\ref{1hveff2} and \ref{12hELl2}.

Based on the $E$--$q$ relation shown in Fig.~\ref{1hEqb2}, we extract the energies associated with selected rational numbers $q$ labeled by $(z,w,v)$ for several values of the interpolation parameter $\epsilon\in(0,1)$, which parametrizes the angular momentum through Eq.~\eqref{amp}. For small $\epsilon$, such as $\epsilon=0.1$, each rational number $q$ admits two distinct energy solutions, denoted by $E_1$ and $E_2$, corresponding to two branches of bound periodic orbits. As $\epsilon$ increases, the second branch gradually disappears and only a single solution remains. Moreover, for fixed $q$, the energy required to sustain a periodic-orbit increases monotonically with the angular momentum.

\begin{figure}[htbp]
\centering
\includegraphics[width=0.85\textwidth]{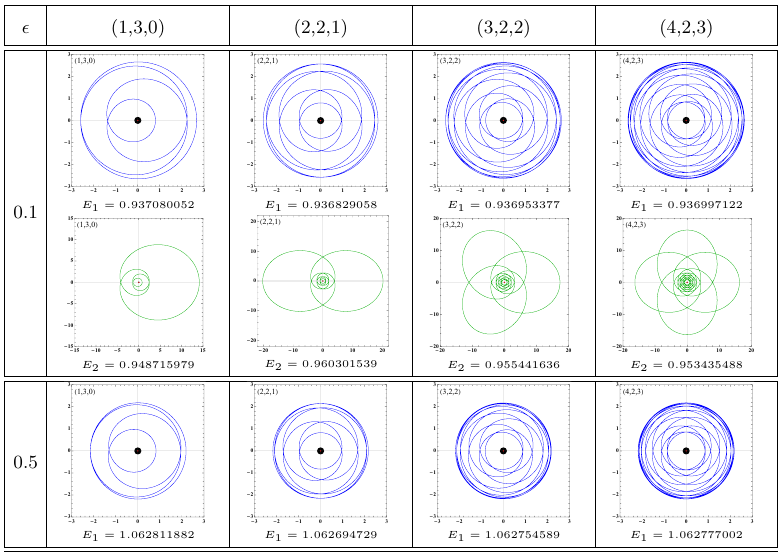}
\caption{Periodic orbits with different $(z,w,v)$ around a dyonic black hole with $\alpha_1 = 1.025755$, for $L \in [L_{\mathrm{ISCO}}, L_{\mathrm{MBO2}}]$. Shown are the cases $\epsilon = 0.1$ ($L = 2.887166748$) and $\epsilon = 0.5$ ($L = 3.520160249$).}
\label{Fig:1hb2tra}
\end{figure}

Using representative values of the angular momentum and the corresponding energies, the particle trajectories are obtained by numerically solving the geodesic equations \eqref{geo}. The resulting periodic orbits are shown in Fig.~\ref{Fig:1hb2tra}, where the motion is projected onto the plane $(r\cos\phi,\, r\sin\phi)$. Even for $E>1$, the trajectories remain closed and periodic. When $\epsilon=0.1$, two distinct orbits coexist for the same rational number $q$, whereas for larger angular momentum only a single periodic-orbit survives.

\begin{figure}[htbp]
	\center{
	\subfigure[$\epsilon=0.1, E_1$]{\label{1hb2pa}
	\includegraphics[width=5.7cm]{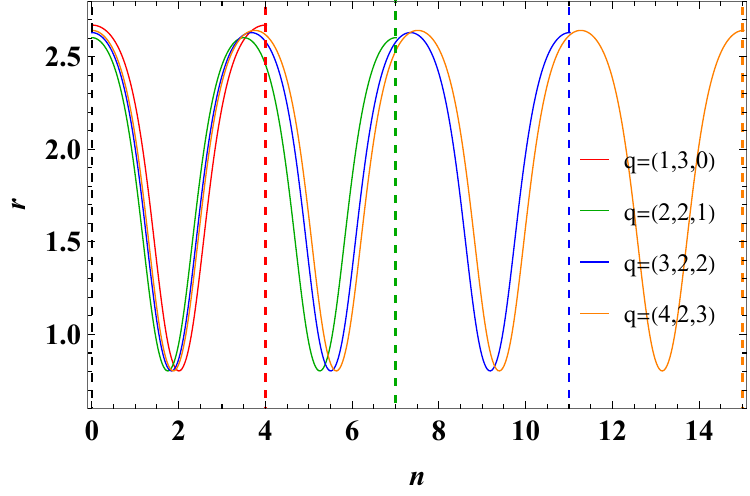}}
	\subfigure[$\epsilon=0.1, E_2$]{\label{1hb2pb}
	\includegraphics[width=5.7cm]{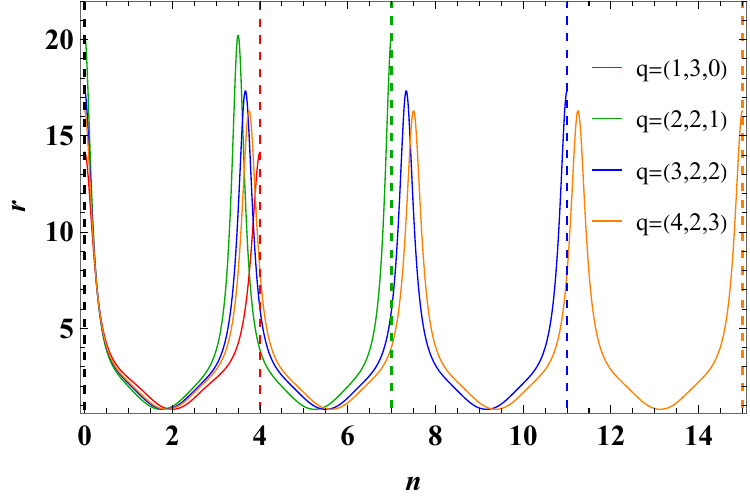}}
	}
\caption{Radial coordinate as a function of the accumulated azimuthal angle
$\Delta\phi/2\pi$ for the periodic orbits shown in Fig.~\ref{Fig:1hb2tra} with $\epsilon=0.1$. Panels (a) and (b) correspond to the two distinct energy
branches $E_1$ and $E_2$ associated with the same rational number $q$.}
\label{Fig:1hb2orbitalphase}
\end{figure}

To clarify the relation between the two branches at $\epsilon=0.1$, Fig.~\ref{Fig:1hb2orbitalphase} shows the radial coordinate $r$ as a function of the accumulated azimuthal angle $\Delta\phi/2\pi$ for the two energy solutions $E_1$ and $E_2$. Although the two branches correspond to different energies, they exhibit the same total azimuthal advance during one complete radial cycle, indicating that they share the same rational number
$q=w+v/z$ and hence the same topological structure. The total azimuthal accumulation satisfies
\begin{equation}\label{loop}
n=\frac{\Delta\phi}{2\pi}=z(1+q),
\end{equation}
where $n$ denotes the winding number. For example, $q=(2,2,1)$ gives $n=7$, while $q=(4,2,3)$ yields $n=15$, consistent with the phase accumulation shown in Fig.~\ref{Fig:1hb2orbitalphase}. Therefore, different energy branches can correspond to topologically equivalent periodic orbits with identical winding numbers.

\begin{enumerate}
  \item[a.2] Periodic orbits for $L\in[L_{\mathrm{ISCO}},L_{\mathrm{MBO3}}]$ in the single-horizon case
\end{enumerate}

We next investigate periodic orbits in the single-horizon dyonic black hole spacetime for the extended angular-momentum range $L\in[L_{\mathrm{ISCO}},L_{\mathrm{MBO3}}]$. As shown in Fig.~\ref{1hEqb3}, for representative rational numbers $q=q(z,w,v)$ and varying angular momentum, parametrized by $\epsilon\in(0,1)$, up to three distinct energy solutions $E_1$, $E_2$, and $E_3$ can exist, reflecting the multi-branch structure of bound motion.
\begin{figure}[htbp]
\centering
\includegraphics[width=0.85\textwidth]{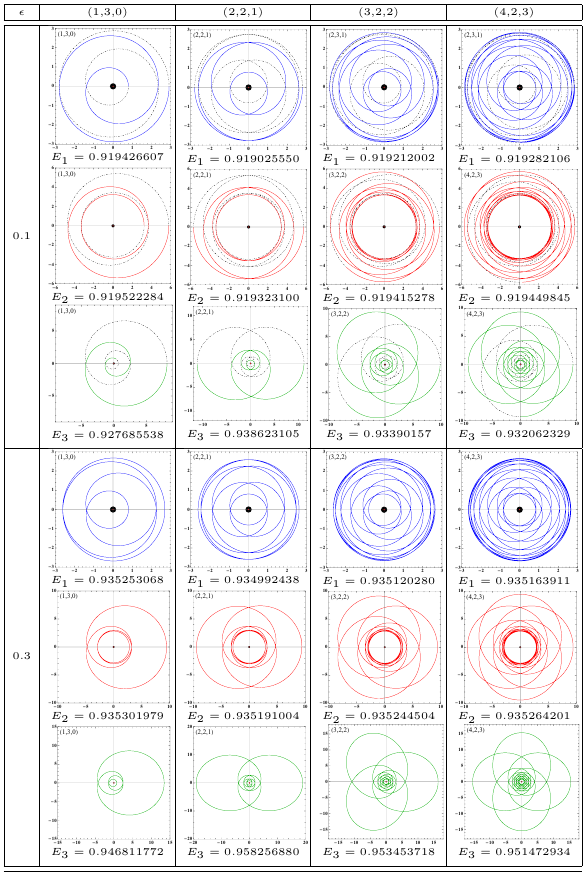}
\caption{Periodic orbits with different $(z,w,v)$ around a dyonic black hole with $\alpha_1 = 1.025755$, for $L \in [L_{\mathrm{ISCO}}, L_{\mathrm{MBO3}}]$. Shown are the cases $\epsilon = 0.1$ ($L = 2.778089089$) and $\epsilon = 0.5$ ($L = 2.876430523$).}
\label{Fig:1hb3tra}
\end{figure}

Figure.~\ref{Fig:1hb3tra} displays representative periodic trajectories for $\epsilon=0.1$ and $\epsilon=0.3$. For a fixed rational number $q$, the corresponding energy increases monotonically with angular momentum, while each $E_i$ labels a distinct branch enabled by the non-monotonic effective potential. For visualization, different plotting styles are adopted, and for $\epsilon=0.1$ black dotted segments indicate one radial period, excluding single-leaf orbits. This representation clearly reveals the zoom–whirl–vertex structure encoded by the integers $(z,w,v)$, following the classification scheme of Ref.~\cite{Levin:2008mq}.
(1) \textbf{Zoom number $z$}: The number of leaves (aphelia) in the orbit. For example, $z=1$ indicates a single-leaf orbit (e.g., $(1,~2,~0)$, $(1,~3,~0)$, while $z=2,~3,~4$ correspond to orbits with 2, 3, or 4 aphelia, respectively. These are clearly visible in the $r-\phi$ plane.
(2) \textbf{Whirl number  $w$}: The number of rapid revolutions around the center between two aphelia. This can be seen from the spiral-like motion near periastron. For example, the orbit labeled $(1,3,0)$ completes three full whirls within one radial period. The dotted line segments in the Figures help track these revolutions.
(3) \textbf{Vertex number $v$}: Indicates the position of the next aphelion (vertex of a polygon) in the sequence, counting from the initial one labeled $v=0$.  When $z\geq 2$, the orbit can be thought of as tracing a regular polygon with $z$ vertices. For example, in the  $(3,2,2)$ orbit, the particle proceeds from the initial vertex to the second (out of three total), while in $(4,2,3)$, it reaches the last vertex in a four-vertex structure within one radial cycle. These definitions align with Eq.~\eqref{vnum}, which constrains $v\in [0,~z-1]$ (For the single leaf case, since $z=1$, then $v$ can only be 0).

\begin{figure}[htbp]
	\center{
	\subfigure[$\epsilon=0.1, E_1$]{\label{1hb3e1pa}
	\includegraphics[width=5.7cm]{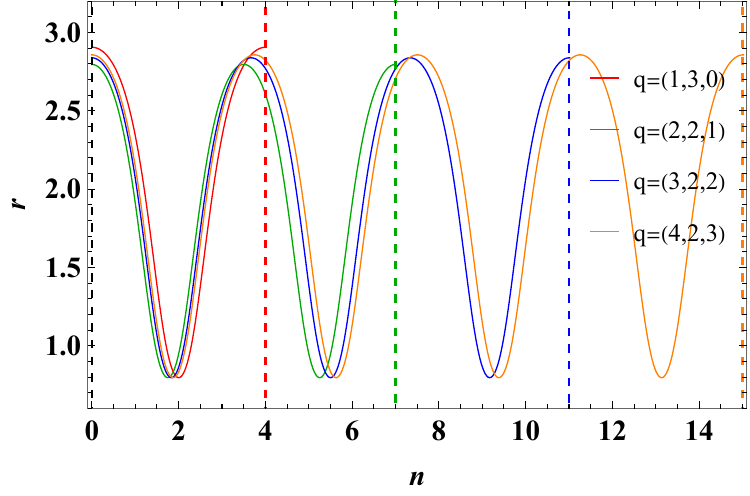}}
	\subfigure[$\epsilon=0.1, E_2$]{\label{1hb3e2pa}
	\includegraphics[width=5.7cm]{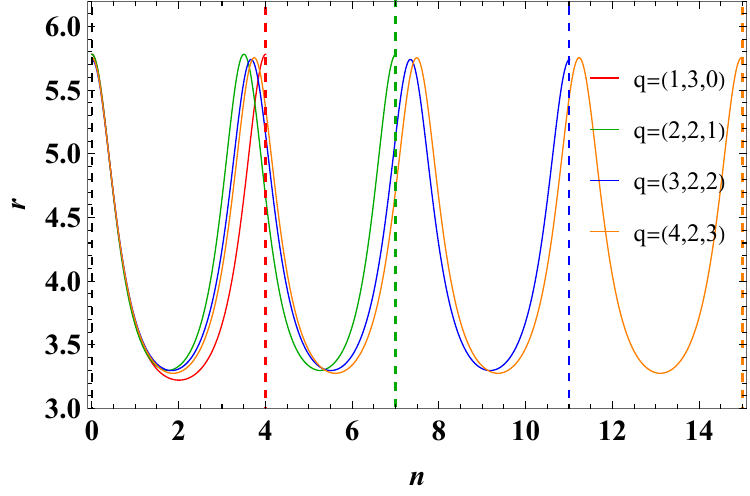}}
    \subfigure[$\epsilon=0.3, E_2$]{\label{1hb3e2pb}
	\includegraphics[width=5.7cm]{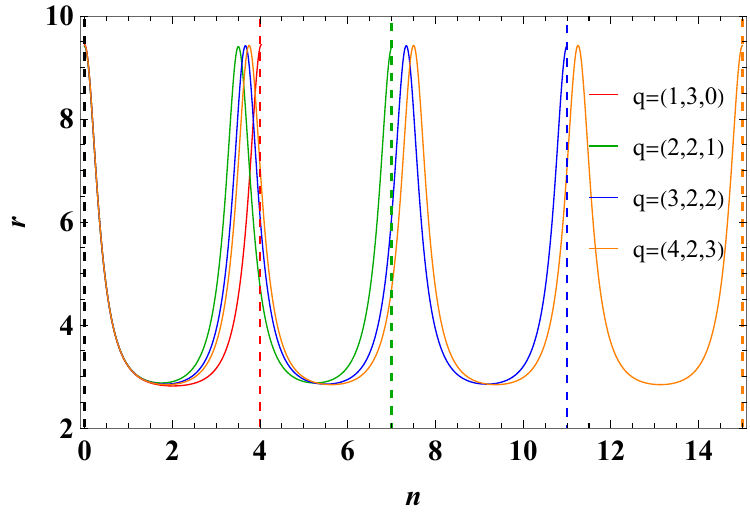}}\\
    \subfigure[$\epsilon=0.3, E_1$]{\label{1hb3e1pb}
	\includegraphics[width=5.7cm]{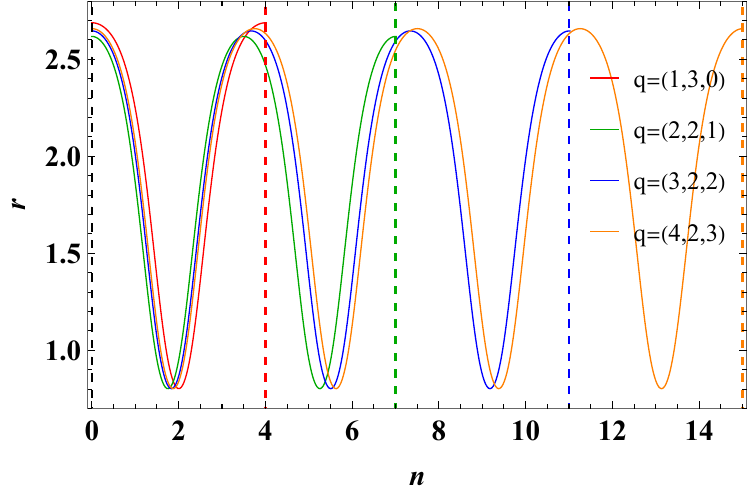}}
    \subfigure[$\epsilon=0.1, E_2$]{\label{1hb3e3pa}
	\includegraphics[width=5.7cm]{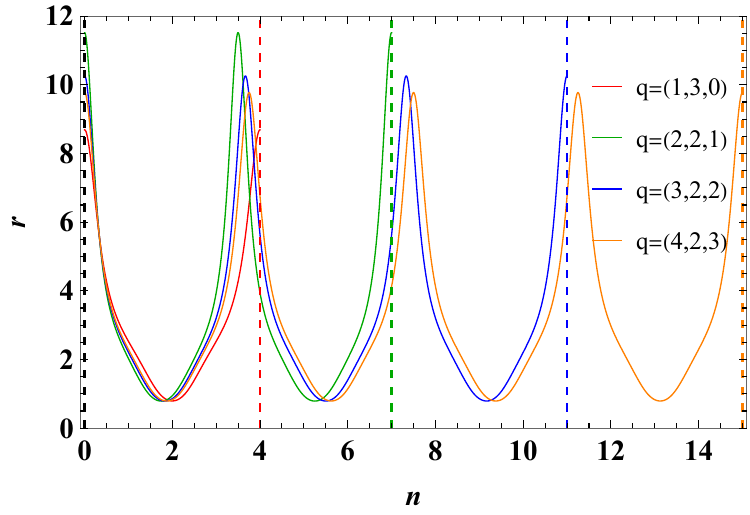}}
    \subfigure[$\epsilon=0.3, E_3$]{\label{1hb3e3pb}
	\includegraphics[width=5.7cm]{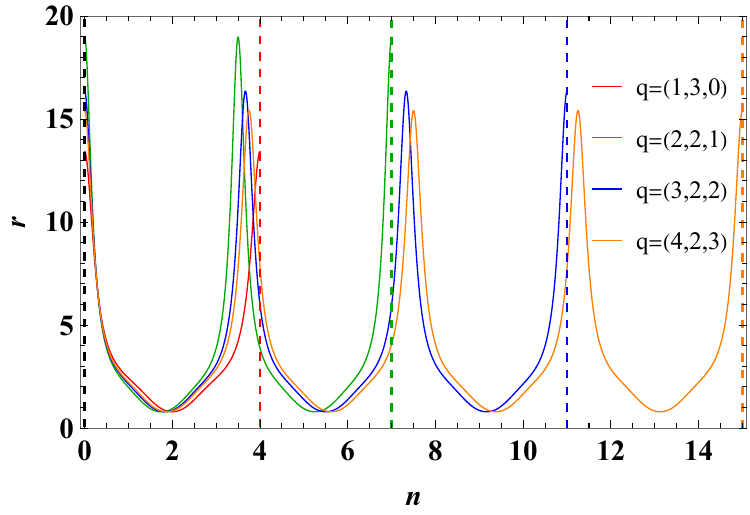}}\\
	}
\caption{Radial coordinate as a function of the accumulated azimuthal angle
$\Delta\phi/2\pi$ for the periodic orbits shown in Fig.~\ref{Fig:1hb3tra}.}
\label{Fig:1hb3orbitalphase}
\end{figure}

Figure.~\ref{Fig:1hb3orbitalphase} further illustrates that periodic orbits belonging to different energy branches can nevertheless share the same rational number $q$. Although the corresponding trajectories differ substantially in geometry and eccentricity, their total azimuthal accumulation during one radial cycle remains identical, implying the same winding number
$n=z(1+q)$ defined in Eq.~\eqref{loop}. Therefore, the three branches represent energetically and dynamically distinct realizations of the same zoom--whirl topological structure. This multi-branch behavior originates from the multi-well effective potential generated by the non-monotonic geometry of the single-horizon dyonic black hole spacetime.

\begin{enumerate}
  \item[a.3]  Periodic orbits at fixed energy $E$ in the single-horizon case
\end{enumerate}

We now investigate the complementary situation in which the particle energy $E$ is fixed while the angular momentum is allowed to vary. This provides an alternative characterization of the multi-branch structure of periodic motion in the single-horizon dyonic black hole spacetime. For each rational number $q=q(z,w,v)$, the corresponding angular momenta are determined at selected energies.

For $E<1$, two distinct angular momenta generally exist for a given rational number $q$, corresponding to periodic orbits supported by the inner and outer potential wells. We denote these two branches by $L_1$ and $L_2$, respectively. In contrast, when $E\geq1$, the outer bound region disappears and only the inner branch survives, leaving a single periodic-orbit solution.

\begin{figure}[H]
\centering
\includegraphics[width=0.835\textwidth]{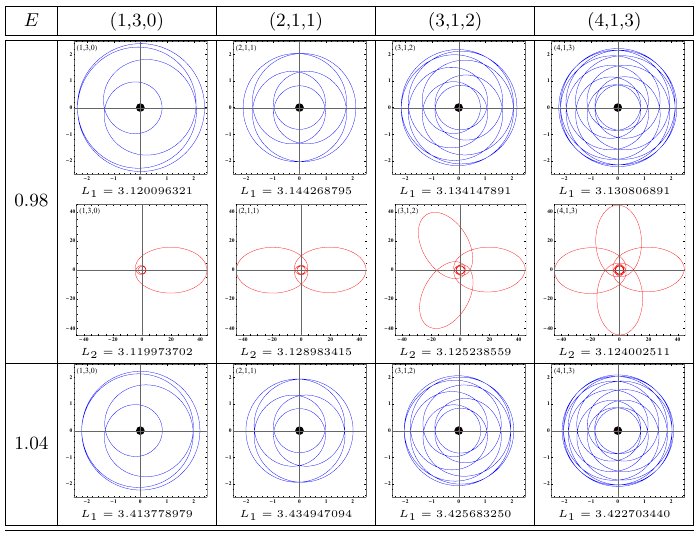}
\caption{Periodic orbits of different $(z,w,v)$ around the dyonic black holes with $\alpha_1=1.025755$ at fixed energy $E$.}
\label{Fig:1hLqtra}
\end{figure}

Representative periodic trajectories are shown in Fig.~\ref{Fig:1hLqtra}. For fixed energy, the two branches exhibit markedly different geometrical structures. The inner-branch orbits, associated with the larger angular momentum $L_1$, remain closer to the black hole and undergo smaller radial oscillations, producing relatively circular trajectories. In contrast, the outer-branch solutions with smaller angular momentum $L_2$ extend to much larger radii and display stronger eccentricity.

\begin{figure}[htbp]
	\center{
    \subfigure[$E=0.98,~L_1$]{\label{1hLq3pa}
	\includegraphics[width=5.7cm]{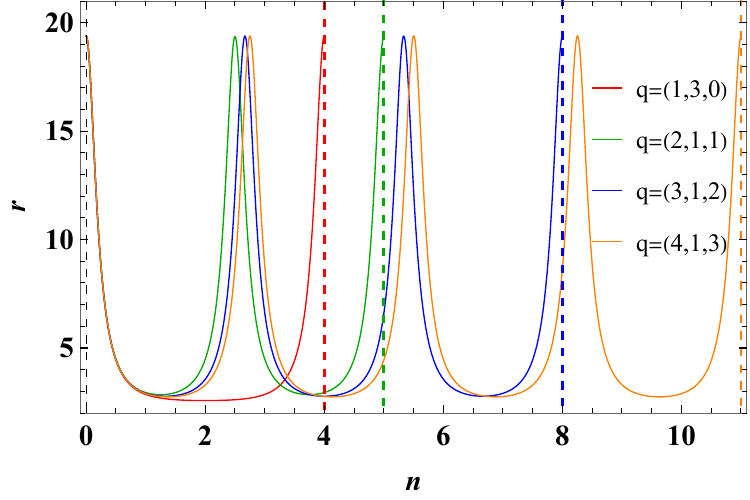}}
	\subfigure[$E=0.98,~L_2$ ]{\label{1hLq3pb}
	\includegraphics[width=5.7cm]{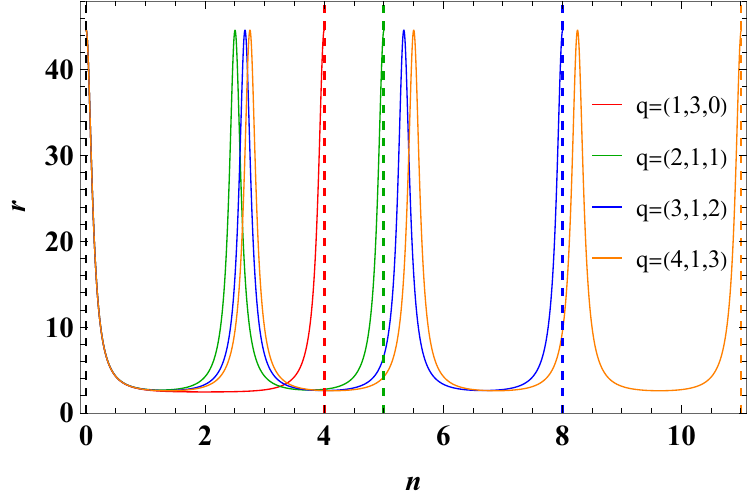}}
	}
\caption{Radial variation versus azimuthal angle for periodic orbits shown in Fig.~\ref{Fig:1hLqtra}.}
\label{Fig:1hLqorbitalphase}
\end{figure}

Figure.~\ref{Fig:1hLqorbitalphase} further shows the radial motion as a function of the accumulated azimuthal angle. Despite the substantial geometrical differences between the two branches, the total azimuthal accumulation during one radial cycle remains identical for fixed $q$, in agreement with Eq.~\eqref{loop}. Thus, the two branches possess the same winding number and identical zoom--whirl topology.

In summary, the double-well effective potential supports two energetically equivalent but dynamically distinct periodic-orbit branches for $E<1$. Although the corresponding trajectories differ significantly in angular momentum, radial extent, and eccentricity, they remain topologically equivalent because they share the same rational number $q$ and winding structure. This demonstrates that the multi-branch nature of periodic motion is an intrinsic consequence of the non-monotonic spacetime geometry rather than a feature tied to a particular parameterization.
\\

\paragraph{Periodic orbits with two horizons}

\begin{enumerate}
  \item[b.1]Periodic orbits for  $L\in[L_{\mathrm{ISCO}},L_{\mathrm{MBO2}}]$ in the two-horizon case
\end{enumerate}

We now extend the analysis to a representative two-horizon dyonic black hole with $\alpha_1=1.0494625$. As shown in Fig.~\ref{solutionscases}, this case still lies in the parameter region where the metric function is non-monotonic outside the outer event horizon. Therefore, the effective potential retains a multi-well structure and is expected to support multi-branch periodic motion, as in the single-horizon case discussed above.

For the angular-momentum interval $L\in[L_{\mathrm{ISCO}},L_{\mathrm{MBO2}}]$,
the $E$--$q$ relation shown in Fig.~\ref{2hEqb2} exhibits multiple energy solutions
for a fixed rational number $q=q(z,w,v)$. The corresponding numerical values are
listed in Tab.~\ref{Tab1}. At low angular momentum, represented by $\epsilon=0.1$,
three distinct branches $E_1<E_2<E_3$ coexist for the same $q$. This is a direct
consequence of the richer multi-well structure of the effective potential in the
two-horizon geometry. When the angular momentum is increased to $\epsilon=0.5$,
only the innermost branch remains, although the corresponding periodic orbits can
still have $E>1$.
For each fixed rational number $q=q(z,w,v)$, the
corresponding energy increases monotonically with angular momentum, in agreement
with the behavior of the effective potential.

\begin{table}[htbp]
\caption{The particle energy $E$ corresponding to orbital trajectories
with various $(z, w, v)$ configurations are computed for angular momentum
values defined by $L = L_{\text{ISCO}} + \epsilon(L_{\text{MBO2}} - L_{\text{ISCO}})$,
 with $\alpha_1$ fixed at 1.049462525. Each $q$ configuration is listed in the column headers, and the corresponding values of $L/M$ are given in the second column.} \label{Tab1}
\centering
\scriptsize
\setlength{\tabcolsep}{4.5pt}
\resizebox{\textwidth}{!}{
\begin{tabular}{c*{13}{c}}
\toprule[0.8pt]
\toprule[0.8pt]
& &
\multicolumn{3}{c}{
  \begin{tabular}{@{}c@{}}
  $(1,3,0)$\\[-0.7ex]
  \rule{5.0cm}{0.45pt}
  \end{tabular}
}
&
\multicolumn{3}{c}{
  \begin{tabular}{@{}c@{}}
  $(2,1,1)$\\[-0.7ex]
  \rule{5.0cm}{0.45pt}
  \end{tabular}
}
&
\multicolumn{3}{c}{
  \begin{tabular}{@{}c@{}}
  $(3,1,2)$\\[-0.7ex]
  \rule{5.0cm}{0.45pt}
  \end{tabular}
}
&
\multicolumn{3}{c}{
  \begin{tabular}{@{}c@{}}
  $(4,1,3)$\\[-0.7ex]
  \rule{5.0cm}{0.45pt}
  \end{tabular}
}
\\[-0.2ex]
$\epsilon$ & $L/M$ & \multicolumn{1}{c}{$E_1$} & \multicolumn{1}{c}{$E_2$} & \multicolumn{1}{c}{$E_3$}
	               & \multicolumn{1}{c}{$E_1$} & \multicolumn{1}{c}{$E_2$} & \multicolumn{1}{c}{$E_3$}
                   & \multicolumn{1}{c}{$E_1$} & \multicolumn{1}{c}{$E_2$} & \multicolumn{1}{c}{$E_3$}
	               & \multicolumn{1}{c}{$E_1$} & \multicolumn{1}{c}{$E_2$} & \multicolumn{1}{c}{$E_3$}\\
\midrule[0.5pt]
0.1 & 3.056459520 & 0.963161895 & 0.963199952 & 0.976458625 & 0.962960199 & 0.963141602 & 0.991538788 & 0.963063815 & 0.963171552 & 0.985150758 & 0.963098545 & 0.963181595 & 0.982538037 \\
0.5 & 4.263081130 & 1.219921675 & \xmark   & \xmark   & 1.219810614 & \xmark & \xmark   & 1.219871637   & \xmark  & \xmark   & 1.219890395 & \xmark   & \xmark \\
\bottomrule[0.8pt]
\bottomrule[0.8pt]
\end{tabular}
}
\end{table}

A particularly illustrative example is provided by the orbit with $(z,w,v)=(2,2,1)$ at $\epsilon=0.1$. The third branch, with $E_3=0.991538788$, extends from a periastron $r_p=0.709311641\,M$ to an apastron $r_a=113.319643083\,M$, showing that the outer branch can occupy a very large radial domain. This example demonstrates that, in the two-horizon case, multi-branch periodic motion persists and may involve highly eccentric orbits extending far from the black hole.

\begin{enumerate}
  \item[b.2]Periodic Orbits for  $L\in[L_{\mathrm{ISCO}},L_{\mathrm{MBO3}}]$ in the two-horizon case
\end{enumerate}

We next consider the same two-horizon configuration with $\alpha_1=1.0494625$, but now focus on the angular-momentum interval $L\in[L_{\mathrm{ISCO}},L_{\mathrm{MBO3}}]$. As indicated by the allowed region in Fig.~\ref{12hELl3}, bound motion in this interval occurs only for $E<1$.

The energy values extracted from the $E$--$q$ relation in Fig.~\ref{2hEqb3} are summarized in Tab.~\ref{Tab2}. For both $\epsilon=0.1$ and $\epsilon=0.3$, three energy branches can coexist for a given rational number $q$. For fixed $q$, the energies generally increase with $\epsilon$, reflecting the dependence of the allowed radial domains on the angular momentum.
\begin{table}[htbp]
\caption{The particle energy $E$ corresponding to orbital trajectories with various $(z, w, v)$ configurations are computed for angular momentum values defined by $L = L_{\text{ISCO}} + \epsilon(L_{\text{MBO3}} - L_{\text{ISCO}})$, with $\alpha_1$ fixed at 1.049462525. Each $q$ configuration is listed in the column headers, and the corresponding values of $L/M$ are given in the second column.}
\label{Tab2}
\centering
\scriptsize
\setlength{\tabcolsep}{4.5pt}
\resizebox{\textwidth}{!}{
\begin{tabular}{c*{13}{c}}
\toprule[0.8pt]
\toprule[0.8pt]
& &
\multicolumn{3}{c}{
  \begin{tabular}{@{}c@{}}
  $(1,3,0)$\\[-0.7ex]
  \rule{5.0cm}{0.45pt}
  \end{tabular}
}
&
\multicolumn{3}{c}{
  \begin{tabular}{@{}c@{}}
  $(2,1,1)$\\[-0.7ex]
  \rule{5.0cm}{0.45pt}
  \end{tabular}
}
&
\multicolumn{3}{c}{
  \begin{tabular}{@{}c@{}}
  $(3,1,2)$\\[-0.7ex]
  \rule{5.0cm}{0.45pt}
  \end{tabular}
}
&
\multicolumn{3}{c}{
  \begin{tabular}{@{}c@{}}
  $(4,1,3)$\\[-0.7ex]
  \rule{5.0cm}{0.45pt}
  \end{tabular}
}
\\[-0.2ex]
$\epsilon$ & $L/M$ & \multicolumn{1}{c}{$E_1$} & \multicolumn{1}{c}{$E_2$} & \multicolumn{1}{c}{$E_3$}
	               & \multicolumn{1}{c}{$E_1$} & \multicolumn{1}{c}{$E_2$} & \multicolumn{1}{c}{$E_3$}
                   & \multicolumn{1}{c}{$E_1$} & \multicolumn{1}{c}{$E_2$} & \multicolumn{1}{c}{$E_3$}
	               & \multicolumn{1}{c}{$E_1$} & \multicolumn{1}{c}{$E_2$} & \multicolumn{1}{c}{$E_3$}\\
\midrule[0.5pt]
0.1 & 2.804294152 & 0.920865882 & 0.920990212 & 0.932080884 & 0.920415550 & 0.920811935 & 0.943546088 & 0.920627138 & 0.920894848 & 0.938640080 & 0.920705270 & 0.920925804 & 0.936654695 \\
0.3 & 2.903274222  & 0.936403295 & 0.936469124 & 0.948903402 & 0.936108757 & 0.936371603& 0.942757533 & 0.936254558 & 0.936419528 & 0.956445532 & 0.936305446 & 0.936436444 & 0.954178025 \\
\bottomrule[0.8pt]
\bottomrule[0.8pt]
\end{tabular}}
\end{table}

\begin{figure}[htbp]
	\subfigure[$\epsilon=0.1, E_1$]{\label{2hb3Eq1phasea}
	\includegraphics[width=5.7cm]{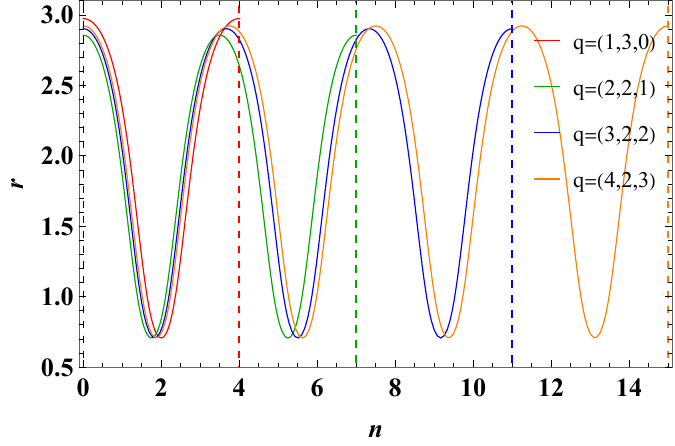}}
	\subfigure[$\epsilon=0.1, E_2$]{\label{2hb3Eq1phaseb}
	\includegraphics[width=5.7cm]{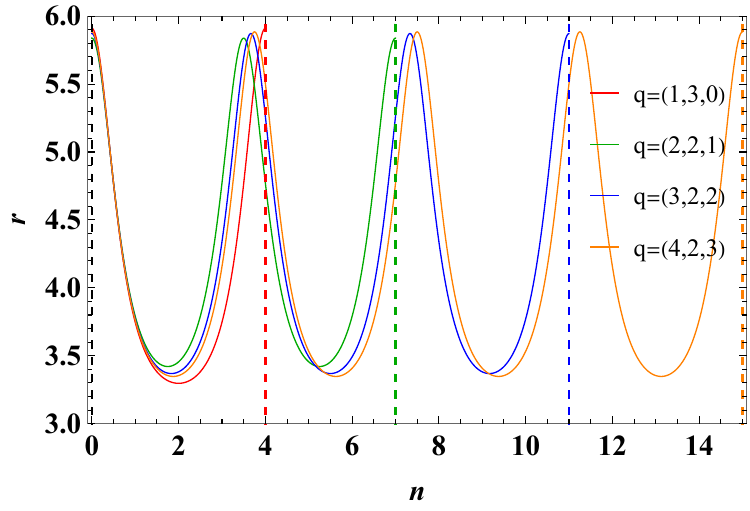}}
	\subfigure[$\epsilon=0.1, E_3$]{\label{2hb3Eq1phasec}
	\includegraphics[width=5.7cm]{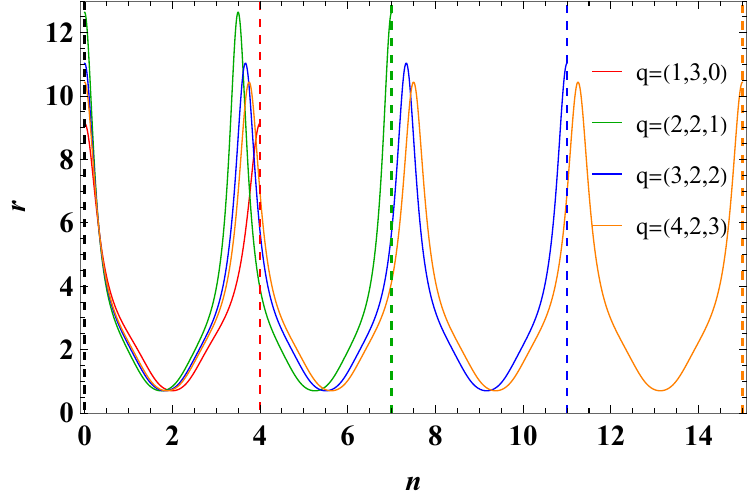}}\\
	\subfigure[$\epsilon=0.3, E_1$]{\label{2hb3Eq2phasea}
	\includegraphics[width=5.7cm]{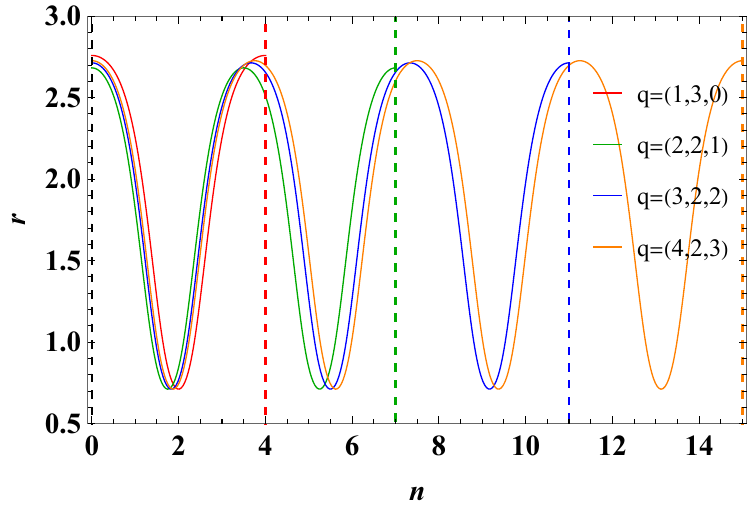}}
	\subfigure[$\epsilon=0.3, E_2$]{\label{2hb3Eq2phaseb}
	\includegraphics[width=5.7cm]{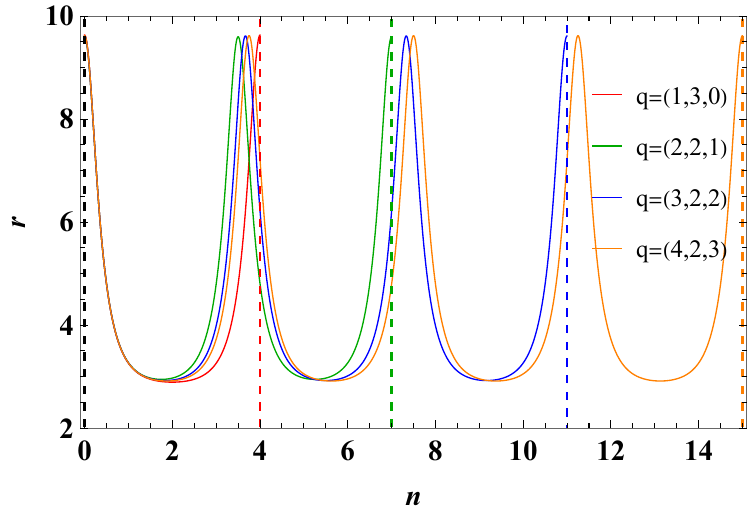}}
	\subfigure[$\epsilon=0.3, E_3$]{\label{2hb3Eq2phasec}
	\includegraphics[width=5.7cm]{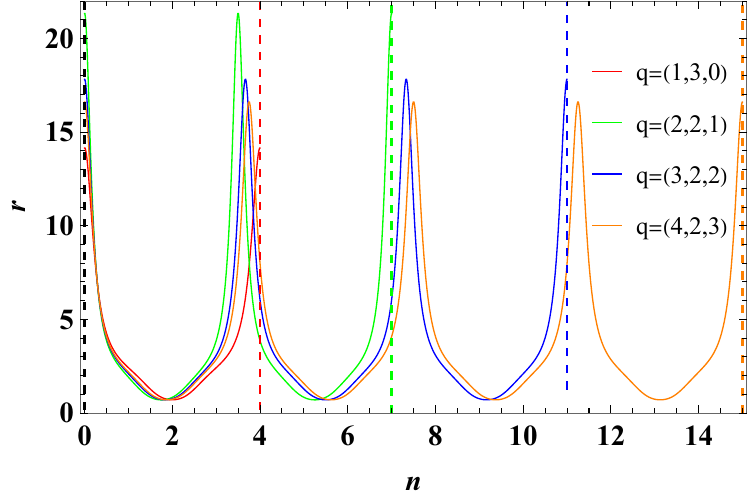}}
\caption{Radial variation versus azimuthal angle for periodic orbits around a dyonic black hole with $\alpha_1 = 1.0494625$, for $L \in [L_{\mathrm{ISCO}}, L_{\mathrm{MBO3}}]$.}
\label{Fig:2hab3orbitalphase}
\end{figure}

The corresponding phase evolution is shown in Fig.~\ref{Fig:2hab3orbitalphase}, where the radial coordinate is plotted as a function of the accumulated azimuthal angle. Although the three branches differ in radial extent and eccentricity, they share the same rational number $q$ and hence the same winding number $n=z(1+q)$. Thus, as in the single-horizon case, the different branches are geometrically distinct but topologically equivalent.

At fixed angular momentum, increasing the energy enlarges the radial excursion of the orbit. Moreover, increasing $\epsilon$ affects the branches nonuniformly: the lowest-energy branch becomes more confined, whereas the higher-energy branches extend over larger radial intervals. This behavior provides a useful comparison with the single-horizon case and confirms that the multi-branch structure is controlled by the non-monotonic exterior geometry rather than by the number of horizons alone.

\begin{enumerate}
  \item[b.3]Periodic orbits at fixed energy $E$ in the two-horizon case
\end{enumerate}

Finally, we analyze the complementary situation in which the particle energy $E$ is fixed and the angular momentum is determined from the $L$--$q$ relation shown in Fig.~\ref{2hLq}. The corresponding angular momenta for several rational numbers $q=q(z,w,v)$ are listed in Tab.~\ref{Tab3}.
\begin{table}[htbp]
\caption{The particle angular momentum $L$ corresponding to orbital trajectories with various $(z, w, v)$ configurations for $\alpha_1 = 1.049462525$. Each configuration $q = q(z, w, v)$ is listed in the column headers, and the corresponding energy values $E$ are shown in the first column.}
\label{Tab3}
\centering
\scriptsize
\resizebox{\textwidth}{!}{
\setlength{\tabcolsep}{4.5pt}
\renewcommand{\arraystretch}{1.05}

\begin{tabular}{c*{8}{c}}
\toprule[0.8pt]
\toprule[0.8pt]
&
\multicolumn{2}{c}{
  \begin{tabular}{@{}c@{}}
  $(1,3,0)$\\[-0.7ex]
  \rule{3.0cm}{0.45pt}
  \end{tabular}
}
&
\multicolumn{2}{c}{
  \begin{tabular}{@{}c@{}}
  $(2,1,1)$\\[-0.7ex]
  \rule{3.0cm}{0.45pt}
  \end{tabular}
}
&
\multicolumn{2}{c}{
  \begin{tabular}{@{}c@{}}
  $(3,1,2)$\\[-0.7ex]
  \rule{3.0cm}{0.45pt}
  \end{tabular}
}
&
\multicolumn{2}{c}{
  \begin{tabular}{@{}c@{}}
  $(4,1,3)$\\[-0.7ex]
  \rule{3.0cm}{0.45pt}
  \end{tabular}
}
\\[-0.2ex]
$E$
& $L_1$ & $L_2$
& $L_1$ & $L_2$
& $L_1$ & $L_2$
& $L_1$ & $L_2$ \\
\midrule[0.5pt]
0.96    & 3.039082022 & 3.038856906 & 3.072064747 & 3.048251620 & 3.040229385 & 3.044406231 & 3.053808659 & 3.043120311   \\
0.98    & 3.146691474 & 3.146527455 & 3.176756286 & 3.154616778 & 3.163850512 & 3.151194916 & 3.159667943 & 3.150069695   \\
1       & 3.249884868 & \xmark   & 3.278075887 & \xmark   & 3.265721158 & \xmark   & 3.261765158 & \xmark     \\
1.04    & 3.447323705 & \xmark   & 3.473371643 & \xmark   & 3.461646717 & \xmark   & 3.457953020 & \xmark     \\
\bottomrule[0.8pt]
\bottomrule[0.8pt]
\end{tabular}}
\end{table}

For $E<1$, two angular-momentum branches, denoted by $L_1$ and $L_2$, generally
exist for the same rational number $q$. These two branches correspond to periodic
orbits supported by different radial wells of the effective potential. By contrast,
for $E\geq1$, the outer bound region disappears and only one periodic-orbit branch
remains.

\begin{figure}[htbp]
    \subfigure[$E=0.96,~L_1$]{\label{2hLq2pa}
	\includegraphics[width=4.1cm]{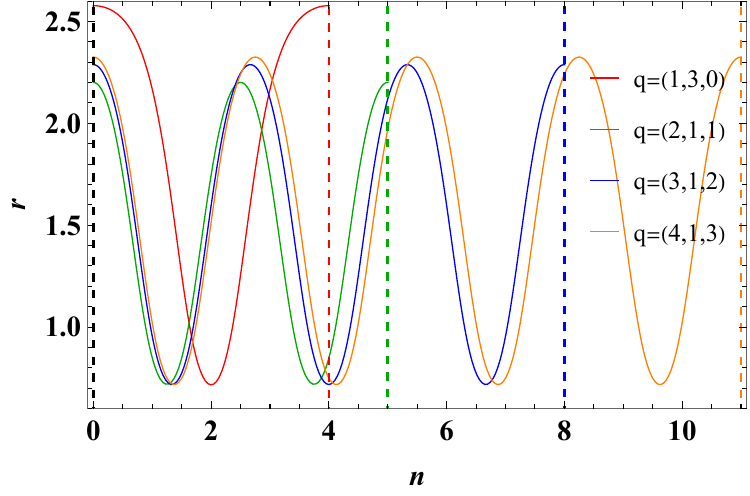}}
    \subfigure[$E=0.98,~L_1$]{\label{2hLq3pa}
	\includegraphics[width=4.1cm]{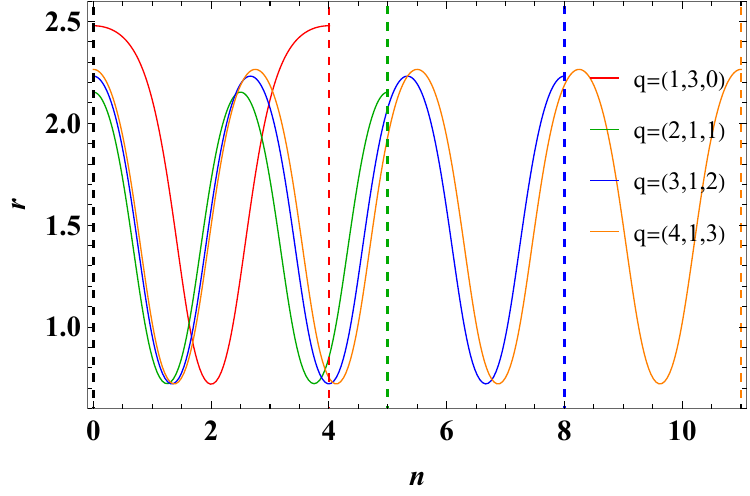}}
	\subfigure[$E=0.96,~L_2$ ]{\label{2hLq2pb}
	\includegraphics[width=4.1cm]{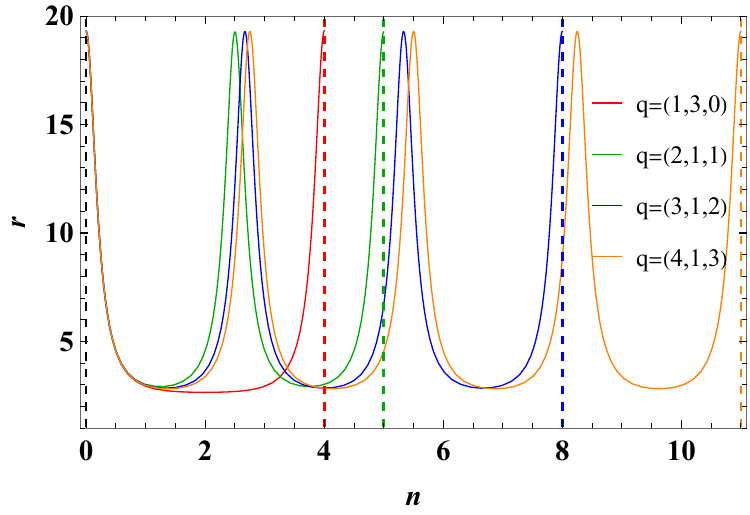}}
	\subfigure[$E=0.98,~L_2$ ]{\label{2hLq3pb}
	\includegraphics[width=4.1cm]{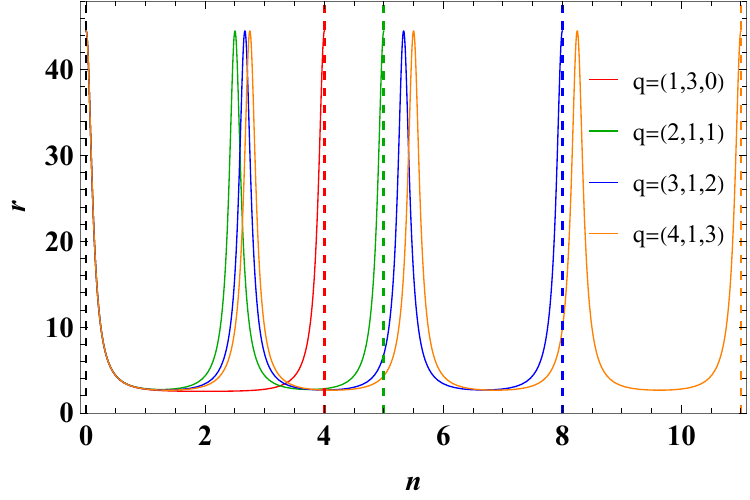}}
\caption{Radial variation versus azimuthal angle for periodic orbits around a dyonic black hole with $\alpha_1 = 1.0494625$ at fixed energy $E$.}
\label{Fig:2haLqorbitalphase}
\end{figure}

The phase evolution in Fig.~\ref{Fig:2haLqorbitalphase} illustrates the topological equivalence of the two branches at fixed energy. Although the branches have different angular momenta and radial amplitudes, their total azimuthal accumulation is fixed by the same rational number $q$, in agreement with Eq.~\eqref{loop}.

A comparison of the two branches shows that the smaller-angular-momentum branch $L_2$
generally extends to larger radii and has a higher eccentricity, while the larger-angular-momentum
branch $L_1$ remains closer to the black hole and is more compact. Furthermore, the radial
response to increasing energy is branch dependent: the radial range of the $L_1$ branch
decreases, whereas that of the $L_2$ branch increases. This behavior mirrors the trends observed earlier at fixed angular momentum and
reflects the non-monotonic structure of the effective potential outside the outer horizon in the two-horizon geometry.
This reversed trend is another
manifestation of the double-well effective potential generated by the non-monotonic
metric function.

\subsubsection{Summary of multi-branch periodic orbits}

We conclude this subsection by summarizing the main properties of periodic orbits in the parameter range
$1.025755\leq\alpha_1<1.07317$, where the metric function $f(r)$ is non-monotonic outside the outer event horizon. This exterior non-monotonicity, rather than the number of horizons itself, is the essential geometric ingredient responsible for the multi-branch structure found above.

The non-monotonic behavior of $f(r)$ produces three MBO branches and generates a multi-well effective potential. By combining the MBO structure with the ISCO, we identified two physically relevant angular-momentum intervals,
$L\in[L_{\mathrm{ISCO}},L_{\mathrm{MBO2}}]$ and
$L\in[L_{\mathrm{ISCO}},L_{\mathrm{MBO3}}]$,
both of which can support multiple disconnected bound regions. These regions provide the radial domains in which periodic orbits can form.

The analysis of both the single-horizon and two-horizon cases shows a consistent picture. For fixed angular momentum, the same rational number $q=w+v/z$ may correspond to several energy branches. Conversely, for fixed energy, the same $q$ may correspond to more than one angular-momentum branch. Therefore, the multi-branch structure is not an artifact of a particular parametrization, but an intrinsic consequence of the multi-well effective potential.

A key feature is that different branches may be topologically equivalent while remaining geometrically distinct. Here the term ``topologically equivalent'' refers to the zoom--whirl topology of the periodic orbit, not to the topology of the event horizon or to a three-dimensional spatial embedding. Since all orbits considered here are equatorial geodesics in a spherically symmetric spacetime, the two-dimensional orbital trajectories and the $r$--$\Delta\phi/2\pi$ phase plots fully encode the topological information specified by $(z,w,v)$ and $n=z(1+q)$. Orbits belonging to different branches can share the same rational number $q$ and the same winding number $n=z(1+q)$, but differ in energy, angular momentum, radial extension, and eccentricity. This means that the standard one-to-one correspondence between a rational periodic-orbit label and a unique geometrical orbit is broken in non-monotonic spacetimes.

This distinction also has observational significance. The non-monotonic exterior geometry affects both null and timelike geodesics. In the null sector, it can generate multiple photon spheres and hence multiple photon rings in black hole images. These photon rings are not direct evidence for multiple horizons, nor do they arise from unstable timelike circular orbits. Rather, they are optical manifestations of the same exterior geometric structure that produces multiple wells in the timelike effective potential. Whether such rings can be resolved by a distant observer depends on observational resolution, but their existence is a property of the spacetime geometry.

In the timelike sector, the same non-monotonic structure gives rise to multi-branch periodic orbits. This provides a natural connection to EMRI systems, where periodic orbits organize zoom--whirl motion and contribute to gravitational-wave emission. In particular, topologically equivalent but geometrically distinct orbits may generate different radiative signatures, as suggested in related studies of EMRI waveforms \cite{Wang:2025hla, Wang:2026ncv}. Therefore, multi-branch periodic orbits offer a possible dynamical counterpart to the optical signatures of multiple photon spheres.

From this perspective, the present results provide theoretical input for future observational tests of non-monotonic black hole geometries. Imaging observations may constrain the null geodesic sector through multiple photon rings, while gravitational wave observations may probe the timelike sector through branch-dependent orbital and radiative features. A detailed quantitative parameter constraint is beyond the scope of this work, but the present analysis identifies the relevant orbital structures and parameter regions where such effects can arise.

We emphasize that the orbital figures shown in this subsection represent equatorial timelike geodesics, not photon-ring images or radiation spectra. Since the background spacetime is static and spherically symmetric, these orbits can be projected onto the $(r\cos\phi,r\sin\phi)$ plane without losing topological information. The multi-branch structure displayed in these figures is caused by the radial non-monotonicity of the metric function outside the event horizon, which generates multiple wells in the effective potential. It should not be interpreted as arising from an angularly non-homogeneous distribution on the horizon cross section. Effects associated with axial rotation, frame dragging, photon-ring spectra, Doppler shifts, and radiative transfer require additional physical ingredients and are beyond the scope of the present work.

Although the present work does not compute radiation reaction or full gravitational waveforms, the multi-branch structure found here is directly relevant to EMRI modeling. Periodic and resonant orbits provide a useful basis for constructing adiabatic EMRI trajectories and snapshot waveforms \cite{Grossman:2011ps}. In multi-well geometries, different branches with the same rational number \(q\) can possess different angular momenta, radial amplitudes, and eccentricities, even when the particle mass and energy are fixed. Recent waveform calculations have shown that such topologically equivalent but geometrically distinct branches can produce distinguishable amplitude modulations and harmonic contents in EMRI signals \cite{Wang:2026ncv}.

\subsection{Single-branch periodic orbits }\label{singlebranch}

We now turn to the parameter regimes in which the metric function $f(r)$ is monotonic outside the event horizon. These cases provide a useful comparison with the multi-branch periodic orbits discussed in the previous subsection. The key point is that the number of horizons characterizes the causal structure of the spacetime, while the exterior monotonicity of $f(r)$ controls the topology of the effective potential. Therefore, even black holes with multiple horizons can support only a single branch of periodic orbits if the exterior metric function remains monotonic.

\subsubsection{Three and Four-Horizon case: $1.07317\leq \alpha_1\leq 1.09859$}\label{3/4horizons}
We now turn to the parameter regions in which the exterior metric function is monotonic and the periodic-orbit structure becomes single-branched. This provides a direct comparison with the multi-branch behavior found in the previous subsection. We first consider the interval
$1.07317\leq \alpha_1\leq 1.09859$, where the spacetime contains three or four horizons. As shown in Fig.~\ref{Fig:frcases}, three-horizon configurations occur at the two endpoints $\alpha_1=1.07317$ and $\alpha_1=1.09859$, while four-horizon black holes appear for $1.07317<\alpha_1<1.09859$. Despite this rich causal structure, the metric function $f(r)$ remains monotonic outside the outer event horizon throughout this interval.

\begin{figure}[htbp]
\includegraphics[width=7cm]{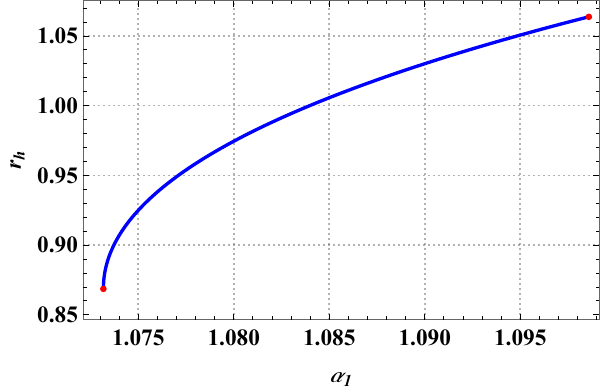}
\caption{
Outer event horizon radius $r_h$ as a function of $\alpha_1$ for
$1.07317\leq\alpha_1\leq1.09859$.
The red points denote the critical three-horizon configurations, while
the blue curve corresponds to the four-horizon branch.
}
\label{34horizons}
\end{figure}

Figure.~\ref{34horizons} shows the dependence of the outer event horizon radius $r_h$ on $\alpha_1$. The red points denote the critical three-horizon configurations at the two endpoints, while the blue curve corresponds to the four-horizon branch. Although the number of horizons changes across this interval, the exterior monotonicity of $f(r)$ is preserved. This distinction is important: the number of horizons characterizes the causal structure, whereas the exterior monotonicity of $f(r)$ controls the topology of the effective potential relevant for bound motion.

\begin{figure}[htbp]
    \subfigure[$r_{\mathrm{MBO}}$ vs $\alpha_1$]{\label{rmbo2b}
	\includegraphics[width=5.7cm]{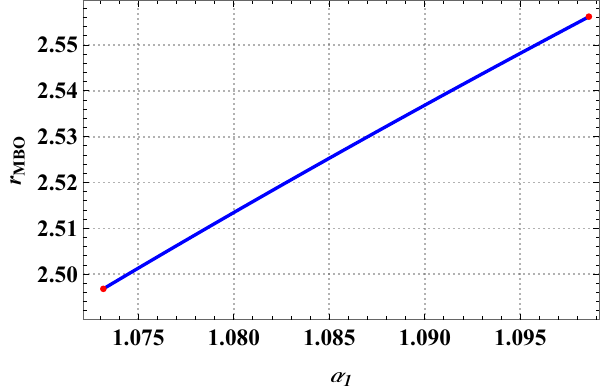}}
	\subfigure[$L_{\mathrm{MBO}}$ vs $\alpha_1$ ]{\label{Lmbo2a}
	\includegraphics[width=5.7cm]{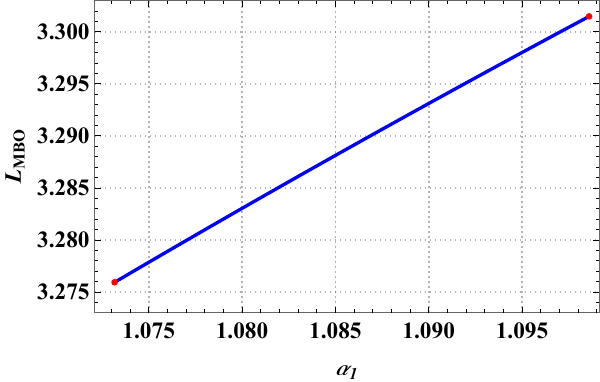}}
\caption{The radius and angular momentum for the MBOs for 3-/4- horizon with $1.07317\leq \alpha_1\leq 1.09859$. }
\label{Fig:34hrLmbo}
\end{figure}

We next examine the characteristic circular orbits in this monotonic-exterior regime. The MBOs, determined by Eq.~\eqref{mbo}, are shown in Fig.~\ref{Fig:34hrLmbo}. Both the MBO radius $r_{\mathrm{MBO}}$ and angular momentum $L_{\mathrm{MBO}}$ increase monotonically with $\alpha_1$. Since the effective potential has a single-well structure in this regime, the MBO defines the outer energy boundary $E=1$ for bound motion, as in standard single-well black hole spacetimes.

\begin{figure*}[htbp]
	\subfigure[$r_{\mathrm{ISCO}}$ vs $\alpha_1$]{\label{risco3a}
	\includegraphics[width=5.7cm]{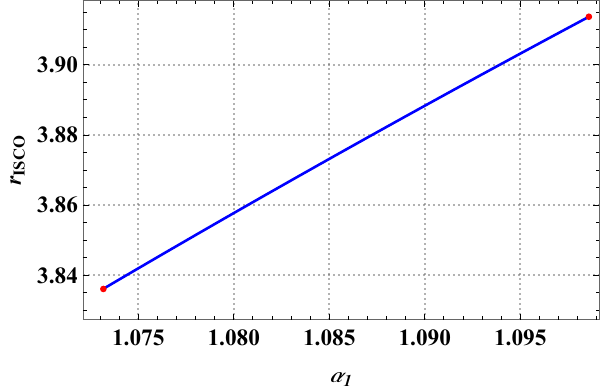}}
	\subfigure[$L_{\mathrm{ISCO}}$ vs $\alpha_1$]{\label{Lisco3b}
	\includegraphics[width=5.7cm]{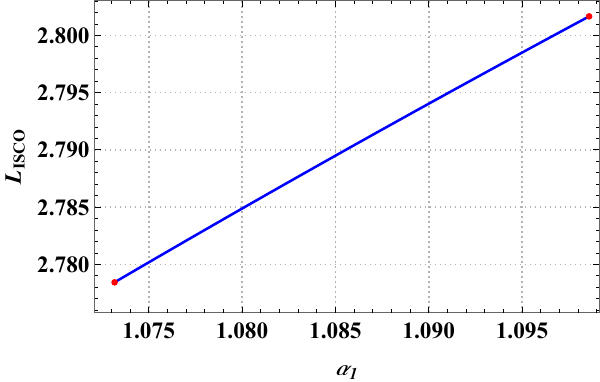}}
	\subfigure[$E_{\mathrm{ISCO}}$ vs $\alpha_1$]{\label{Eisco3c}
	\includegraphics[width=5.7cm]{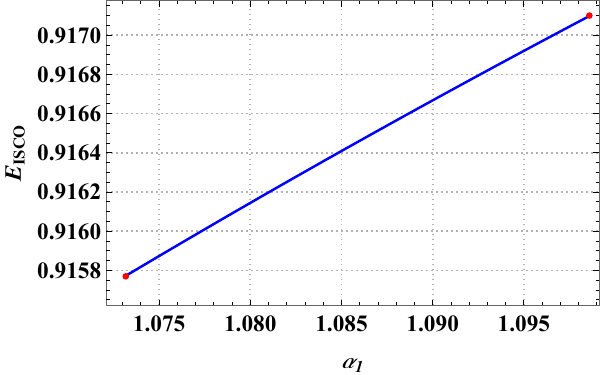}}
\caption{The radius, angular momentum, and energy of ISCOs as a function of $\alpha_1$ within the range $1.07317 \leq \alpha_1 < 1.09859$.}
\label{Fig:34hrLEriso}
\end{figure*}

The ISCO quantities are shown in Fig.~\ref{Fig:34hrLEriso}. The radius $r_{\mathrm{ISCO}}$, angular momentum $L_{\mathrm{ISCO}}$, and energy $E_{\mathrm{ISCO}}$ also increase monotonically with $\alpha_1$. Together, the MBO and ISCO determine the allowed angular-momentum interval
$L\in[L_{\mathrm{ISCO}},L_{\mathrm{MBO}}]$
and hence the physically relevant $(E,L)$ domain for bound periodic orbits.

\begin{figure}[htbp]
\includegraphics[width=7cm]{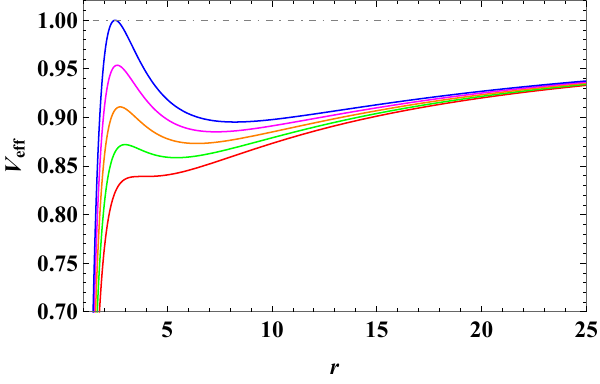}
\caption{
Effective potential $V_{\mathrm{eff}}(r)$ for different angular momenta in the monotonic-exterior three-/four-horizon regime. Here $\alpha_1=1.08$ and $L$ varies from $L_{\mathrm{ISCO}}$ to $L_{\mathrm{MBO}}$. The potential contains a single bound well, in contrast to the double-well structure found in the non-monotonic case.
}
\label{34hveff}
\end{figure}

The corresponding effective potentials are displayed in Fig.~\ref{34hveff}. In contrast to the double-well potentials generated by non-monotonic exterior metric functions, the potential here contains only a single bound well for each admissible angular momentum. The single maximum associated with the MBO separates bound and unbound motion, while the stable region is bounded from below by the ISCO. Consequently, only one radial oscillation interval exists for a given pair $(E,L)$.

\begin{figure}[htbp]
\includegraphics[width=7cm]{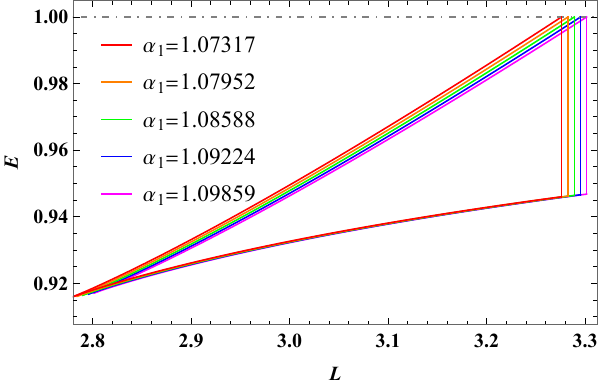}
\caption{Allowed regions in the $(L, E)$ plane for the bound orbits around dyonic black holes for different values of $\alpha_1$.}
\label{Fig:34hELp}
\end{figure}

The allowed regions for bound motion in the $(E,L)$ plane are shown in Fig.~\ref{Fig:34hELp}. As $\alpha_1$ increases from $1.07317$ to $1.09859$, the admissible angular-momentum range expands, whereas the allowed energy range becomes narrower. These quantitative changes modify the size of the bound-orbit domain but do not change its topology: the allowed region remains single-connected, consistent with a single branch of periodic motion.

\begin{figure}[htbp]
	\center{
	\subfigure[$\alpha_1=1.08$]{\label{4hEq4}
	\includegraphics[width=5.7cm]{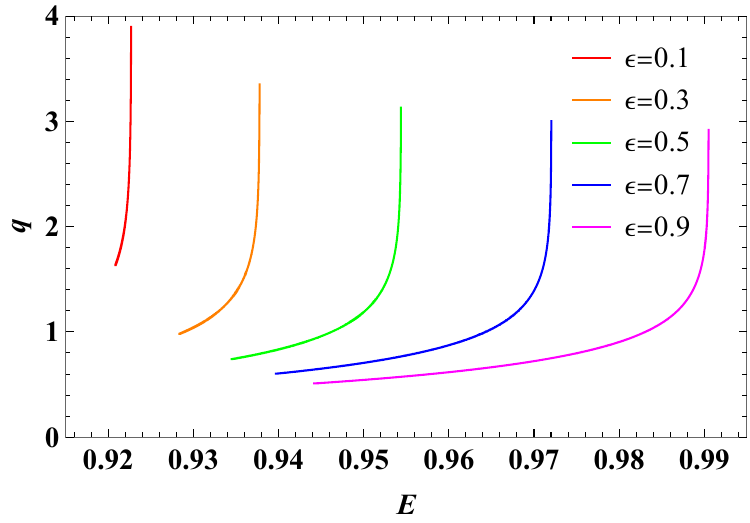}}
	\subfigure[$\alpha_1=1.08$]{\label{4hLq4}
	\includegraphics[width=5.7cm]{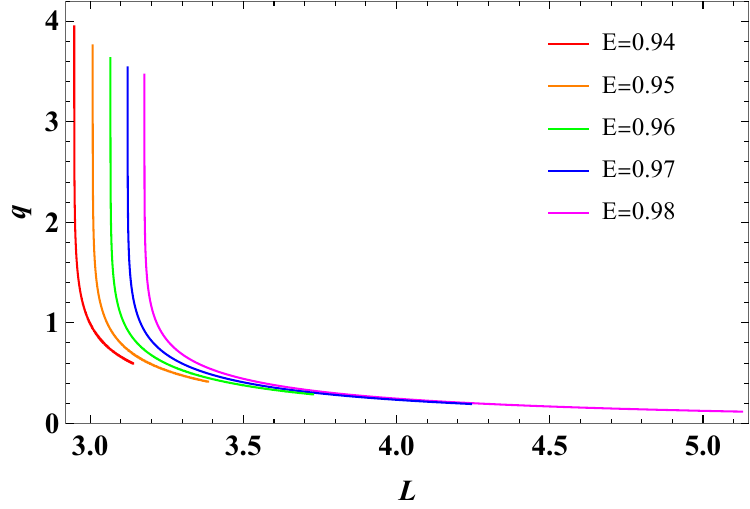}}
	}
\caption{The rational number q as a function of the energy  and angular momentum  of
periodic orbits around dyonic black holes.}
\label{Fig:34hELq}
\end{figure}

We then compute the periodic-orbit parameter $q$. Figure.~\ref{Fig:34hELq} shows the dependence of $q$ on the energy $E$ for fixed angular momentum parameter $\epsilon$, and on the angular momentum $L$ for fixed energy. For fixed $\epsilon$, $q$ increases with $E$ and diverges as the orbit approaches the separatrix. For fixed $E$, $q$ decreases monotonically with increasing $L$ and diverges near the lower angular-momentum boundary. This behavior is characteristic of a single-well effective potential and yields only one periodic-orbit branch.

Representative values of $E$ and $L$ for selected rational numbers $q=q(z,w,v)$ are listed in Tab.~\ref{Tab4}. For fixed $\epsilon=0.5$, the energy associated with a given $q$ increases slightly with $\alpha_1$. Similarly, for fixed $E=0.98$, the corresponding angular momentum also increases with $\alpha_1$. Since each rational number $q$ corresponds to a unique admissible energy or angular momentum, no multi-branch degeneracy appears in this regime.
\begin{table*}[htbp]
\caption{The orbital energy $E$ and angular momentum $L$ for trajectories
with different $(z, w, v)$ configurations. The first three rows show the values of $E$ for $\alpha_1 = 1.07317$, $1.08$, and $1.09859$,
respectively. The last three rows show the values of $L$ at $E = 0.98$ for
the same $\alpha_1$. $L_{\alpha_1=1.07317}=3.027183930$, $L_{\alpha_1=1.08}=3.033956704$ and  $L_{\alpha_1=1.09859}=3.051567153$ with $\epsilon=0.5$.}
\label{Tab4}
\centering
\scriptsize
\resizebox{\textwidth}{!}{
\begin{tabular}{cccccccccccc}
\toprule[0.5pt]\toprule[0.5pt]
$\epsilon$   &  $E_{(1,1,0)}$    & $E_{(1,2,0)}$   & $E_{(1,3,0)}$  & $E_{(2,1,1)}$  &  $E_{(2,2,1)}$  & $E_{(3,1,2)}$  & $E_{(3,2,2)}$  & $E_{(4,0,3)}$ & $E_{(4,1,3)}$ \\
\midrule[0.5pt]
0.5  & 0.945878325   & 0.953847528 & 0.954177589 & 0.952542213 & 0.954119707 & 0.953216520 & 0.954149305 & 0.934681118 & 0.953440466 \\
0.5  & 0.946215325   & 0.954054116 & 0.954373572 & 0.952779068 & 0.954317989 & 0.953438136 & 0.954346452 & 0.935131614 & 0.953657415 \\
0.5  & 0.947065086   & 0.954582061 & 0.954875166 & 0.953378385 & 0.954824890 & 0.954003376 & 0.954850770 & 0.936265965 & 0.954210483 \\
\toprule[0.5pt]\toprule[0.5pt]
$E$  &   $L_{(1,1,0)}$    & $L_{(1,2,0)}$  & $L_{(1,3,0)}$  & $L_{(2,1,1)}$  &  $L_{(2,2,1)}$  & $L_{(3,1,2)}$  & $L_{(3,2,2)}$  & $L_{(4,0,3)}$ & $L_{(4,1,3)}$ \\
\midrule[0.5pt]
0.98 & 3.211221386 & 3.171898546 & 3.170540253 & 3.177923360 & 3.170755536 & 3.174749562 & 3.170642860 & 3.269318098 & 3.173709928 \\
0.98 & 3.217062611 & 3.178348135 & 3.177050180 & 3.184270619 & 3.177253951 & 3.181149616 & 3.177147939 & 3.274821300 & 3.180129621 \\
0.98 & 3.232663461 & 3.195143399 & 3.193950308 & 3.200784718 & 3.194138008 & 3.197783780 & 3.194038503 & 3.289379631 & 3.196824862 \\
\bottomrule[0.5pt] \bottomrule[0.5pt]
\end{tabular}}
\end{table*}

In summary, the three-/four-horizon regime provides an important counterexample to any simple connection between the number of horizons and the number of periodic-orbit branches. Although the spacetime possesses several horizons, the exterior metric function remains monotonic, and the effective potential contains only a single well. Consequently, each rational number $q$ labels a unique periodic-orbit, and no multi-branch structure arises.

\subsubsection{One and Two-Horizon case B: $1.09859< \alpha_1 \leq 43.31447$}\label{12horizonsb}

We next consider another monotonic-exterior regime, namely
$1.09859<\alpha_1\leq43.31447$. This interval contains the second two-horizon branch and terminates at a degenerate one-horizon configuration. Although the horizon radius varies non-monotonically with $\alpha_1$, the metric function $f(r)$ remains monotonic outside the outer event horizon. Therefore, this case provides a useful contrast with the one-/two-horizon non-monotonic regime discussed in Subsec.~\ref{Multibranch}.

\begin{figure}[htbp]
\center{
\includegraphics[width=7cm]{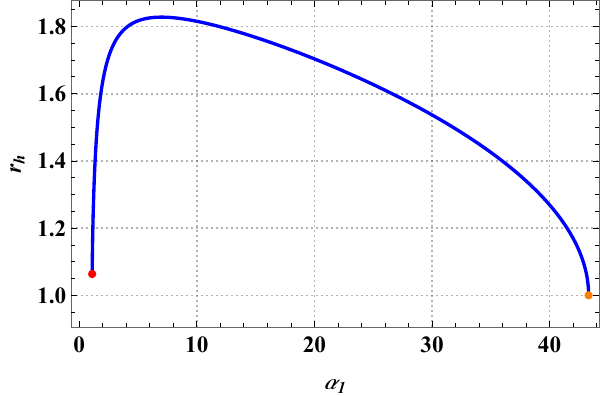}}
\caption{
Outer event horizon radius $r_h$ as a function of $\alpha_1$ for case B with fixed $\alpha_2=2.76$, $P=0.15$, and $Q=1.05$.
The red point denotes the critical three-horizon configuration at $\alpha_1=1.09859$, the blue curve corresponds to the two-horizon branch $1.09859<\alpha_1<43.31447$, and the orange point denotes the degenerate one-horizon endpoint at $\alpha_1=43.31447$.
}\label{123horizons}
\end{figure}

Figure.~\ref{123horizons} shows the outer event horizon radius $r_h$ as a function of $\alpha_1$ in case B. The red point at $\alpha_1=1.09859$ denotes the critical three-horizon configuration. For $1.09859<\alpha_1<43.31447$, the blue curve represents the two-horizon black hole branch. The orange point at $\alpha_1=43.31447$ denotes the degenerate one-horizon endpoint. Along the two-horizon branch, $r_h$ first increases, reaches a maximum near $\alpha_1=6.995555544$, and then decreases toward the one-horizon endpoint. This non-monotonic variation of the event horizon radius reflects changes in the causal structure, but it does not imply a multi-well effective potential because $f(r)$ remains monotonic outside the outer horizon throughout this parameter range.

\begin{figure}[htbp]
	\center{
	\subfigure[$r_{\mathrm{MBO}}$ vs $\alpha_1$]{\label{2hbLMBO2b}
	\includegraphics[width=5.7cm]{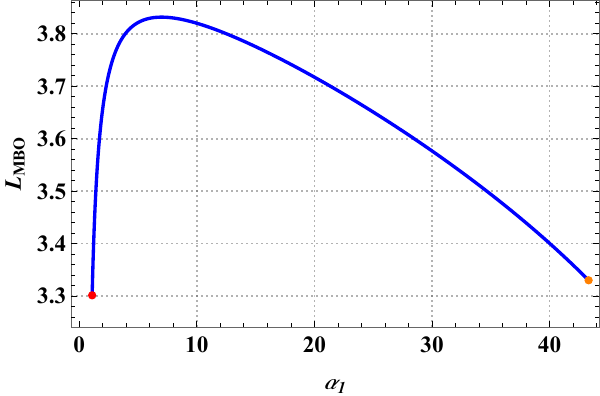}}
	\subfigure[$L_{\mathrm{MBO}}$ vs $\alpha_1$ ]{\label{2hbrMBO2b}
	\includegraphics[width=5.7cm]{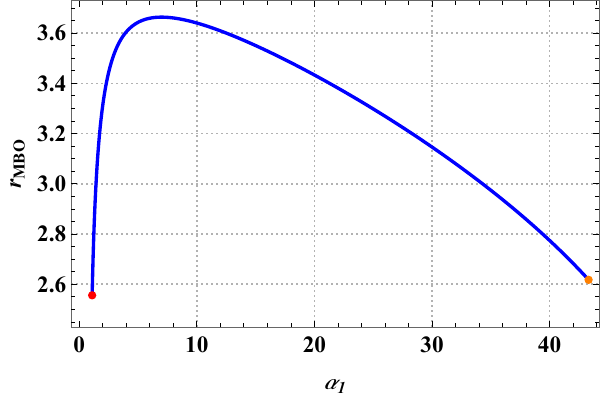}}
	}
\caption{The radius and angular momentum for the MBOs for 1-/2-/3- horizon.
Here $1.09859 \leq \alpha_1 \leq 43.31447$. }
\label{Fig:123hrLmbo}
\end{figure}

\begin{figure*}[htbp]
	\center{
	\subfigure[$r_{\mathrm{ISCO}}$ vs $\alpha_1$ ]{\label{rISCO2b}
	\includegraphics[width=5.7cm]{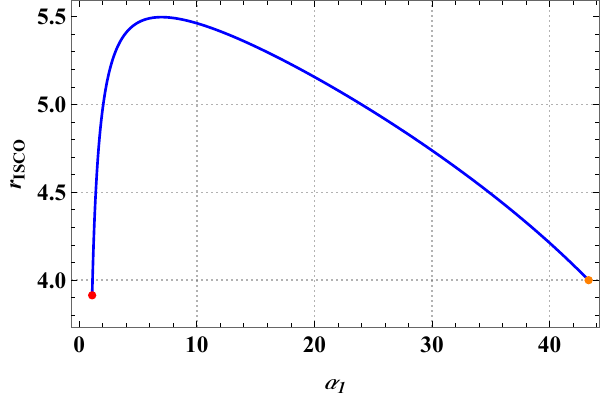}}
	\subfigure[$L_{\mathrm{ISCO}}$ vs $\alpha_1$]{\label{LISCO2b}
	\includegraphics[width=5.7cm]{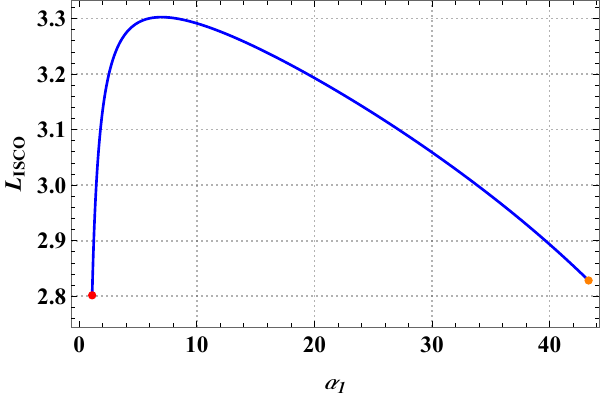}}
	\subfigure[$E_{\mathrm{ISCO}}$ vs $\alpha_1$]{\label{EISCO2b}
	\includegraphics[width=5.7cm]{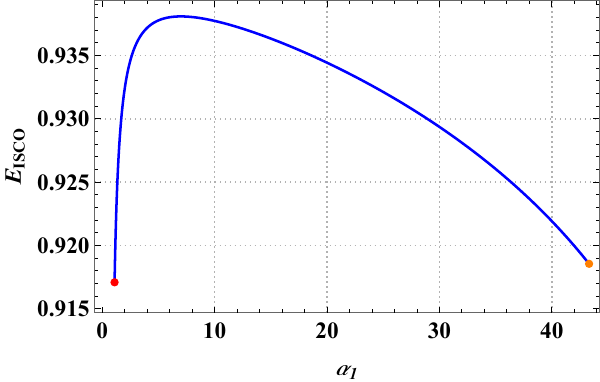}}
	}
\caption{The radius, energy, and angular momentum  for the ISCO with  $1.09859 \leq \alpha_1 \leq 43.31447$.}
\label{Fig:123hrLEriso}
\end{figure*}

The corresponding MBO quantities are shown in Fig.~\ref{Fig:123hrLmbo}. Both $r_{\mathrm{MBO}}$ and $L_{\mathrm{MBO}}$ follow the same qualitative non-monotonic trend as the horizon radius. A similar behavior appears for the ISCO quantities, as shown in Fig.~\ref{Fig:123hrLEriso}. The radius $r_{\mathrm{ISCO}}$, angular momentum $L_{\mathrm{ISCO}}$, and energy $E_{\mathrm{ISCO}}$ first increase and then decrease as $\alpha_1$ varies across this interval.

These variations indicate that the characteristic scales of bound motion are sensitive to the black hole parameters. However, they do not change the topology of the effective potential. Since $f(r)$ is monotonic outside the horizon, the radial potential retains a single-well structure throughout this parameter range.

\begin{figure}[htbp]
	\center{
	\subfigure[$r_{\mathrm{MBO}}$ vs $\alpha_1$]{\label{2hbELl}
	\includegraphics[width=5.7cm]{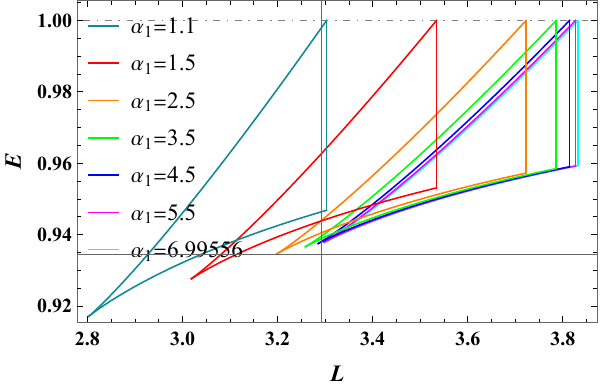}}
	\subfigure[$L_{\mathrm{MBO}}$ vs $\alpha_1$ ]{\label{2hbELr}
	\includegraphics[width=5.7cm]{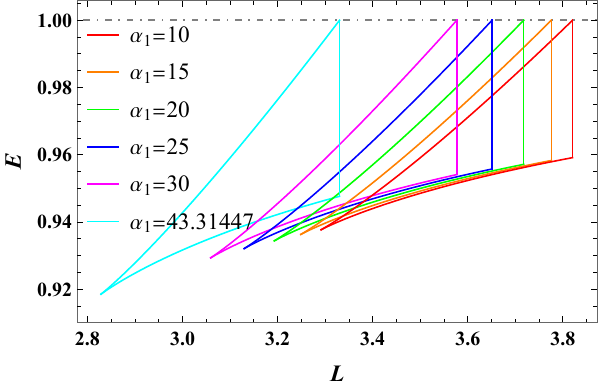}}
	}
\caption{Allowed regions in the $(L, E)$ plane for the bound orbits around dyonic black holes for different values of $\alpha_1$.}
\label{Fig:123hELp}
\end{figure}

The allowed regions for bound motion in the $(E,L)$ plane are shown in Fig.~\ref{Fig:123hELp}. For $1.1\leq\alpha_1\leq6.99556$, the allowed region contracts as $\alpha_1$ increases, whereas for $10\leq\alpha_1\leq43.31447$ it expands again. This behavior follows the non-monotonic variation of the characteristic circular-orbit scales. Nevertheless, the allowed region remains single-connected, reflecting the single-well nature of the effective potential.

Representative periodic-orbit parameters are listed in Tab.~\ref{Tab5} for $\alpha_1=1.5$ and $\alpha_1=10$. For both fixed $\epsilon=0.5$ and fixed $E=0.98$, each rational number $q=q(z,w,v)$ corresponds to a unique admissible value of $E$ or $L$. Thus, despite the non-monotonic dependence of the horizon radius, MBO, and ISCO on $\alpha_1$, periodic motion remains single-branched in this regime. The resulting trajectories are qualitatively similar to standard single-well periodic orbits and are therefore not shown separately.
\begin{table*}[htbp]
\caption{Orbital energy $E$ and specific angular momentum $L$ for trajectories with different $(z, w, v)$ configurations
under a fixed parameter. The cases $\alpha_1 = 1.5$ and
$\alpha_1 = 10$, corresponding to four-horizon black holes, are
listed in the first and second rows, respectively.
$L_{\alpha_1=1.5}=3.276343756$ and $L_{\alpha_1=10}=3.555852800$ with $\epsilon=0.5$.} \label{Tab5}
\centering
\scriptsize
\resizebox{\textwidth}{!}{
\begin{tabular}{cccccccccccc}
\toprule[0.5pt]\toprule[0.5pt]
$\epsilon$   &  $E_{(1,1,0)}$    & $E_{(1,2,0)}$   & $E_{(1,3,0)}$  & $E_{(2,1,1)}$  &  $E_{(2,2,1)}$  & $E_{(3,1,2)}$  & $E_{(3,2,2)}$  & $E_{(4,0,3)}$ & $E_{(4,1,3)}$ \\
\midrule[0.5pt]
0.5  & 0.955410041 & 0.960297323 & \xmark & 0.959623897 & 0.960411281 & 0.959987512 & 0.960421785 & 0.947327593 & 0.960102089 \\
\midrule[0.5pt]
0.5  & 0.962210601 & 0.965660402 & \xmark & 0.965228841 & 0.965725559 & 0.965466948 & \xmark & 0.955976069 & 0.965540382 \\
\toprule[0.5pt]\toprule[0.5pt]
$E$  &   $L_{(1,1,0)}$    & $L_{(1,2,0)}$  & $L_{(1,3,0)}$  & $L_{(2,1,1)}$  &  $L_{(2,2,1)}$  & $L_{(3,1,2)}$  & $L_{(3,2,2)}$  & $L_{(4,0,3)}$ & $L_{(4,1,3)}$ \\
\midrule[0.5pt]
0.98 & 3.435789732 & 3.408150797 & 3.407547608 & 3.411616491 & 3.407624314 & 3.409710362 & 3.415311555 & 3.483832955 & 3.409103862  \\
\midrule[0.5pt]
0.98 & 3.693518265 & 3.670467517 & \xmark      & 3.673110376 & 3.670113317 & 3.671590231 & 3.670087012 & 3.736994345 & 3.671159526  \\
\bottomrule[0.5pt] \bottomrule[0.5pt]
\end{tabular}}
\end{table*}

\subsubsection{Summary of single-branch periodic orbits}

The two monotonic-exterior regimes discussed above demonstrate that the number of horizons alone does not determine the multiplicity of periodic-orbit branches. In the three-/four-horizon regime
$1.07317\leq\alpha_1\leq1.09859$, the spacetime may contain several horizons, but $f(r)$ remains monotonic outside the event horizon. In case B,
$1.09859<\alpha_1\leq43.31447$, the horizon radius and circular-orbit scales vary non-monotonically with $\alpha_1$, but the exterior metric function is again monotonic.

In both cases, the effective potential contains only a single bound well. Consequently, the allowed $(E,L)$ region is single-connected, and each rational number $q$ corresponds to only one periodic-orbit. This behavior is qualitatively the same as
in standard Schwarzschild \cite{Levin:2008mq, Lim:2024mkb}, Reissner--Nordstr\"om \cite{Misra:2010pu}, and Kerr-like \cite{Glampedakis:2002ya} single-well geometries.

These results confirm the central conclusion of this section: multi-branch periodic motion is controlled not by the number of horizons, but by the exterior non-monotonicity of the metric function. When $f(r)$ is monotonic outside the event horizon, the system supports only a single branch of periodic orbits, even if the black hole possesses multiple horizons or if the horizon radius varies non-monotonically with the model parameters.

\section{Conclusions and discussions}\label{sec5}
In this work, we systematically investigated periodic orbits of massive test particles in dyonic black hole spacetimes arising from quasi-topological electrodynamics.
By varying the dimensionless coupling parameter $\alpha_1$ within the range
$1.025755\leq\alpha_1\leq43.31447$ along a representative parameter-space slice with fixed $M=1$, $P=0.15$, $Q=1.05$, and $\alpha_2=2.76$, we explored black hole configurations with different horizon structures and analyzed how the exterior geometry governs the properties of bound periodic motion.

Restricting the dynamics to the equatorial plane, we formulated the radial motion through the effective potential $V_{\mathrm{eff}}(r)$ and determined the corresponding MBOs and ISCOs. Periodic orbits were classified using the rational number
$q=q(z,w,v)=w+v/z$ introduced by Levin \textit{et al.}~\cite{Levin:2008mq}, where the integers $(z,w,v)$ characterize the zoom, whirl, and vertex structure of the trajectory. The geodesic equations were then solved numerically to construct the corresponding periodic motions.

A central result of this work is that the topology of periodic orbits is controlled not by the number of horizons, but by the monotonicity of the metric function $f(r)$ outside the event horizon. In the parameter range
$1.025755\leq \alpha_1<1.07317$, the exterior metric function becomes non-monotonic and develops multiple extrema. This produces double-well effective potentials and multiple MBOs, allowing several energetically distinct periodic orbits to coexist for the same rational number $q$. In this regime, bound periodic motion with $E>1$ can also occur. Although the corresponding trajectories differ substantially in radial extent and eccentricity, their winding number
$n=z(1+q)$ remains identical, implying that they belong to the same topological class.

In contrast, for
$1.07317\leq \alpha_1\leq1.09859$
and
$1.09859<\alpha_1\leq43.31447$,
the metric function is monotonic outside the event horizon even though the spacetime may still possess multiple horizons. The effective potential then reduces to a single-well structure, and only one periodic-orbit branch exists for each rational number $q$. This demonstrates that the emergence of multi-branch periodic motion is directly associated with the non-monotonic exterior geometry rather than with the horizon number itself.

From a physical perspective, the non-monotonic geometry considered here may lead to several potentially observable strong-field signatures. Since the extrema of the effective potential correspond to stable and unstable circular geodesics, a non-monotonic metric function outside the horizon naturally implies the existence of multiple circular orbit families and, for null geodesics, multiple photon spheres \cite{Liu:2019rib, Wei:2020rbh}. Consequently, unlike Schwarzschild or Kerr black holes, which possess only a single standard photon ring outside the horizon, these dyonic black holes may support multiple photon ring structures associated with different unstable photon orbits. Such features are determined by the exterior geometry and are therefore independent of the observer location, although resolving them observationally would require sufficiently high angular resolution.

The existence of multiple periodic-orbit branches may also have important implications for gravitational-wave astronomy. In particular, different branches sharing the same rational number $q$ are topologically equivalent but geometrically distinct, leading to different radial amplitudes and eccentricities. As recently discussed in Ref.~\cite{Wang:2026ncv}, such branch-dependent orbital structures can generate distinguishable gravitational wave signals in EMRI systems, providing a possible observational probe of non-monotonic black hole geometries beyond Kerr. In this sense, periodic orbits may serve as a direct bridge connecting the global structure of spacetime, photon-ring phenomenology, and strong-field gravitational wave observations.
We have treated the dyonic black hole as a fixed background and focused on test-particle geodesic dynamics. At this level, the multi-branch periodic-orbit structure is governed by the exterior effective potential, or equivalently by the radial behavior of the metric function, and is not directly controlled by horizon thermodynamic quantities such as temperature and entropy. Thermodynamic analyses of related dyonic black holes \cite{Panahiyan:2018gzv} may nevertheless provide complementary constraints on the admissible parameter space, and their connection with multi-branch orbital dynamics deserves separate investigation.

Finally, although the present analysis focuses on equatorial timelike geodesics, the framework developed here can be extended in several directions. It would be interesting to investigate nonequatorial motion, spinning particles, charged particle dynamics, and the corresponding null geodesic structure in the same spacetime. The relation between multi-branch periodic motion, multiple photon spheres, black hole shadows, and EMRI waveforms may provide further insight into observational signatures of non-monotonic strong-field geometries.

\section*{Acknowledgements}
This work was supported by the National Natural Science Foundation of China (Grants No. 12475055, and No. 12247101), the Basic Research Foundation of Central Universities (Grant No. lzujbky-2024-jdzx06), the Natural Science Foundation of Gansu Province (Grant No. 22JR5RA389), and the `111 Center' under Grant No. B20063. Chao-Hui Wang was supported by Gansu Provincial Department of Education: 2026 Graduate Innovation Star Project "2026CXZX-154". Yu-Peng Zhang was supported by the ``Talent Scientific Fund of Lanzhou University''. Tao Zhu is supported by the National Natural Science Foundation of China under Grants No.~12275238, the Zhejiang Provincial Natural Science Foundation of China under Grants No.~LR21A050001 and No.~LY20A050002, the National Key Research and Development Program of China under Grant No. 2020YFC2201503, and the Fundamental Research Funds for the Provincial Universities of Zhejiang in China under Grant No.~RF-A2019015.


\begin{thebibliography}{100}

\bibitem{LIGOScientific:2016aoc} B.~P.~Abbott \textit{et al.} [LIGO Scientific and Virgo],
{\it{Observation of Gravitational Waves from a Binary Black Hole Merger}},
\href{https://journals.aps.org/prl/abstract/10.1103/PhysRevLett.116.061102}{Phys. Rev. Lett. \textbf{116}, 061102 (2016)},
[\href{https://journals.aps.org/prl/abstract/10.1103/PhysRevLett.116.061102}{arXiv:1602.03837 [gr-qc]}].

\bibitem{LIGOScientific:2016sjg} B.~P.~Abbott \textit{et al.} [LIGO Scientific and Virgo],
{\it{GW151226: Observation of Gravitational Waves from a 22-Solar-Mass Binary Black Hole Coalescence}},
\href{https://journals.aps.org/prl/abstract/10.1103/PhysRevLett.116.241103}{Phys. Rev. Lett. \textbf{116}, 241103 (2016)},
[\href{https://arxiv.org/abs/1706.01812}{arXiv:1606.04855 [gr-qc]}].

\bibitem{EventHorizonTelescope:2019dse} K.~Akiyama \textit{et al.} [Event Horizon Telescope],
{\it{First M87 Event Horizon Telescope Results. I. The Shadow of the Supermassive Black Hole}},
\href{https://iopscience.iop.org/article/10.3847/2041-8213/ab0ec7}{ Astrophys. J. Lett. \textbf{875}, L1 (2019)},
[\href{https://arxiv.org/abs/1906.11238}{ arXiv:1906.11238 [astro-ph.GA]}].

\bibitem{EventHorizonTelescope:2022wkp} K.~Akiyama \textit{et al.} [Event Horizon Telescope],
{\it{First Sagittarius A* Event Horizon Telescope Results. I. The Shadow of the Supermassive Black Hole in the Center of the Milky Way}},
\href{https://iopscience.iop.org/article/10.3847/2041-8213/ac6674 }{ Astrophys. J. Lett. \textbf{930},  L12 (2022)},
[\href{https://arxiv.org/abs/2311.08680 }{ arXiv:2311.08680 [astro-ph.HE]}].


\bibitem{Boyer:1966qh}
R.~H.~Boyer and R.~W.~Lindquist,
{\it{Maximal analytic extension of the Kerr metric}},
\href{https://pubs.aip.org/aip/jmp/article/8/2/265/233738/Maximal-Analytic-Extension-of-the-Kerr-Metric}{J. Math. Phys. \textbf{8}, 265 (1967)}.

\bibitem{Carter:1968ks}
B.~Carter,
{\it{Hamilton-Jacobi and Schrodinger separable solutions of Einstein's equations}},
\href{https://link.springer.com/article/10.1007/BF03399503}{Commun. Math. Phys. \textbf{10},  280-310 (1968)}.

\bibitem{Bardeen:1972fi}
J.~M.~Bardeen, W.~H.~Press, and S.~A.~Teukolsky,
{\it{Rotating black holes: Locally nonrotating frames, energy extraction, and scalar synchrotron radiation}},
\href{https://ui.adsabs.harvard.edu/abs/1972ApJ...178..347B/abstract}{Astrophys. J. \textbf{178}, 347 (1972)}.

\bibitem{Schmidt:2002qk}
W.~Schmidt,
{\it{Celestial mechanics in Kerr space-time}},
\href{https://iopscience.iop.org/article/10.1088/0264-9381/19/10/314}{Class. Quant. Grav. \textbf{19}, 2743 (2002)},
[\href{https://arxiv.org/abs/gr-qc/0202090}{arXiv:gr-qc/0202090 [gr-qc]}].

\bibitem{Wilkins:1972rs}
D.~C.~Wilkins,
{\it{Bound Geodesics in the Kerr Metric}},
\href{https://journals.aps.org/prd/abstract/10.1103/PhysRevD.5.814}{Phys. Rev. D \textbf{5}, 814-822 (1972)}.

\bibitem{Fujita:2009bp}
R.~Fujita and W.~Hikida,
{\it{Analytical solutions of bound timelike geodesic orbits in Kerr spacetime}},
\href{https://iopscience.iop.org/article/10.1088/0264-9381/26/13/135002}{Class. Quant. Grav. \textbf{26}, 135002 (2009)},
[\href{https://arxiv.org/abs/0906.1420}{arXiv:0906.1420 [gr-qc]}].

\bibitem{Momynov:2024bhn}
S.~Momynov, K.~Boshkayev, H.~Quevedo, F.~Belissarova, A.~Dalelkhankyzy, A.~Taukenova, A.~Urazalina, and D.~Utepova,
{\it{Gravitational capture cross-section in Zipoy-Voorhees spacetimes}},
\href{https://link.springer.com/article/10.1007/s10714-025-03426-w}{Gen. Rel. Grav. \textbf{57},  91 (2025)},
[\href{https://arxiv.org/abs/2412.06598}{arXiv:2412.06598 [gr-qc]}].

\bibitem{Boshkayev:2021chc}
K.~Boshkayev, T.~Konysbayev, E.~Kurmanov, O.~Luongo, D.~Malafarina, and H.~Quevedo,
{\it{Luminosity of accretion disks in compact objects with a quadrupole}},
\href{https://journals.aps.org/prd/abstract/10.1103/PhysRevD.104.084009}{Phys. Rev. D \textbf{104}, 084009 (2021)},
[\href{https://arxiv.org/abs/2106.04932}{arXiv:2106.04932 [gr-qc]}].

\bibitem{Healy:2009zm}
J.~Healy, J.~Levin, and D.~Shoemaker,
{\it{Zoom-Whirl Orbits in Black Hole Binaries}},
\href{https://journals.aps.org/prl/abstract/10.1103/PhysRevLett.103.131101}{Phys. Rev. Lett. \textbf{103}, 131101 (2009)},
[\href{https://arxiv.org/abs/0907.0671}{arXiv:0907.0671 [gr-qc]}].


\bibitem{Misner:1973prb}
C.~W.~Misner, K.~S.~Thorne, and J.~A.~Wheeler,
{\it{Gravitation}},
W. H. Freeman, 1973,
ISBN 978-0-7167-0344-0, 978-0-691-17779-3.

\bibitem{Chandrasekhar:1984siy}
S.~Chandrasekhar,
{\it{The Mathematical Theory of Black Holes}},
Oxford University Press, New York, (1992).

\bibitem{Pugliese:2010ps}
D.~Pugliese, H.~Quevedo, and R.~Ruffini,
{\it{Circular motion of neutral test particles in Reissner-Nordstr{\"o}m spacetime}},
\href{https://journals.aps.org/prd/abstract/10.1103/PhysRevD.83.024021}{Phys. Rev. D \textbf{83}, 024021 (2011)},
[\href{https://arxiv.org/abs/1012.5411}{arXiv:1012.5411 [astro-ph.HE]}].

\bibitem{Pugliese:2011py}
D.~Pugliese, H.~Quevedo, and R.~Ruffini,
{\it{Motion of charged test particles in Reissner-Nordstr{\"o}m spacetime}},
\href{https://journals.aps.org/prd/abstract/10.1103/PhysRevD.83.104052}{Phys. Rev. D \textbf{83}, 104052 (2011)},
[\href{https://arxiv.org/abs/1103.1807}{arXiv:1103.1807 [gr-qc]}].

\bibitem{Pugliese:2011xn}
D.~Pugliese, H.~Quevedo, and R.~Ruffini,
{\it{Equatorial circular motion in Kerr spacetime}},
\href{https://journals.aps.org/prd/abstract/10.1103/PhysRevD.84.044030}{Phys. Rev. D \textbf{84}, 044030 (2011)},
[\href{https://arxiv.org/abs/1105.2959}{arXiv:1105.2959 [gr-qc]}].

\bibitem{Pugliese:2013zma}
D.~Pugliese, H.~Quevedo, and R.~Ruffini,
{\it{Equatorial circular orbits of neutral test particles in the Kerr-Newman spacetime}},
\href{https://journals.aps.org/prd/abstract/10.1103/PhysRevD.88.024042}{Phys. Rev. D \textbf{88},  024042 (2013)},
[\href{https://arxiv.org/abs/1303.6250}{arXiv:1303.6250 [gr-qc]}].

\bibitem{Grandclement:2014msa} P.~Grandclement, C.~Som\'e, and E.~Gourgoulhon,
{\it{Models of rotating boson stars and geodesics around them: new type of orbits}},
\href{https://journals.aps.org/prd/abstract/10.1103/PhysRevD.90.024068 }{ Phys. Rev. D \textbf{90}, 024068 (2014)},
[\href{ https://arxiv.org/abs/1405.4837}{ arXiv:1405.4837 [gr-qc]}].

\bibitem{Grould:2017rzz} M.~Grould, Z.~Meliani, F.~H.~Vincent, P.~Grandcl\'ement, and E.~Gourgoulhon,
{\it{Comparing timelike geodesics around a Kerr black hole and a boson star}},
\href{ https://iopscience.iop.org/article/10.1088/1361-6382/aa8d39}{ Class. Quant. Grav. \textbf{34}, 215007 (2017)},
[\href{https://arxiv.org/abs/1709.05938}{ arXiv:1709.05938 [astro-ph.HE]}].

\bibitem{Teodoro:2020kok} M.~C.~Teodoro, L.~G.~Collodel, and J.~Kunz,
{\it{Retrograde Polish Doughnuts around Boson Stars}},
\href{ https://iopscience.iop.org/article/10.1088/1475-7516/2021/03/063}{ JCAP \textbf{03}, 063 (2021)},
[\href{https://arxiv.org/abs/2011.10288 }{ arXiv:2011.10288 [gr-qc]}].

\bibitem{Muller:2008zza} T.~Muller,
{\it{Exact geometric optics in a Morris-Thorne wormhole spacetime}},
\href{https://journals.aps.org/prd/abstract/10.1103/PhysRevD.77.044043}{ Phys. Rev. D \textbf{77}, 044043 (2008)}.

\bibitem{Teodoro:2021ezj} M.~C.~Teodoro, L.~G.~Collodel, D.~Doneva, J.~Kunz, P.~Nedkova, and S.~Yazadjiev,
{\it{Thick toroidal configurations around scalarized Kerr black holes}},
\href{https://journals.aps.org/prd/abstract/10.1103/PhysRevD.104.124047}{ Phys. Rev. D \textbf{104},  124047 (2021)},
[\href{https://arxiv.org/abs/2108.08640}{ arXiv:2108.08640 [gr-qc]}].

\bibitem{Cunha:2015yba} P.~V.~P.~Cunha, C.~A.~R.~Herdeiro, E.~Radu, and H.~F.~Runarsson,
{\it{Shadows of Kerr black holes with scalar hair}},
\href{https://journals.aps.org/prl/abstract/10.1103/PhysRevLett.115.211102}{ Phys. Rev. Lett. \textbf{115},  211102 (2015)},
[\href{https://arxiv.org/abs/1509.00021}{ arXiv:1509.00021 [gr-qc]}].

\bibitem{Gibbons:1999uv} G.~W.~Gibbons and C.~A.~R.~Herdeiro,
{\it{Supersymmetric rotating black holes and causality violation}},
\href{https://iopscience.iop.org/article/10.1088/0264-9381/16/11/311}{ Class. Quant. Grav. \textbf{16}, 3619-3652 (1999)},
[\href{https://arxiv.org/abs/hep-th/9906098 }{ arXiv:hep-th/9906098 [hep-th]}].

\bibitem{Herdeiro:2000ap} C.~A.~R.~Herdeiro,
{\it{Special properties of five-dimensional BPS rotating black holes}},
\href{https://www.sciencedirect.com/science/article/pii/S0550321300003357?via\%3Dihub }{ Nucl. Phys. B \textbf{582}, 363-392 (2000)},
[\href{ https://arxiv.org/abs/hep-th/0003063}{ arXiv:hep-th/0003063 [hep-th]}].

\bibitem{Diemer:2013fza} V.~Diemer and J.~Kunz,
{\it{Supersymmetric rotating black hole spacetime tested by geodesics}},
\href{ https://journals.aps.org/prd/abstract/10.1103/PhysRevD.89.084001}{ Phys. Rev. D \textbf{89},  084001 (2014)},
[\href{https://arxiv.org/abs/1312.6540 }{ arXiv:1312.6540 [gr-qc]}].

\bibitem{Delgado:2021jxd} J.~F.~M.~Delgado, C.~A.~R.~Herdeiro, and E.~Radu,
{\it{Equatorial timelike circular orbits around generic ultracompact objects}},
\href{https://journals.aps.org/prd/abstract/10.1103/PhysRevD.105.064026 }{ Phys. Rev. D \textbf{105},  064026 (2022)},
[\href{ https://arxiv.org/abs/2107.03404}{ arXiv:2107.03404 [gr-qc]}].

\bibitem{Lehebel:2022yyz} A.~Leh\'ebel and V.~Cardoso,
{\it{Fate of observers in circular motion}},
\href{https://journals.aps.org/prd/abstract/10.1103/PhysRevD.105.064014}{ Phys. Rev. D \textbf{105}, 064014 (2022)},
[\href{https://arxiv.org/abs/2202.08850 }{ arXiv:2202.08850 [gr-qc]}].

\bibitem{Levin:2008mq} J.~Levin and G.~Perez-Giz,
{\it{A Periodic table for Black Hole Orbits}},
\href{https://journals.aps.org/prd/abstract/10.1103/PhysRevD.77.103005}{Phys. Rev. D \textbf{77}, 103005 (2008)},
[\href{https://arxiv.org/abs/0802.0459}{arXiv:0802.0459 [gr-qc]}].

\bibitem{Levin:2008ci} J.~Levin and B.~Grossman,
{\it{Multi-horizon and Critical Behavior in Gravitational Collapse of Massless Scalar Dynamics of Black Hole Pairs. I. Periodic tables}},
\href{https://journals.aps.org/prd/abstract/10.1103/PhysRevD.79.043016}{Phys. Rev. D \textbf{79}, 043016 (2009)},
[\href{https://arxiv.org/abs/0809.3838}{arXiv:0809.3838 [gr-qc]}].

\bibitem{Levin:2008yp} J.~Levin and G.~Perez-Giz,
{\it{Homoclinic Orbits around Spinning black holes. I. Exact Solution for the Kerr Separatrix}},
\href{https://journals.aps.org/prd/abstract/10.1103/PhysRevD.79.124013}{Phys. Rev. D \textbf{79}, 124013 (2009)},
[\href{https://arxiv.org/abs/0811.3814}{arXiv:0811.3814 [gr-qc]}].

\bibitem{Levin:2009sk} J.~Levin,
{\it{Energy Level Diagrams for black hole Orbits}},
\href{https://iopscience.iop.org/article/10.1088/0264-9381/26/23/235010}{Class. Quant. Grav. \textbf{26}, 235010 (2009)},
[\href{https://arxiv.org/abs/0907.5195}{arXiv:0907.5195 [gr-qc]}].

\bibitem{Grossman:2011ps} R.~Grossman, J.~Levin, and G.~Perez-Giz,
{\it{The harmonic structure of generic Kerr orbits}},
\href{https://arxiv.org/abs/1105.5811}{Phys. Rev. D \textbf{85}, 023012 (2012)},
[\href{https://arxiv.org/abs/1105.5811}{arXiv:1105.5811 [gr-qc]}].

\bibitem{Lim:2024mkb} Y.~K.~Lim and Z.~C.~Yeo,
{\it{Energies and angular momenta of periodic Schwarzschild geodesics}},
\href{https://journals.aps.org/prd/abstract/10.1103/PhysRevD.109.024037}{Phys. Rev. D \textbf{109}, 024037 (2024)},
[\href{https://arxiv.org/abs/2401.13894}{arXiv:2401.13894 [gr-qc]}].

\bibitem{Glampedakis:2002ya} K.~Glampedakis and D.~Kennefick,
{\it{Zoom and whirl: Eccentric equatorial orbits around spinning black holes and their evolution under gravitational radiation reaction}},
\href{https://journals.aps.org/prd/abstract/10.1103/PhysRevD.66.044002}{Phys. Rev. D \textbf{66}, 044002 (2002)},
[\href{https://arxiv.org/abs/gr-qc/0203086}{arXiv:gr-qc/0203086 [gr-qc]}].

\bibitem{Grossman:2011im} R.~Grossman, J.~Levin, and G.~Perez-Giz,
{\it{Faster computation of adiabatic extreme mass-ratio inspirals using resonances}},
\href{https://journals.aps.org/prd/abstract/10.1103/PhysRevD.88.023002}{Phys. Rev. D \textbf{88}, 023002 (2013)},
[\href{https://arxiv.org/abs/1108.1819}{arXiv:1108.1819 [gr-qc]}].

\bibitem{Wang:2025hla} C.~H.~Wang, X.~C.~Meng, Y.~P.~Zhang, T.~Zhu, and S.~W.~Wei,
{\it{Equatorial periodic orbits and gravitational waveforms in a black hole free of Cauchy horizon}},
\href{https://iopscience.iop.org/article/10.1088/1475-7516/2025/07/021}{JCAP \textbf{07}  021 (2025)},
[\href{https://arxiv.org/abs/2502.08994}{arXiv:2502.08994 [gr-qc]}].

\bibitem{Misra:2010pu} V.~Misra and J.~Levin,
{\it{Rational Orbits around Charged Black Holes}},
\href{https://journals.aps.org/prd/abstract/10.1103/PhysRevD.82.083001}{Phys. Rev. D \textbf{82}, 083001 (2010)},
[\href{https://arxiv.org/abs/1007.2699}{arXiv:1007.2699 [gr-qc]}].

\bibitem{Zhou:2020zys} T.~Y.~Zhou and Y.~Xie,
{\it{Precessing and periodic motions around a black-bounce/traversable wormhole}},
\href{https://link.springer.com/article/10.1140/epjc/s10052-020-08661-w}{Eur. Phys. J. C \textbf{80}, 1070 (2020)}.

\bibitem{Wei:2019zdf} S.~W.~Wei, J.~Yang, and Y.~X.~Liu,
{\it{Geodesics and periodic orbits in Kehagias-Sfetsos black holes in deformed Ho\u{r}ava-Lifshitz gravity}},
\href{https://journals.aps.org/prd/abstract/10.1103/PhysRevD.99.104016}{Phys. Rev. D \textbf{99}, 104016 (2019)},
[\href{https://arxiv.org/abs/1904.03129}{arXiv:1904.03129 [gr-qc]}].

\bibitem{Jiang:2024cpe} H.~Jiang, M.~Alloqulov, Q.~Wu, S.~Shaymatov, and T.~Zhu,
{\it{Periodic orbits and plasma effects on gravitational weak lensing by self-dual black hole in loop quantum gravity}},
\href{https://doi.org/10.1016/j.dark.2024.101627}{Phys. Dark Univ. \textbf{46}, 101627 (2024)}.

\bibitem{Yang:2024lmj} S.~Yang, Y.~P.~Zhang, T.~Zhu, L.~Zhao, and Y.~X.~Liu,
{\it{Gravitational waveforms from periodic orbits around a quantum-corrected black hole}},
\href{https://doi.org/10.1088/1475-7516/2025/01/091}{JCAP \textbf{01}, 091 (2025)},
[\href{https://arxiv.org/abs/2407.00283}{arXiv:2407.00283 [gr-qc]}].

\bibitem{Tu:2023xab} Z.~Y.~Tu, T.~Zhu, and A.~Z.~Wang,
{\it{Periodic orbits and their gravitational wave radiations in a polymer black hole in loop quantum gravity}},
\href{https://journals.aps.org/prd/abstract/10.1103/PhysRevD.108.024035}{Phys. Rev. D \textbf{108}, 2 (2023)},
[\href{https://arxiv.org/abs/2407.00283}{arXiv:2304.14160 [gr-qc]}].

\bibitem{Babar:2017gsg} G.~Z.~Babar, A.~Z.~Babar, and Y.~K.~Lim,
{\it{Periodic orbits around a spherically symmetric naked singularity}},
\href{https://journals.aps.org/prd/abstract/10.1103/PhysRevD.96.084052}{Phys. Rev. D \textbf{96}, 084052 (2017)},
[\href{https://arxiv.org/abs/1710.09581}{arXiv:1710.09581 [gr-qc]}].

\bibitem{Chen:2025aqh} J.~W.~Chen and J.~S.~Yang,
{\it{Timelike periodic orbits and gravitational waveforms around quantum-corrected black holes}},
\href{https://link.springer.com/article/10.1140/epjc/s10052-025-14457-7}{Eur. Phys. J. C \textbf{85}, 726 (2025)},
[\href{https://arxiv.org/abs/2505.02660}{arXiv:2505.02660 [gr-qc]}].

\bibitem{Meng:2025kzx} L.~P.~Meng, Z.~Y.~Xu, and M.~Tang,
{\it{Periodic Orbits and Gravitational Wave Radiation of Black Holes in 4D-EGB gravity}},
\href{https://www.sciencedirect.com/science/article/pii/S0370269326003370?via%3Dihub}{Phys. Lett. B \textbf{877} (2026) 140484},
[\href{https://arxiv.org/abs/2506.05015}{arXiv:2506.05015 [gr-qc]}].

\bibitem{Meng:2024cnq}
L.~P.~Meng, Z.~Y.~Xu, and M.~R.~Tang,
{\it{Bound orbits and gravitational wave radiation around the hairy black hole}},
\href{https://link.springer.com/article/10.1140/epjc/s10052-025-14032-0}{Eur. Phys. J. C \textbf{85},  306 (2025)},
[\href{https://arxiv.org/abs/2411.01858}{arXiv:2411.01858 [gr-qc]}].


\bibitem{Lu:2025cxx}
S.~Lu and T.~Zhu,
{\it{Gravitational radiations from periodic orbits around Einstein-{\AE}ther black holes}},
\href{https://www.sciencedirect.com/science/article/pii/S2212686425003346?via\%3Dihub}{Phys. Dark Univ. \textbf{50}, 102141 (2025)},
[\href{https://arxiv.org/abs/2505.00294}{arXiv:2505.00294 [gr-qc]}].

\bibitem{Alloqulov:2025bxh}
M.~Alloqulov, S.~Shaymatov, B.~Ahmedov, and T.~Zhu,
{\it{Regular black hole{\textquoteright}s impact on the gravitational waveforms from periodic orbits}},
\href{https://link.springer.com/article/10.1140/epjc/s10052-025-15251-1}{Eur. Phys. J. C \textbf{86},  117 (2026)},
[\href{https://arxiv.org/abs/2508.05245}{arXiv:2508.05245 [gr-qc]}].

\bibitem{Deng:2025wzz}
W.~K.~Deng, S.~Long, Q.~Tan, and J.~L.~Jing,
{\it{Gravitational waveforms from periodic orbits around a charged black hole with scalar hair*}},
\href{https://iopscience.iop.org/article/10.1088/1674-1137/ae28eb}{Chin. Phys. \textbf{50},  035103 (2026)},
[\href{https://arxiv.org/abs/2510.24468}{arXiv:2510.24468 [gr-qc]}].

\bibitem{Haroon:2025rzx}
S.~Haroon and T.~Zhu,
{\it{Periodic orbits and their gravitational wave radiations in a black hole with a dark matter halo}},
\href{https://journals.aps.org/prd/abstract/10.1103/ckdt-wtsl}{Phys. Rev. D \textbf{112}, 4 (2025)},
[\href{https://arxiv.org/abs/2502.09171}{arXiv:2502.09171 [gr-qc]}].

\bibitem{Gogoi:2026obd}
D.~J.~Gogoi, J.~Bora, and A.~{\"O}vg{\"u}n,
{\it{Equatorial periodic orbits and gravitational wave signatures in Euler-Heisenberg black holes surrounded by perfect fluid dark matter}},
[\href{https://arxiv.org/abs/2604.11866}{arXiv:2604.11866 [gr-qc]}].


\bibitem{Cunha:2020azh}
P.~V.~P.~Cunha and C.~A.~R.~Herdeiro,
{\it{Stationary black holes and light rings}},
\href{https://journals.aps.org/prl/abstract/10.1103/PhysRevLett.124.181101}{Phys. Rev. Lett. \textbf{124},  181101 (2020)},
[\href{https://arxiv.org/abs/2003.06445}{arXiv:2003.06445 [gr-qc]}].

\bibitem{Wei:2022mzv}
S.~W.~Wei and Y.~X.~Liu,
{\it{Topology of equatorial timelike circular orbits around stationary black holes}},
\href{https://journals.aps.org/prd/abstract/10.1103/PhysRevD.107.064006}{Phys. Rev. D \textbf{107},  064006 (2023)},
[\href{https://arxiv.org/abs/2207.08397}{arXiv:2207.08397 [gr-qc]}].


\bibitem{Gan:2021pwu} Q.~Y.~Gan, P.~Wang, H.~W.~Wu, and H.~T.~Yang,
{\it{Photon spheres and spherical accretion image of a hairy black hole}},
\href{https://journals.aps.org/prd/abstract/10.1103/PhysRevD.104.024003}{Phys. Rev. D \textbf{104}, 024003 (2021)},
[\href{https://arxiv.org/abs/2104.08703}{arXiv:2104.08703 [gr-qc]}].

\bibitem{Gan:2021xdl} Q.~Y.~Gan, P.~Wang, H.~W.~Wu, and H.~T.~Yang,
{\it{Photon ring and observational appearance of a hairy black hole}},
\href{https://journals.aps.org/prd/abstract/10.1103/PhysRevD.104.044049}{Phys. Rev. D \textbf{104}, 044049 (2021)},
[\href{https://arxiv.org/abs/2105.11770}{arXiv:2105.11770 [gr-qc]}].

\bibitem{Guo:2022muy} G.~Z.~Guo, X.~Jiang, P.~Wang, and H.~W.~Wu,
{\it{Gravitational lensing by black holes with multiple photon spheres}},
\href{https://journals.aps.org/prd/abstract/10.1103/PhysRevD.105.124064}{Phys. Rev. D \textbf{105},  124064 (2022)},
[\href{https://arxiv.org/abs/2204.13948}{arXiv:2204.13948 [gr-qc]}].

\bibitem{Zhou:2023yaj} X.~Zhou, S.~B.~Chen, and J.~L.~Jing,
{\it{Geometrically thick equilibrium tori around a dyonic black hole with quasi-topological electromagnetism}},
\href{https://link.springer.com/article/10.1140/epjc/s10052-024-12483-5}{Eur. Phys. J. C \textbf{84},  130 (2024)},
[\href{https://arxiv.org/abs/2307.01996}{arXiv:2307.01996 [gr-qc]}].

\bibitem{Ramadhan:2026ekb}
H.~S.~Ramadhan, M.~F.~Fauzi, D.~A.~Witjaksana, and A.~Sulaksono,
{\it{Effective null geodesics and black hole images in Kruglov nonlinear electrodynamics}},
[\href{https://arxiv.org/abs/2604.22309}{arXiv:2604.22309 [gr-qc]}].

\bibitem{Hui:2023ibl}
S.~Y.~Hui, B.~R.~Mu, and P.~Wang,
{\it{Echoes from charged black holes influenced by quintessence}},
\href{https://www.sciencedirect.com/science/article/pii/S2212686423002297?via\%3Dihub}{Phys. Dark Univ. \textbf{43}, 101396 (2024)},
[\href{https://arxiv.org/abs/2305.11200}{arXiv:2305.11200 [gr-qc]}].


\bibitem{Wang:2026ncv}
C.~H.~Wang, S.~W.~Wei, T.~Zhu, and Y.~X.~Liu,
{\it{Topologically equivalent yet radiatively distinct orbits in EMRI system}},
[\href{https://arxiv.org/abs/2604.13564}{arXiv:2604.13564 [gr-qc]}].




\bibitem{Liu:2019rib} H.~S.~Liu, Z.~F.~Mai, Y.~Z.~Li, and H.~L\"u,
{\it{Quasi-topological Electromagnetism: Dark Energy, Dyonic Black Holes, Stable Photon Spheres and Hidden Electromagnetic Duality}},
\href{https://link.springer.com/article/10.1007/s11433-019-1446-1}{Sci. China Phys. Mech. Astron. \textbf{63}, 240411 (2020)},
[\href{https://arxiv.org/abs/1907.10876}{arXiv:1907.10876 [hep-th]}].


\bibitem{Abishev:2015pqa}
M.~E.~Abishev, K.~A.~Boshkayev, V.~D.~Dzhunushaliev, and V.~D.~Ivashchuk,
{\it{Dilatonic dyon black hole solutions}},
\href{https://iopscience.iop.org/article/10.1088/0264-9381/32/16/165010}{Class. Quant. Grav. \textbf{32}, 165010 (2015)},
[\href{https://arxiv.org/abs/1504.07657}{arXiv:1504.07657 [gr-qc]}].

\bibitem{Abishev:2017hmd}
M.~E.~Abishev, K.~A.~Boshkayev, and V.~D.~Ivashchuk,
{\it{Dilatonic dyon-like black hole solutions in the model with two Abelian gauge fields}},
\href{https://link.springer.com/article/10.1140/epjc/s10052-017-4749-1}{Eur. Phys. J. C \textbf{77}, 180 (2017)},
[\href{https://arxiv.org/abs/1701.02029}{arXiv:1701.02029 [gr-qc]}].

\bibitem{Bronnikov:2026kgl}
K.~A.~Bronnikov, S.~V.~Bolokhov, G.~S.~Nurbakova, and B.~Tynyshbay,
{\it{On gravitating dyonic configurations in nonlinear electrodynamics}},
[\href{https://arxiv.org/abs/2604.23741}{arXiv:2604.23741 [gr-qc]}].


\bibitem{Wei:2020rbh} S.~W.~Wei,
{\it{Topological Charge and Black Hole Photon Spheres}},
\href{https://journals.aps.org/prd/abstract/10.1103/PhysRevD.102.064039}{Phys. Rev. D \textbf{102}, 064039 (2020)},
[\href{https://arxiv.org/abs/2006.02112}{arXiv:2006.02112 [gr-qc]}].


\bibitem{Ivashchuk:2025rnc}
V.~D.~Ivashchuk, U.~S.~Kayumov, A.~N.~Malybayev, and G.~S.~Nurbakova,
{\it{Photon Sphere for a Dilatonic Dyonic Black Hole in a Model with an Abelian Gauge Field and a Scalar Field}},
\href{https://link.springer.com/article/10.1134/S0202289325700471}{Grav. Cosmol. \textbf{31},  591-599 (2025)},
[\href{https://arxiv.org/abs/2603.19505}{arXiv:2603.19505 [gr-qc]}].

\bibitem{Ferreira:2026ldq}
J.~Ferreira, G.~Bozzola, C.~A.~R.~Herdeiro, V.~Paschalidis, and M.~Zilh{\~a}o,
{\it{Electromagnetic duality degeneracy in dynamical black hole mergers}},
[\href{https://arxiv.org/abs/2605.20493}{arXiv:2605.20493 [gr-qc]}].



\bibitem{Wei:2023bgp} S.~W.~Wei, Y.~P.~Zhang, Y.~X.~Liu, and R.~B.~Mann,
{\it{Static spheres around spherically symmetric black hole spacetime}},
\href{https://journals.aps.org/prresearch/abstract/10.1103/PhysRevResearch.5.043050}{Phys. Rev. Res. \textbf{5}, 043050 (2023)},
[\href{https://arxiv.org/abs/2303.06814}{arXiv:2303.06814 [gr-qc]}].

\bibitem{Wei:2023fkn} S.~W.~Wei and Y.~X.~Liu,
{\it{Aschenbach effect and circular orbits in static and spherically symmetric black hole backgrounds}},
\href{https://www.sciencedirect.com/science/article/pii/S221268642300242X?via\%3Dihub}{Phys. Dark Univ. \textbf{43}, 101409 (2024)},
[\href{https://arxiv.org/abs/2308.11883}{arXiv:2308.11883 [gr-qc]}].

\bibitem{Ye:2023gmk} X.~Ye and S.~W.~Wei,
{\it{Distinct topological configurations of equatorial timelike circular orbit for spherically symmetric (hairy) black~holes}},
\href{https://iopscience.iop.org/article/10.1088/1475-7516/2023/07/049}{JCAP \textbf{07}, 049 (2023)},
[\href{https://arxiv.org/abs/2301.04786}{arXiv:2301.04786 [gr-qc]}].

\bibitem{Cardoso:2008bp}
V.~Cardoso, A.~S.~Miranda, E.~Berti, H.~Witek, and V.~T.~Zanchin,
{\it{Geodesic stability, Lyapunov exponents and quasinormal modes}},
\href{https://journals.aps.org/prd/abstract/10.1103/PhysRevD.79.064016}{Phys. Rev. D \textbf{79},  064016 (2009)},
[\href{https://arxiv.org/abs/0812.1806}{arXiv:0812.1806 [hep-th]}].

\bibitem{Diaz-Guerra:2025jip}
D.~D{\'\i}az-Guerra, A.~Rincon, and D.~Rubiera-Garcia,
{\it{Observational appearance and photon rings of non-singular black holes from anisotropic fluids}},
[\href{https://arxiv.org/abs/2511.21183}{arXiv:2511.21183 [gr-qc]}].



\bibitem{Boshkayev:2024zxq}
K.~Boshkayev, G.~Takey, V.~Ivashchuk, A.~Malybayev, G.~Nurbakova and A.~Urazalina,
{\it{Stability analysis of circular geodesics in dyonic dilatonic black hole spacetimes}},
\href{https://www.sciencedirect.com/science/article/pii/S2212686425000573?via\%3Dihub}{Phys. Dark Univ. \textbf{48}, 101862 (2025)},
[\href{https://arxiv.org/abs/2411.15006}{arXiv:2411.15006 [gr-qc]}].


\bibitem{Lei:2020clg} Y.~Q.~Lei, X.~H.~Ge, and C.~Ran,
{\it{Chaos of particle motion near a black hole with quasitopological electromagnetism}},
\href{https://journals.aps.org/prd/abstract/10.1103/PhysRevD.104.046020}{Phys. Rev. D \textbf{104},  046020 (2021)},
[\href{https://arxiv.org/abs/2008.01384}{arXiv:2008.01384 [hep-th]}].

\bibitem{Huang:2021qwe} H.~Huang, M.~Y.~Ou, M.~Y.~Lai, and H.~Lu,
{\it{Echoes from classical black holes}},
\href{https://journals.aps.org/prd/abstract/10.1103/PhysRevD.105.104049}{Phys. Rev. D \textbf{105}, 104049 (2022)},
[\href{https://arxiv.org/abs/2112.14780}{arXiv:2112.14780 [hep-th]}].

\bibitem{Li:2022vcd} M.~D.~Li, H.~M.~Wang, and S.~W.~Wei,
{\it{Triple points and novel phase transitions of dyonic AdS black holes with quasitopological electromagnetism}},
\href{https://journals.aps.org/prd/abstract/10.1103/PhysRevD.105.104048}{Phys. Rev. D \textbf{105},  104048 (2022)},
[\href{https://arxiv.org/abs/2201.09026}{arXiv:2201.09026 [gr-qc]}].

\bibitem{Chen:2024sow} Z.~Q.~Chen and S.~W.~Wei,
{\it{Thermodynamical topology with multiple defect curves for dyonic AdS black holes}},
\href{https://link.springer.com/article/10.1140/epjc/s10052-024-13620-w}{Eur. Phys. J. C \textbf{84}, 1294 (2024)},
[\href{https://arxiv.org/abs/2405.07525}{arXiv:2405.07525 [hep-th]}].


%
\bibitem{Qiao:2024qsf}
C.~K.~Qiao,
{\it{The Existence and Distribution of Photon Spheres Near Spherically Symmetric Black Holes -- A Geometric Analysis}},
\href{https://link.springer.com/article/10.1140/epjc/s10052-025-13912-9}{Eur. Phys. J. C \textbf{85}, 191 (2025)},
[\href{https://arxiv.org/abs/2407.14035}{arXiv:2407.14035 [gr-qc]}].

\bibitem{Qiao:2025ojr}
C.~K.~Qiao, P.~Su, and Y.~Huang,
{\it{A general discussion on photon spheres in different categories of spacetimes}},
\href{https://link.springer.com/article/10.1140/epjc/s10052-025-14435-z}{Eur. Phys. J. C \textbf{85}, 709 (2025)},
[\href{https://arxiv.org/abs/2602.04573}{arXiv:2602.04573 [gr-qc]}].


\bibitem{Panahiyan:2018gzv}
S.~Panahiyan,
{\it{Nonlinearly charged dyonic black holes}},
\href{https://www.sciencedirect.com/science/article/pii/S0550321319303177?via\%3Dihub}{Nucl. Phys. B \textbf{950}, 114831 (2020)},
[\href{https://arxiv.org/abs/1812.07374}{arXiv:1812.07374 [physics.gen-ph]}].


\end{thebibliography}
\end{document}